\documentclass{aastex61}
\usepackage{graphicx}
\shorttitle{Environmental effects on core formation }
\shortauthors{ Yi et al.}

\begin{document}

\title{PLANCK COLD CLUMPS IN THE $\lambda$ ORIONIS COMPLEX. II. \\
    ENVIRONMENTAL EFFECTS ON CORE FORMATION}

\correspondingauthor{Jeong-Eun Lee}
\email{jeongeun.lee@khu.ac.kr}

\author[0000-0002-0786-7307]{Hee-Weon Yi}
\author{Jeong-Eun Lee}
\affil{School of Space Research, Kyung Hee University, 1732, Deogyeong-daero, Giheung-gu, Yongin-si, Gyeonggi-do 17104, Korea }

\author{Tie Liu}
\affiliation{Korea Astronomy and Space Science Institute, 776 Daedeokdaero, Yuseong-gu, Daejeon 34055, Republic of Korea}
\affiliation{East Asian Observatory, 660 N. A'ohoku Place, Hilo, HI 96720, USA}

\author{Kee-Tae Kim}
\affiliation{Korea Astronomy and Space Science Institute, 776 Daedeokdaero, Yuseong-gu, Daejeon 34055, Republic of Korea}

\author{Minho Choi}
\affiliation{Korea Astronomy and Space Science Institute, 776 Daedeokdaero, Yuseong-gu, Daejeon 34055, Republic of Korea}

\author{David Eden}
\affiliation{Astrophysics Research Institute, Liverpool John Moores University, IC2, Liverpool Science Park, 146 Brownlow Hill, Liverpool L3 5RF, UK}

\author{Neal J. Evans II}
\affiliation{Korea Astronomy and Space Science Institute, 776 Daedeokdaero, Yuseong-gu, Daejeon 34055, Republic of Korea}
\affiliation{Department of Astronomy, The University of Texas at Austin, 2515 Speedway, Stop C1400, Austin, TX 78712-1205}
\affiliation{Humanitas College, Global Campus, Kyung Hee University, 1732, Deogyeong-daero, Giheung-gu, Yongin-si, Gyeonggi-do 17104, Korea}

\author{James Di Francesco}
\affiliation{NRC Herzberg Astronomy and Astrophysics, 5071 West Saanich Rd, Victoria, BC V9E 2E7, Canada}
\affiliation{Department of Physics and Astronomy, University of Victoria, Victoria, BC V8P 1A1, Canada}

\author{Gary Fuller}
\affil{UK ALMA Regional Centre Node, Jodrell Bank Centre for Astrophysics, School of Physics and Astronomy, The University of Manchester, Oxford Road, Manchester M13 9PL, UK}

\author{N. Hirano}
\affiliation{Institute of Astronomy and Astrophysics, Academia Sinica. 11F of Astronomy-Mathematics Building, AS/NTU No.1, Sec. 4, Roosevelt Rd, Taipei 10617, Taiwan, R.O.C.}

\author{Mika Juvela}
\affil{Department of Physics, P.O.Box 64, FI-00014, University of Helsinki, Finland}

\author{Sung-ju Kang}
\affiliation{Korea Astronomy and Space Science Institute, 776 Daedeokdaero, Yuseong-gu, Daejeon 34055, Republic of Korea}

\author{Gwanjeong Kim}
\affiliation{Korea Astronomy and Space Science Institute, 776 Daedeokdaero, Yuseong-gu, Daejeon 34055, Republic of Korea}

\author{Patrick M. Koch}
\affiliation{Institute of Astronomy and Astrophysics, Academia Sinica. 11F of Astronomy-Mathematics Building, AS/NTU No.1, Sec. 4, Roosevelt Rd, Taipei 10617, Taiwan, R.O.C.}

\author{Chang Won Lee}
\affiliation{Korea Astronomy and Space Science Institute, 776 Daedeokdaero, Yuseong-gu, Daejeon 34055, Republic of Korea}
\affiliation{University of Science \& Technology, 176 Gajeong-dong, Yuseong-gu, Daejeon, Republic of Korea}

\author{Di Li}
\affiliation{CAS Key Laboratory of FAST, National Astronomical Observatories, Chinese Academy of Sciences, Beijing}
\affiliation{University of Chinese Academy of Sciences, Beijing}

\author{H.-Y. B. Liu}
\affiliation{European Southern Observatory, Karl-Schwarzschild-Str.2, D-85748 Garching bei M\"{u}nchen, Germany}

\author{Hong-Li Liu}
\affiliation{Department of Physics, The Chinese University of Hong Kong, Shatin, NT, Hong Kong SAR}
\affiliation{Departamento de Astronom\'ia, Universidad de Concepci\'on, Av. Esteban Iturra s/n, Distrito Universitario, 160-C, Chile}
\affiliation{Chinese Academy of Sciences South America Center for Astronomy}

\author{Sheng-Yuan Liu}
\affiliation{Institute of Astronomy and Astrophysics, Academia Sinica. 11F of Astronomy-Mathematics Building, AS/NTU No.1, Sec. 4, Roosevelt Rd, Taipei 10617, Taiwan, R.O.C.}

\author{Mark G. Rawlings}
\affiliation{East Asian Observatory, 660 N. A'ohoku Place, Hilo, HI 96720, USA}

\author{I. Ristorcelli}
\affiliation{IRAP, CNRS (UMR5277), Universit\'{e} Paul Sabatier, 9 avenue du Colonel Roche, BP 44346, 31028, Toulouse Cedex 4, France}

\author{Patrico Sanhueza}
\affiliation{National Astronomical Observatory of Japan, National Institutes of Natural Sciences, 2-21-1 Osawa, Mitaka, Tokyo 181-8588, Japan}

\author{Archana Soam}
\affiliation{Korea Astronomy and Space Science Institute, 776 Daedeokdaero, Yuseong-gu, Daejeon 34055, Republic of Korea} 

\author{Ken'ichi Tatematsu}
\affil{National Astronomical Observatory of Japan, National Institutes of Natural Sciences, 2-21-1 Osawa, Mitaka, Tokyo 181-8588, Japan}

\author{Mark Thompson}
\affiliation{Centre for Astrophysics Research, School of Physics Astronomy \& Mathematics, University of Hertfordshire, College Lane, Hatfield, AL10 9AB, UK}

\author{L. V. Toth}
\affil{E\"{o}tv\"{o}s Lor\'{a}nd University, Department of Astronomy, P\'{a}zm\'{a}ny P\'{e}ter s\'{e}t\'{a}ny 1/A, H-1117, Budapest, Hungary}
\affil{Konkoly Observatory, Research Centre for Astronomy and Earth Sciences, HAS, Konkoly-Thege M. ut 15-17, 1121 Budapest, Hungary}

\author{Ke Wang}
\affiliation{European Southern Observatory, Karl-Schwarzschild-Str.2, D-85748 Garching bei M\"{u}nchen, Germany}

\author{Glenn J. White}
\affil{Department of Physics and Astronomy, The Open University, Walton Hall, Milton Keynes, MK7 6AA, UK}
\affil{RAL Space, STFC Rutherford Appleton Laboratory, Chilton, Didcot, Oxfordshire, OX11 0QX, UK}

\author{Yuefang Wu}
\affiliation{Department of Astronomy, Peking University, 100871, Beijing China}

\author{Yao-Lun Yang}
\affiliation{Department of Astronomy, The University of Texas at Austin, 2515 Speedway, Stop C1400, Austin, TX 78712-1205}

\author{the JCMT Large Program``SCOPE'' Collaboration; TRAO Key Science Program``TOP'' Collaboration}



\begin{abstract}
Based on the 850 $\micron$ dust continuum data from SCUBA-2 at James Clerk Maxwell Telescope (JCMT), 
we compare overall properties of Planck Galactic Cold Clumps (PGCCs)
 in the $\lambda$ Orionis cloud to those of PGCCs in the Orion A and B clouds. 
 The Orion A and B clouds are well known active star-forming regions, while the $\lambda$ Orionis 
 cloud has a different environment as a consequence of the interaction with a prominent 
 OB association and a giant H{\sc ii} region. PGCCs in the $\lambda$ Orionis cloud have higher 
 dust temperatures ($T_{\rm d} = 16.13 \pm 0.15 $ K) and lower values of dust emissivity spectral 
 index ($ \beta  = 1.65 \pm 0.02$) than PGCCs in the Orion A ($T_{\rm d} = 13.79 \pm 0.21$ K, $ \beta  = 2.07 \pm 0.03 $) 
 and Orion B ($T_{\rm d} = 13.82 \pm 0.19 $ K, $ \beta  = 1.96 \pm 0.02$) clouds. We find 119 sub-structures
  within the 40 detected PGCCs and identify them as cores. Of total 119 cores, 15 cores are discovered in the 
  $\lambda$ Orionis cloud, while 74 and 30 cores are found in the Orion A and B clouds, respectively. 
  The cores in the $\lambda$ Orionis cloud show much lower mean values of  size R = 0.08 pc, 
  column density $N(\rm H_{2})$ = $(9.5 \pm 1.2) \times 10^{22}$ cm$^{-2}$, number density
  $n(\rm H_{2})$ = $(2.9 \pm 0.4) \times 10^{5}$ cm$^{-3}$, and mass $M_{core}$ = $1.0 \pm 0.3$ M$_{\sun}$
  compared to the cores in the Orion A (R = 0.11 pc, $N(\rm H_{2})$ = $(2.3 \pm 0.3) \times 10^{23}$ cm$^{-2}$, 
  $n(\rm H_{2})$ = $(3.8 \pm 0.5) \times 10^{5}$ cm$^{-3}$, and  $M_{core}$ = $2.4 \pm 0.3$ M$_{\sun}$)
  and Orion B (R = 0.16  pc, $N(\rm H_{2})$ = $(3.8 \pm 0.4) \times 10^{23}$ cm$^{-2}$,  
  $n(\rm H_{2})$ = $(15.6 \pm 1.8) \times 10^{5}$ cm$^{-3}$, and  $M_{core}$= $2.7 \pm 0.3$ M$_{\sun}$) clouds. 
  These core properties in the $\lambda$ Orionis cloud can be attributed to the photodissociation
  and external heating by the nearby H{\sc ii} region, which may prevent the PGCCs from forming 
   gravitationally bound structures and eventually disperse them. 
   These results support the idea of negative stellar feedback on core formation.
\end{abstract}




\section{Introduction} \label{sec:intro}

 Molecular clouds commonly show hierarchical structures, from clumps ($n \sim 10^3 - 10^4$ cm$^{-3}$, 0.3 -- 3 pc) down to dense cores ($n \sim 10^4 - 10^5$ cm$^{-3}$, 0.03 -- 0.2 pc) \citep{Williams00, BT07}. Since stars form via the gravitational collapse of the dense cores, it is important to identify the gravitationally unstable cores to understand the initial conditions of star formation. Nevertheless, the details of the core formation process are still poorly understood because local environmental conditions, such as turbulence, magnetic field, and radiation can significantly affect the process of core formation. 
Therefore, in order to understand the formation of cores, which are formed under these diverse environments, we must perform a statistical study in various environments and in various evolutionary stages.

Orion Molecular Cloud Complex is the largest (extend up to $25\arcdeg$ or 187 pc) and most massive molecular clouds within 500 pc of the Sun. This region is an important laboratory for investigating  core and star formation in a range of environments and hierarchical structures from extended features to isolated objects.
There are three clouds from south to north in the Orion complex; the Orion A and B clouds, and the $\lambda$ Orionis cloud. The $\lambda$ Orionis cloud, a region known as the ``head'' of the Orion complex with a distance 380 $\pm$ 30 pc \citep{ESA1997} and the total molecular mass of $1.4\times10^{4} \, \rm M_{\sun}$ \citep{Lang2000}, has different environments, including local radiation fields and star-forming activity, from the Orion A and B clouds.
At the center of the $\lambda$ Orionis cloud, there is one of the nearest OB associations known as Collinder 69, including at least one O star, $\lambda$ Ori, with the spectral type of O8 III and an age of about 5.5 Myr \citep{CS96, DM01, DM02}. Stars in this association are unbound due to the rapid removal of molecular gas by a supernova explosion that occurred about 1 Myr ago \citep{DM01}. This might have subsequently led to the formation of the dusty and gaseous ring by making one of the nearest large H {\sc ii} regions. The molecular clumps in the $\lambda$ Orionis cloud show clear velocity and temperature gradients \citep{Liu12, Liu16A, Gold16}, hinting at external compression by the H {\sc ii} region. \citet{Liu16A} suggested that star formation in PGCC G192.32-11.88, which is located in the $\lambda$ Orionis cloud, has been greatly suppressed because of stellar feedback.  

 The Orion A and B clouds have been extensively studied and are well known active star-forming regions, which contain thousands of young stellar objects (YSOs) \citep{Megeath12, Polychroni13, Buckle12} with same averaged distances of 420 pc \citep{Sand07, Jeff07} and a total mass greater than $2\times10^{5}  \, M_{\sun}$ \citep{Wilson05}. Supernova explosions and H{\sc ii} regions are not discovered in these two clouds. Therefore, comparisons of core properties in the Orion molecular cloud complex will provide an important opportunity to study the effect of the strong radiation field on the next generation of star formation.  
 
We have been carrying out a legacy survey toward about 1000 Planck Galactic Cold Clumps (PGCCs; \citealt{Planck15}) in the 850 $\micron$ dust continuum, ``SCUBA-2 Continuum Observations of Pre-protostellar Evolution" (SCOPE, \citealt{Liu18A}; Eden et al. in preparation), to investigate physical conditions of the PGCCs with a high angular resolution of 14$\arcsec$.0. The PGCC catalog lists 13,188 Galactic sources over the whole sky, identified by the $\it Planck$ survey, which have lower temperatures (6 -- 20 K) than their surrounding environments and the interstellar medium and thus, provides a wealth of sources which may be in the early stage of star formation. It was built using the Planck data at 353, 545, and 857 GHz combined with the {\it Infrared Astronomical Satellite} (IRAS) data \citep{Planck15}. Thousands of dense cores within the PGCCs have been identified in the ``SCOPE" survey, and most of them are either starless cores or very young protostellar objects (Class 0/I) \citep{Liu16A, Liu18B, Tatematsu17, Kim17, Tang18}. The submillimeter dust continuum images of PGCCs can provide an opportunity to study the formation and evolution of cores because the dust continuum is a good tracer of dense and cold regions. Thus, we can explore how dense cores form and how star formation varies as a function of environment.

We present a detailed comparative statistical study of the 850 $\micron$ continuum data of 96 PGCCs in the three clouds of the Orion complex incorporated with archival data. 
 We compare their overall features using the 850 $\micron$ continuum data and physical properties (e.g. dust temperature and dust opacity spectral index) using the PGCC catalog. Also, we investigate physical properties (size, column density, mass, and number density) of  newly identified cores within detected clumps and discuss the differences among three clouds.
 In Section 2, observations and data are described. In Section 3, we present the results of our analyses. We discuss the environmental effect of the $\lambda$ Orionis cloud and the overall properties of cores in the Orion complex in Section 4. The summary is presented in Section 5.

\section{Data} \label{sec:data}

\subsection{JCMT/SCUBA-2} 
 As part of the legacy survey, SCOPE (proposal code: M16AL003), we observed 58 PGCCs in the ``CV Daisy" mapping mode \citep{Bintley14}, which is suitable for small and compact sources with sizes of less than 3$\arcmin$ at 450 and 850 $\micron$ with the Submillimetre Common-User Bolometer Array 2 (SCUBA-2) at the 15-m JCMT \citep{Holland13}. The observed PGCCs consist of 50 in the $\lambda$ Orionis cloud and 8 in the Orion A and B clouds. Our observations were carried out under weather band 3/4, which maintains 225 GHz opacity ranging from 0.12  to 0.2. This weather condition is not sufficient to obtain 450 $\micron$ continuum data due to poor atmospheric transmission. The 450 $\micron$ continuum emission thus was not detected toward any of the 58 PGCCs. The map size is $\thicksim$ 12$\arcmin \times 12 \arcmin$ and the beam size of SCUBA-2  is 14$\arcsec$.0 at 850 $\micron$ \citep{Holland13}.
The observed continuum data were reduced using an iterative map-making technique, ORAC-DR, in the STARLINK package, developed by the Joint Astronomy Centre. The reduction is tailored to filter out scales larger than 200$\arcsec$ on a 4$\arcsec$ pixel size. We also included archival data of 38 PGCCs located in the Orion A and B clouds from the JCMT Science Archive (JSA) hosted by the Canadian Astronomical Data Centre (CADC). Table~\ref{tab:list_PGCCs} lists information of the 96 PGCCs including coordinates, notes for detection, and rms noise level of each SCUBA-2 image.

\subsection{Auxiliary infrared data}
We included the data from the {\it Wide-Field Infrared Survey Explorer} ({\it WISE }) - ALLWISE catalog \citep{Wright10}, which mapped the whole sky in four mid-infrared bands at 3.4, 4.6, 12, and 22 $\micron$  to study an association of any infrared (IR) source within each PGCC. From the data, embedded protostars were identified and their evolutionary stages were classified as described in Section 3.3.1.

\section{RESULTS} \label{sec:results}

\subsection{SCUBA-2 observational results: Detection and Morphology} 
We obtained the 850 $\micron$ maps of 96 PGCCs in the Orion molecular cloud complex either by our own observations or from the JCMT archive, CADC. Fifty PGCCs are located in the $\lambda$ Orionis cloud, thirty in Orion A and other sixteen in Orion B clouds, respectively. To avoid spurious detections, we adopted 3$\sigma_{rms}$ as the minimum significance required for detection.
 The detected and non-detected PGCCs are marked differently in Figure~\ref{fig:Lambda_cloud} and Figure~\ref{fig:OrionAandB_cloud} as red circles and orange triangles, respectively.  Figure~\ref{fig:Lambda_cloud}  marks the locations of selected PGCCs with H$_{2}$ column densities greater than 5 $ \times 10^{20}$ cm$^{-2}$ (A$_{\rm V}$ $\thicksim$ 0.5) on top of the Planck composite image in the $\lambda$ Orionis cloud.  Figure~\ref{fig:OrionAandB_cloud} provides PGCCs in the Orion A and B clouds.
 We cover all 50 PGCCs with column densities greater than 5$\times$10$^{20}$ cm$ ^{-2}$ in the  $\lambda$ Orionis cloud from the PGCC catalog, while the PGCCs in the Orion A and B clouds are not fully investigated. The 38 out of 46 PGCCs obtained from the archive have column densities greater than 5 $\times$ 10$^{20}$ cm$ ^{-2}$ but not all PGCCs with the column densities greater than 5$\times10^{20}$ cm$ ^{-2}$ were  observed with SCUBA-2. The total numbers of PGCCs with the $\rm H_{2}$ column densities greater than 5$\times$10$^{20}$ cm$ ^{-2}$ are 66 and 68 in the Orion A and B clouds, respectively. Therefore, it is not appropriate to compare detection rates of SCUBA-2 in the three clouds. However, the H$_{2}$ column densities of PGCCs are much higher in the Orion A and B clouds (see Table~\ref{tab:PGCCs}), which means these regions are much denser than the  $\lambda$ Orionis cloud. 

The overall morphology of  PGCCs located in the $\lambda$ Orionis cloud can be characterized as isolated and compact, while most PGCCs in the Orion A and B clouds are extended and have filamentary structures. The 850 $\micron$ dust continuum images also reveal hierarchical structures (see Figure~\ref{fig:Lambda_PGCCs} to Figure~\ref{fig:OrionB_PGCCs}). Most PGCCs fragment into several substructures except for four PGCCs (G192.12-11.10, G215.44-16.38, G201.52-11.08, and G206.12-15.76). We performed source fitting to identify and classify the substructures of PGCCs as will be described in the next section.

\subsection{Source Fitting}
Source fitting was  performed using ClumpFind in the STARLINK package. It is an automatic routine for analyzing clumpy structures. The 5$\sigma_{rms}$ detection threshold was used to identify substructures in PGCCs and to determine their sizes. This process contours the data array at many different levels from the peak, and then follows down to a specific minimum contour level defined by users \citep{WB94}. Any substructure with values above 5$\sigma_{rms}$ is passed to the fitting routine and then the set of pixels is identified as a source.  

Finally, we obtained 119 cores within the 40 detected PGCCs in the Orion complex. In the $\lambda$ Orionis cloud, there are 8 detected PGCCs, from which 15 cores are identified. The Orion A cloud has  74 cores in 23 PGCCs, and the Orion B cloud has 30 cores in 9 PGCCs. The list of cores in each cloud is found in Tables \ref{tab:list_cores_Lambda} to \ref{tab:list_cores_OrionB}.

\subsection{Properties of cores}

\subsubsection{ The H$_{2}$ column density and the total mass}
For analysis, we assumed that the dust continuum emission is optically thin; thus the continuum emission traces the total mass along the line of sight in the dense parts of PGCCs. 
The beam-averaged column density is estimated by the equation: 
\begin{equation}
 N(H_{2}) = \frac{S_{\nu}'}  {\Omega \mu m_{H} \kappa_{\nu} B_{\nu}(\it T_{\rm dust})}
 \end{equation}
 where {$S_{\nu}'$ is the beam-averaged flux density per beam, which is integrated over the solid angle defined by  $\Omega = (\pi {\theta^{2}}_{HPBW}) / (4 ln2)$ for a gaussian aperture, where $\theta_{HPBW}$ is the half-power beam width, $\mu$ is the mean molecular weight, and $m_{H}$ is the mass unit of atomic hydrogen, and $B_{\nu}(\it T_{\rm dust})$ is the Planck function of dust temperature $\it T_{\rm dust}$.  We adopted the dust opacity per gram of gas from Beckwith et al. (1990) as $\kappa_{\nu} = 0.1(\nu/1\rm \, THz)^{\beta} \, cm^{2} \, g^{-1}$ which is appropriate for cores with intermediate and high densities. The $\it T_{\rm dust}$ and the dust emissivity spectral index ${\beta}$ were obtained from the PGCC catalog.

 The mean density of the core was calculated using the beam-averaged hydrogen column density
 \begin{equation}
n_{\rm H_{2}} = \frac{N\rm(H_2)}  {R}
 \end{equation}
where the core diameter R is defined as $R = \sqrt{a\cdot b}$, again a and b are the major and minor axes (full width at half maximum) of the fitting in Sect. 3.2. The calculated parameters for each source are presented in Tables \ref{tab:list_cores_Lambda} to~\ref{tab:list_cores_OrionB}.

We estimated the total mass as $M_{\rm total}= S_{\nu}D^{2} / \kappa_{\nu}B_{\nu}(\it T_{\rm dust})$, where $S_{\nu}$ is the continuum flux density at 850 $\micron$, D is the distance.  
The median values of column density, number density, mass and size of cores in each cloud are summarized in Table~\ref{tab:statistics_cores}. The median value of column density is the highest in the Orion B cloud (38.4 $\times 10^{22}$ cm$^{-2}$) and the lowest in the $\lambda$ Orionis cloud (8.2 $\times 10^{22}$ cm$^{-2}$). The lowest value is found in the $\lambda$ Orionis cloud (2.5$\times 10^{22}$ cm$^{-2}$), which is about 46.4 times lower than the highest value in the Orion A cloud (116.6 $\times 10^{22}$ cm$^{-2}$). The derived core masses range from 0.06 to 12.25 $\rm M_{\odot}$. The core with the minimum mass is located in the $\lambda$ Orionis cloud, while the core with the maximum mass is found in the Orion A cloud. In the median mass  the $\lambda$ Orionis cloud has the lowest value (0.77 $\rm M_{\sun}$) and the Orion B cloud has the highest value (1.81 $\rm M_{\sun}$).

To quantify the differences between cores in the $\lambda$ Orionis cloud and cores in the Orion A or B clouds, we have calculated the probability whether these two distributions arise from the same population using the Kolmogorov-Smirnov (K-S) test. The K-S test is a more robust method for measuring the similarity between two distributions than the comparison of the median and/or mean values. We tested the difference between two distributions under a significance level of 0.05. Two distributions are statistically different when the K-S test gives larger distance values ($D_{n}$) than a critical value and smaller p-values than the significance level ($\alpha = 0.05$). Figure~\ref{fig:K-S} shows the results of the K-S test of core masses (upper panels) and column densities (lower panels) of the $\lambda$ Orionis and Orion A/B clouds in the form of cumulative distribution function. The probability that core masses in the $\lambda$ Orionis and the Orion A clouds  are derived from the same parent population is 0.36\%, and only 0.10\% for the $\lambda$ Orionis and Orion B clouds. The probability of the column density distribution of both the Orion A and B clouds originating from the same parent population as the $\lambda$ Orionis is even lower, at 0.01\% (lower panels of Figure~\ref{fig:K-S}). As the p-values are lower than the significance level ($\alpha = 0.05$), we can conclude that the distributions of core masses and column densities in the $\lambda$ Orionis cloud and in the Orion A/B clouds originate from different parent populations.

\subsubsection{IR sources associated with cores}
Distinguishing protostellar cores from starless cores depends on the detection of embedded infrared sources. Bright emission at 24 $\micron$ traces material accreting from the core onto the central protostar; the dust around the protostar can be heated by protostellar luminosity which mainly arise from accretion shocks \citep{Chambers09} . 
For that reason, if no emission is detected in a core in any of the four {\it WISE}  bands (3.4 - 22 $\micron$), it is classified as ``starless", otherwise, it is regarded as a candidate for a ``protostellar core". 
However, there are a few difficulties faced when attempting to identify protostellar cores: deeply embedded sources are too faint to be detected, and many IR sources can be unresolved external galaxies or bright Galactic giant stars, which are positioned by chance along the same line of sight as cores. To avoid these issues, we adopted the ``Young Stellar Object (YSO) finding scheme" with the {\it WISE} data by \citet{KL14}; we first excluded extragalactic contaminants such as star-forming galaxies (SFG) and PAH (Polycyclic Aromatic Hydrocarbons)-feature emissions following their method. As a result, one IR source, which is placed in the inner region of the dashed lines in Figure~\ref{fig:color-color} was removed. 

We also attempted to eliminate Active Galactic Nucleus (AGN) contamination using the color-magnitude cut in \citet{KL14}, but found that known YSOs frequently meet this criterion and are hence often misclassified as AGNs. Deeply embedded protostars show faint emission and they could be removed by the color-magnitude cut. \citet{Jorgensen06, Jorgensen07} suggested that the possibility of coincidence between an AGN and a core within 15$\arcsec$ is only $\thicksim 1 \%$ by a random distribution of AGNs.

All IR sources associated with our cores are located within 10$\arcsec$ from the dust continuum peaks. 
Therefore, we did not apply the AGN cut to identify protostars. Based on this scheme, we identified 5, 22, and 8 YSOs in the $\lambda$ Orionis, Orion A and Orion B clouds, respectively. In total, 35 YSOs are associated with PGCCs in the Orion complex.

\subsubsection{Classification of YSOs}
To study the differences of star formation status among the three clouds, we classified evolutionary stages of the identified 35 YSOs. There are a few physical parameters used in classifying the evolutionary stages of YSOs. First, the most useful one is the bolometric temperature $(T_{\rm bol})$ which is defined as the temperature of a blackbody with the same flux-weighted mean frequency in the observed SED \citep{ML93}. It is expected that Class 0 protostar has a $T_{\rm bol} < $ 70 K, a Class I protostar has 70 K $ \le  T_{\rm bol} \le 650 $ K, and a Class II pre-main-sequence star has 650 K $ < T_{\rm bol} \le 2800$ K \citep{Chen95}. 

The second one is the near- to mid-infrared spectral index $\alpha$, which is given as 
 \begin{equation}
\alpha = \frac {\rm d \, log(\lambda S(\lambda))} {\rm d \, log(\lambda)}   .
\end{equation}
The spectral index calculated in the wavelength range from $\thicksim$ 2 to 20 $\micron$ has traditionally been used  
\citep{Adams87, Lada87,  AM94, Evans09, Dunham14}. The index is $\alpha \ge$ 0.3 for Class I, $-0.3  <\alpha < 0.3$ for a flat-spectrum source, and $-1.6 < \alpha < -0.3$ for Class II. To classify the evolutionary stages of YSOs, we used the combination of the 4.5 - 24 $\micron$ spectral index $(\alpha_{4.5 - 24})$ with $T_{\rm bol}$ as proposed in the Furlan et al. (2016): $\alpha_{4.5 - 24} > 0.3$ and $T_{\rm bol} <$ 70 K for Class 0 protostars, $\alpha_{4.5 - 24} > 0.3$ and 70 K $< T_{\rm bol} <$ 650 K for Class I protostars, $-0.3 < \alpha_{4.5 - 24} < 0.3$ for flat-spectrum sources, and $\alpha_{4.5 - 24} <$ -0.3 and $T_{\rm bol} > $ 650 K for Class II pre-main-sequence stars.

We used {\it WISE} 4.6 -- 22 $\micron$ data and SCUBA-2 850 $\micron$ peak flux to calculate $T_{\rm bol}$.  In case that a source has a wide range of data including near-infrared data of MIPS of the {\it Spitzer} and also far-infrared data of {\it AKARI}, {\it IRAS}, and {\it Herschel}, these data points are also included to get more reliable  $T_{\rm bol}$ \citep{Adams87, Lada87,  AM94, Evans09, Dunham14}. 

Based on this method, 5 YSOs in the $\lambda$ Orionis cloud were classified as 2 Class 0, 1 Class I, and 2 flat-spectrum sources. We found 6 Class 0, 11 Class I, and 5 flat-spectrum sources in the Orion A cloud and 2 Class 0, 4 Class I, and 1 flat-spectrum sources in the Orion B cloud (Figure~\ref{fig:classification}). However, because of a lack of far-infrared data for 3 YSOs in the Orion A cloud, the $T_{\rm bol}$ is probably overestimated, which could result in the misclassification of these sources. The sources that do not have the far-infrared data are marked with asterisks in Figure~\ref{fig:classification}. Table~\ref{tab:YSOs} summarizes the numbers of YSOs in each class and the median values of $L_{\rm bol}$, $T_{\rm bol}$, and $\alpha$ in each cloud.

To confirm whether our classifications of identified YSOs are really reliable, we use the {\it Spitzer} and {\it Herschel} protostar catalogs. {\it Spitzer Space telescope} survey identified 3479 YSOs with disks or infalling envelopes in the Orion A and B clouds \citep{Megeath12}, but did not carry out observations toward the $\lambda$ Orionis cloud. These YSOs were also observed at 70 and 160 $\micron$ with the Photodetector Array Camera and Spectrometer (PACS) as part of the Herschel Orion Protostar Survey (HOPS) and 16 YSOs  were newly identified \citep{Stutz13, Tobin15}. We confirmed that all of the 22 YSOs in the Orion A cloud and 7 out of the 8 YSOs in the Orion B cloud identified from the {\it WISE} data are matched with the YSOs in the {\it Spitzer} and {\it Herschel} catalogs within 3$\arcsec$. The one YSO that is not listed in the {\it Herschel} catalog is thought to be an Asymptotic Giant Branch(AGB) star or a Classical Be (CBe) star because it was positioned on the boarder of color-magnitude criteria of AGB and CBe stars. AGB stars at great distance are really similar to YSOs and CBe stars can mimic blue transition disk objects  in color-color space \citep{KL14}. Thus, the YSO, which reside in a core Middle3 of G205.46-14.56, have high probability of misclassification, and we do not account this source as a YSO hereafter. This source is marked with a green triangle in Figure~\ref{fig:classification}. Finally, in this study, we identified 5, 22, and 7 YSOs in the $\lambda$ Orionis, Orion A, and Orion B clouds, respectively from the {\it WISE} data. 10 and 6 YSOs listed in the {\it Spitzer} and {\it Herschel} protostar catalogs in the Orion A and B clouds are missed due to lower signal-to-noise ratios and high uncertainties of the gaussian profile fitting of the {\it WISE} data. These two parameters given in the AllWISE photometric catalog are important to discriminate between real and fake point sources \citep{KL14}.

In total, from the $\it{WISE}$ data, 10 YSOs were identified as Class 0, 16 as Class I, and 8 as flat-spectrum sources. 
The median $L_{\rm bol}$, $T_{\rm bol}$, and $\alpha_{3.6 - 22}$ values for the total 34 YSOs are 3.0 $L_{\sun}$, 150 K, and 1.05, respectively. 
 In the protostellar luminosity study (Kryukova et al. 2012), using the $L_{\rm bol}$ from the c2d survey, the luminosity of YSOs in the Orion A and B clouds peaks near 1 $L_{\sun}$, while the bolometric luminosity ($L_{\rm bol}$) peaks in this study around 3 $L_{\sun}$ probably due to the lower sensitvity of {\it WISE}. 
 
The comparison among three clouds shows that the median $T_{\rm bol}$ of the YSOs in the $\lambda$ Orionis cloud is 101.7 K, which is lower than that in the Orion A (152.6 K) cloud, and the YSOs in the Orion B cloud have the highest median $T_{\rm bol}$ of 182.8 K (Table~\ref{tab:YSOs}).
 This may indicate that the protostars in the $\lambda$ Orionis cloud are less evolved compared to those in the Orion A and B clouds.
$L_{\rm bol}$, $T_{\rm bol}$, $\alpha_{3.6 - 22}$, and the evolutionary stages of each YSOs are listed in the last four columns of Table~\ref{tab:list_cores_Lambda} to Table~\ref{tab:list_cores_OrionB}.

\subsubsection{Gravitational Instability of cores}
 Cores may become gravitationally unstable and subsequently collapse. We investigated whether the starless cores are gravitationally bound and have the ability to form protostars. The Jeans mass is the minimum mass of a spherical portion of a uniform medium that can trigger collapse by its own gravity. The equation of the Jeans mass is adopted from Wang et al. (2014). 
\begin{equation}
M_{\rm J} = \frac{\pi^{5/2}c_{s}^3} {6\sqrt{G^3\rho}} = 0.877\, \bigg( {\frac{T}{10\, \rm K}} \bigg)^{3/2} \bigg( {\frac{n}{10^5\, \rm cm^{-3}}} \bigg)^{-1/2} \rm M_{\odot},
\end{equation}
where $T$ is the temperature and n is the H$_{2}$ volume density, respectively. We assumed that the gas and dust is well coupled in cores with high densities (n $\ge 10^{5} \rm \, cm^{-3}$) via collisions \citep{GL74}, and thus, their temperatures are the same \citep{Galli02}. The core mass and its Jeans mass, {\it M} and ${M_{\rm J}}$, are listed in Table~\ref{tab:list_cores_Lambda} to Table~\ref{tab:list_cores_OrionB}.

 Cores with masses greater than their Jeans masses will collapse when we do not consider the effect by the magnetic field or turbulence. Cores with masses less than their Jeans masses are not gravitationally bound and will be dispersed by their own internal motion. In the $\lambda$ Orionis cloud, there are four starless cores whose Jeans masses are significantly larger than their core masses, suggesting that these starless cores may disperse in the future. The other six starless cores have larger core masses than their Jeans masses. In the Orion A cloud, 37 out of 52 starless cores have masses larger than their Jeans masses, and they may collapse to form protostars. In case of the Orion B cloud, all the 23 starless cores have masses larger than their Jeans masses. These results are presented in Figure~\ref{fig:jeans_mass}. In total, 65 out of 85 starless cores may collapse, while 20 cores remain as starless. The highest and lowest fractions of gravitationally unstable starless cores are found in the Orion B cloud (100$\%$) and in the $\lambda$ Orionis cloud (60$\%$), respectively.

\subsection{Dense gas fraction}
 To compare the potential of star formation in the three clouds of the Orion complex, we determine the fraction of dense gas, from which stars form, in PGCCs. 
Submillimeter dust continuum observations with ground-based bolometer arrays filter out the large-scale diffuse gas.  
 Therefore, the ratio between the masses derived from the $\it {Planck}$ data (at 857, 545, and 353 GHz) combined with the 3 THz IRAS data and from the SCUBA-2 850 $\micron$ emission represents the dense gas fraction in the cold clump. The core masses are calculated from the SCUBA-2 850 $\micron$ images as described in section 3.3.1 and the clump masses are adopted from the PGCC catalog. Since some PGCCs have no information about masses in the catalog, we adopt a 3$\sigma$ limit for the flux at 353 GHz and estimate the upper limit of the clump masses. As shown in Figure~\ref{fig:clump_mass}, the dense gas fraction is the highest in the Orion B cloud (0.20) and the lowest in the $\lambda$ Orionis cloud (0.10). In the Orion A cloud, the fraction is about 0.12. 

 In the right panel of Figure~\ref{fig:clump_mass}, PGCCs that are not detected at 850 $\micron$ are plotted together; the detected PGCCs and non-detected PGCCs are clearly divided into two groups. For the non-detected PGCCs, we calculated the upper limit of core mass adopting 3$\sigma$ noise of 850 $\micron$ images. The median clump masses of detected PGCCs are higher than those of non-detected PGCCs  by factors of 1.6 to 2.2 in the three clouds. The difference in clump mass between detected and non-detected PGCCs is the greatest (a factor of 2.2) in the $\lambda$ Orionis cloud. 
The median column density ($4.9\times10^{21}$ cm$^{-2}$) and number density ($2.8\times10^3$ cm$^{-3}$) of the detected PGCCs are higher by factors of 4.9 and 5.4, respectively, compared to those of non-detected PGCCs. Other parameters provided by the PGCC catalog, such as the warm background temperature, do not show a clear difference.

\section{DISCUSSION} \label{sec:discussion}

\subsection{The effect of stellar feedback on PGCCs in the $\lambda$ Orionis cloud}
The $\lambda$ Orionis cloud provides a good example for showing the effect of a nearby massive star on the core properties. The decrease of dust temperatures of PGCCs along the projected distance from $\lambda$ Ori, suggests that the central star has great influence on the PGCCs, as presented in Figure~\ref{fig:projected}. In Table~\ref{tab:PGCCs}, the physical properties of PGCCs from the PGCC catalog are listed; compared to the PGCCs in the Orion A and Orion B clouds, the PGCCs in the $\lambda$ Orionis cloud have smaller column and volume densities, and clump masses. Therefore, the lower values of physical properties such as core mass, density, and size in the $\lambda$ Orionis cloud, compared to the Orion A and B clouds, may indicate a strong stellar feedback such as erosion or even destruction of the cores by the photoionizing radiation. As shown in Figure~\ref{fig:Lambda_cloud}, the $\lambda$ Orionis cloud is dominated by H$\alpha$ emission, indicating that it is shaped by the ionized hydrogen content of a gas cloud. The median dust temperature (16.08 K) of PGCCs in the $\lambda$ Orionis cloud is higher than those (13 -- 14 K) in the other two clouds. This also suggests the PGCCs in the $\lambda$ Orionis cloud are externally heated. 

The other effect of stellar feedback in the $\lambda$ Orionis cloud is the unbounding state of the region. Dolan $\&$ Mathieu (2001, 2002) interpreted that the shock wave from the supernova explosion dispersed gas at the center of the $\lambda$ Orionis cloud and left the region unbound. Although most of the PGCCs are located along the ring shaped structure (see Figure~\ref{fig:Lambda_cloud}), it remains unclear whether they are bound or not. However, we can expect that this region was also affected by supernovae about 1 Myr ago, which may have disrupted the formation of dense and gravitationally-bound structures.  The recent result showing that the star formation efficiency (SFE) decreases greatly in less-bound clouds \citep{Lucas17} is supported  by our result.

Additionally, the dust emissivity spectral index $\beta$ in the Orion A and B clouds are close to 2, while $\beta$ in the $\lambda$ Orionis cloud is much smaller as 1.65 (see Table~\ref{tab:PGCCs}). \citet{Juvela15} estimated the $\beta$ using $\it Herschel$, $\it Planck$, and IRAS data and found median values over all fields of the {\it Planck} clumps are 1.84. The $\beta$ values are well known to be dependent on the physical environments such as grain growth or composition of dust grain and evolution of the carbonaceous component \citep{Jones13}. \citet{Li2017}, however, reported that grain growth may reduce $\beta$ in circumstellar disks or envelopes only from the late Class 0 stage to the end of the Class I stage of YSOs. \citet{Forbrich15} also found that $\beta$ in the inner region of a starless core FeSt 1-457, shows no significant difference from the value for the local cloud, indicating that grain growth does not occur significantly in the very early stage of star formation.  The $\lambda$ Orionis cloud mostly harbors starless cores, and thus, $\beta$ may represent their initial dust property. This result might indicate that the $\lambda$ Orionis cloud was initially very dense, resulting in vigorous grain growth, and could trigger the massive star formation. 

 After the explosion of one of O type stars, the PGCCs can be affected by strong radiation. The dust temperature of PGCCs decreases with the projected distance from  $\lambda$ Ori. (see Figure~\ref{fig:projected}). The errors are large owing to the sensitivity of  Planck, but we can see the tendency for dust temperature to decrease with distance. As seen in Table~\ref{tab:PGCCs}, the $\lambda$ Orionis cloud has a higher median/mean dust temperature and lower median/mean density than other clouds. This may be due to the external heating from the  H $_{\rm II}$ region, which can increase the temperature of the molecular cloud, making it hard to cool down to form dense cores. \citet{Juvela18} investigated properties of $\beta$ and $T_{\rm d}$ of PGCCs with IRAS, $\it {Planck}$, SCUBA-2, and $\it {Herschel}$ data and suggested that the heating effect of a protostar on its environment is very small, because of its small volume to the surrounding material. Thus, the embedded protostars should not significantly contribute to raise the dust temperature of their parent clumps either in the $\lambda$ Orionis cloud, supporting the idea that the higher dust temperatures in the $\lambda$ Orionis cloud are mainly influenced by the external heating. This may also account for the lower column densities of PGCCs in the $\lambda$ Orionis cloud because the external radiation can disperse gas via photo-evaporation as well as heating.

\subsection{Two types of sub-clouds in the $\lambda$ Orionis cloud}
\citet{DM02} suggested that the original shape of the $\lambda$ Orionis cloud was like a linear string of dense molecular clouds. The most massive region was the central part of the initial cloud including B30, B35, the location of today's ``$\lambda$ Ori", and B223 (at the southwest of these). The rest of the cloud was filled by lower density molecular gas. As mentioned above, the supernova explosion quickly dispersed most of the parent cloud and then swept up molecular gas and dust, creating the ring structure that is seen today. \citet{Lang2000} suggested that the sub-clouds in the $\lambda$ Orionis cloud were divided into two classes after the supernova explosion, based on their CO J=1-0 survey; (1) dense and massive sub-clouds and (2) diffuse and low mass sub-clouds. In the following section, we explore the characteristics of each sub-cloud in the context of the above scenarios and summarize the results in Table~\ref{tab:Lambda_clouds}.

\subsubsection{Dense and massive regions} \label{sec:dense}
\citet{DM02}  suggested that dense and high-mass regions of the $\lambda$ Orionis cloud are remaining parts of the initial parent cloud.  
B30 and B35 support this scenario. They were accompanied by a massive cloud core which became the birthplace of OB stars, called ``$\lambda$ Ori" today. These two clouds may be massive enough to survive from the supernova explosion. 

On the other hand, B223 might not be part of the $\lambda$ Orionis cloud because it is located at the southwest region away from the central part (``$\lambda$ Ori", B30 and B35). The pre-main sequence stars in B223 have different radial velocities from those in B30 and B35 (\citealt{DM02}). In our survey, there are 18 PGCCs in this region, but none of them were detected at 850 $\micron$. This supports the idea that B223 was not originally associated with the $\lambda$ Orionis cloud, but is just coincidentally projected to the same region of sky as the $\lambda$ Orionis cloud. 
 
Thirteen PGCCs were observed in two dense clouds (B30 and B35) and the detection rate is about 38$\%$ (5 out of 13), about 2.5 times higher than the total detection rate (16$\%$, 8 out of 50) in the $\lambda$ Orionis cloud. The median mass and column density of the 13 PGCCs are 13.8 M$_{\sun}$ and $1.3\times10^{21}$ cm$^{-2}$, respectively. The dense gas fraction, already mentioned earlier, is 0.19, which is comparable to the value for the Orion B cloud. The median column density of detected cores is $1.1\times10^{23}$ cm$^{-2}$, which is similar to those of Orion A ($1.5 \times10^{23}$ cm$^{-2}$) and lower than those of Orion B ($3.8 \times10^{23}$ cm$^{-2}$).
This result suggests that core formation in dense and massive regions is much less affected by the stellar feedback.

\subsubsection{Diffuse and low mass regions} \label{sec:diffuse}
The low-mass and low-density regions may result from the supernova blast and subsequent fragmentation of the molecular ring \citep{Lang2000, DM01}. 
We find several indications that core formation is much more depressed in the diffuse and low-mass regions. All properties are lower than those of the dense and massive regions of the $\lambda$ Orionis cloud and other star-forming regions of the Orion A and B clouds. 

 The detection rate of the diffuse and low-mass region is about 8$\%$ (3 out of 36), which is five times lower than that of dense regions (38 $\%$) in the $\lambda$ Orionis cloud. The dense gas fraction is also the lowest at 0.05. It is lower by a factor of four compared to the dense and massive regions of the $\lambda$ Orionis cloud. The median mass and column density of the 36 PGCCs are 13.9 M$_{\sun}$ and $9.9\times10^{20}$ cm$^{-2}$, respectively, which are not very different from those of PGCCs in the dense and massive regions (see Section \ref{sec:dense}). However, the median mass and column density of cores in this low density region are 0.96 M$_{\sun}$ and $5.7 \times10^{22}$  cm$^{-2}$, which are lower than those in the high density region by a factor of two (Table~\ref{tab:Lambda_clouds}). These results imply that after the supernova explosion, the diffuse and low-mass regions are significantly affected by massive stars; core formation was suppressed, and thus, the fraction of dense gas became low.

\subsection{Multiplicity of the three clouds}
We have identified a total of 119 cores within 40 PGCCs as previously discussed.
The averaged multiplicity (number of cores in one PGCC) for the three clouds is nearly 3. Figure~\ref{fig:histogram_PGCC} shows histograms of multiplicity, column density, and mass of PGCCs, respectively. The multiplicity in the $\lambda$ Orionis is less than 2 (15 cores within 8 PGCCs), but for Orion A (74 cores within 23 PGCCs) and B (30 cores within 9 PGCCs), the multiplicities are significantly larger than 3 (see also Table~\ref{tab:list_cores_Lambda} to Table~\ref{tab:list_cores_OrionB}). This clear difference is obvious in  Figure~\ref{fig:Lambda_PGCCs} to Figure~\ref{fig:OrionB_PGCCs}. In Figure~\ref{fig:histogram_PGCC}, the column densities and masses of PGCCs in the $\lambda$ Orionis cloud occupy the bins of lower values, while PGCCs in the Orion A and B clouds show wide distributions. 
The difference of multiplicity in the $\lambda$ Orionis cloud can arise from the strong radiation field. 
The UV photons from the OB association in the  $\lambda$ Orionis cloud heat the clumps externally and hinder them from cooling. Therefore, the clumps are difficult to collapse and fragment. 

In Figure~\ref{fig:K-S}, we plot the cumulative mass and column density fraction of cores in the three clouds. The two profiles of each panel are strikingly different. The $\lambda$ Orionis cloud has substantially lower fractions of core mass and column density than those of Orion A and B clouds. We also found that the mean and median clump masses and volume densities are the lowest in the $\lambda$ Orionis cloud. This is in line with results of  \citealt{Polychroni13} that clumps within filaments have higher masses  than clumps outside of the filaments. The entire structure of molecular clouds of Orion A and B are filamentary, while $\lambda$ Orionis is ring-shaped structure (Figure~\ref{fig:Lambda_cloud} and ~\ref{fig:OrionAandB_cloud}). 
This discrepancy can provide a significant clue concerning the physical origin of the environmental conditions between the $\lambda$ Orionis cloud and Orion A and B clouds.  

The fragmentation of clumps, which increases multiplicity, also might be caused by non-thermal motions, such as turbulence, which could be manifestedd by line broadening by bulk motions. We conducted single dish observations (Yi et al. in prep) using Korean VLBI Network (KVN) and found that the line widths are meaningfully broader in cores of Orion A and B than cores in the $\lambda$ Orionis cloud. The median full width at half maximum (FWHM) of both HCN ($J = 1 \rightarrow 0$) and H$_{2}$CO (2$_{1,2}$ $\rightarrow$ 1$_{1,1}$) are the largest in  Orion B as 1.40 km s$\rm ^{-1}$, and the FWHM of C$_{2}$H ($N = 1 \rightarrow 0$) is the largest in the Orion A as 1.14 km s$\rm ^{-1}$. The median FWHMs of HCN, H$_{2}$CO, and C$_{2}$H in the $\lambda$ Orionis clouds are 0.87, 1.00, and 0.78 km s$\rm ^{-1}$, respectively. Theses are the lowest median values among the three clouds. This result may support that greater internal motions of turbulence actively led to fragmentation of clumps down to cores in the Orion A and B clouds.

\section{SUMMARY}
  We investigated the physical properties of all the 96 PGCCs in three clouds of the Orion complex using the JCMT/SCUBA-2 850 $\micron$ continuum data. In total, 58 out of 96 PGCCs were observed as part of the SCOPE survey and rest of 38 PGCCs were studied with  archival data. 

We identified 119 cores within the detected 40 PGCCs; 34 cores are protostellar, while 85 cores are starless. To examine the gravitational instability of starless cores, we estimated their Jeans masses.
  In the Orion A cloud, 37 out of 52 starless cores have masses larger than their Jeans masses, and they may collapse to form protostars. In case of the Orion B cloud, all the 23 starless cores have masses larger than their Jeans masses.
  In the $\lambda$ Orionis cloud, 40$\%$ (four out of ten) of the starless cores remain gravitationally unbound and thus, may be finally dispersed. The rest of 60$\%$ starless cores may collapse to form protostars. 
 
 Additionally, we explore the characteristics of each sub-cloud in the $\lambda$ Orionis cloud following a scenario of Lang et al (2000). 
The cores in the dense and massive regions show comparable column density ($ 1.1 \times 10^{23}$ cm$^{-2}$) to that of Orion A cloud ($  1.5 \times 10^{23}$ cm$^{-2}$) while the cores in the diffuse and low mass regions show the lowest median column density of $5.7 \times 10^{22}$ cm$^{-2}$. 
This difference suggests that after the supernova explosion, the diffuse and low-mass regions are significantly affected by massive stars; core formation was suppressed, and thus, the fraction of dense gas became low.

 We compared the overall properties of PGCCs in the $\lambda$ Orionis cloud with those of PGCCs in the Orion A and Orion B clouds and we found PGCCs in the $\lambda$ Orionis cloud usually have  higher $T_{\rm dust}$, suggesting that PGCCs in the $\lambda$ Orionis cloud are externally heated. In addition, high energy photons from the massive star ($\lambda$ Ori) can photo-evaporate the dense material and lead to lower column densities in the $\lambda$ Orionis cloud. This effect can also cause the lower multiplicity in the $\lambda$ Orionis cloud (1.8) compared to that of Orion A (3.2) and B (3.3) clouds. Among three clouds, the cores in the $\lambda$ Orionis cloud have the lowest median values of the mass (0.77 M$_{\sun}$), size (0.09 pc), column density ($  8.2\times 10^{22}$ cm$^{-2}$), volume density ($2.5\times10^{5}$ cm$^{-3}$), and dense gas fraction (0.10).  Therefore, we conclude that a massive star gave negative feedback to the core formation and evolution in the $\lambda$ Orionis cloud.

This work was supported by the Basic Science Research Program through the National Research Foundation of  Korea (NRF) (grant No. NRF-2018R1A2B6003423) and the Korea Astronomy and Space Science Institute under the R$\&$D program supervised by the Ministry of Science, ICT and Future Planning and the BK21 plus program through the National Research Foundation (NRF) funded by the Ministry of Education of Korea. 
Tie Liu is supported by KASI fellowship and EACOA fellowship. This research was partly supported by the OTKA grant NN-111016 and NSFC No. 11725313.
PS was financially supported by Grant-in-Aid for Scientific Research (KAKENHI Number 18H01259) of Japan Society for the Promotion of Science (JSPS).
CWL was supported by Basic Science Research Program though the National Research Foundation of Korea (NRF) funded by the Ministry of Education, Science, and Technology (NRF-2016R1A2B4012593). ``The James Clerk Maxwell Telescope is operated by the East Asian Observatory on behalf of The National Astronomical Observatory of Japan; Academia Sinica Institute of Astronomy and Astrophysics; the Korea Astronomy and Space Science Institute; the Operation, Maintenance and Upgrading Fund for Astronomical Telescopes and Facility Instruments, budgeted from the Ministry of Finance (MOF) of China and administrated by the Chinese Academy of Sciences (CAS), as well as the National Key R\&D Program of China (No. 2017YFA0402700). Additional funding support is provided by the Science and Technology Facilities Council of the United Kingdom and participating universities in the United Kingdom and Canada."

\clearpage
\begin{figure}
\epsscale{0.9}
\plotone{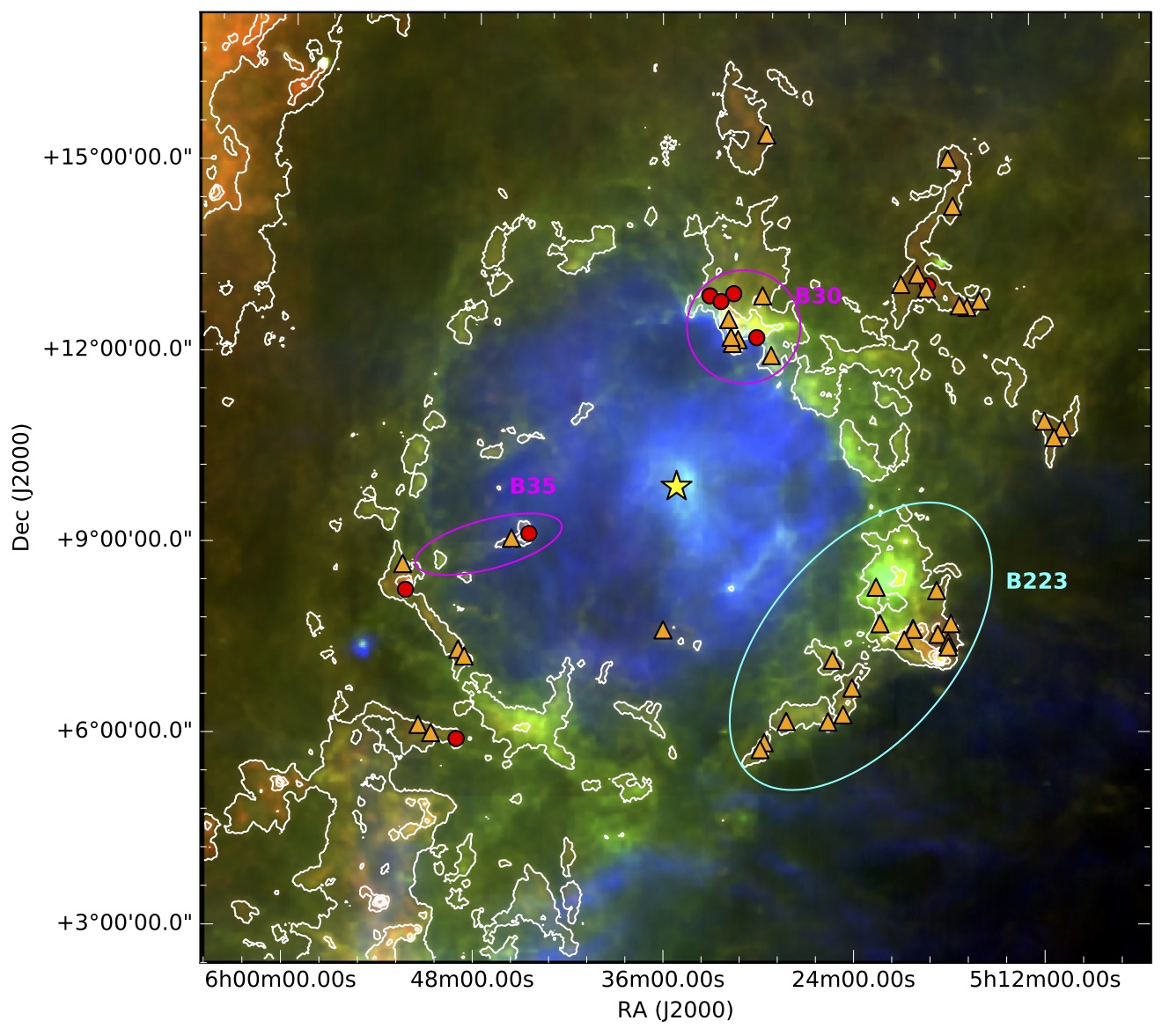}
\caption{\label{fig:Lambda_cloud} Three-color composite image (red: Planck 353 GHz; green: IRAS 100 $\micron$; blue: H$\alpha$) of the $\lambda$ Orionis cloud. The white contours show the flux density of Planck 857 GHz continuum emission from 20$\%$ to 100$\%$ in a step of 20$\%$ of the peak value 156.4 MJy sr$^{-1}$. The yellow star indicates the position of the ``$\lambda$ Ori$"$ OB binary. The red circles and orange triangles are all 50 PGCCs with N(H$_{2}$) $>$ 5$\times$10$^{20}$ cm$^{-2}$ in this region. The former are detected at  850 $\micron$, while the latter are not detected. The two types of sub-clouds B30, B35, and B223 are marked with pink ellipses and a cyan ellipse, respectively (see Section 4.2).}
\end{figure}

\clearpage
\begin{figure}
\epsscale{1.0}
\plotone{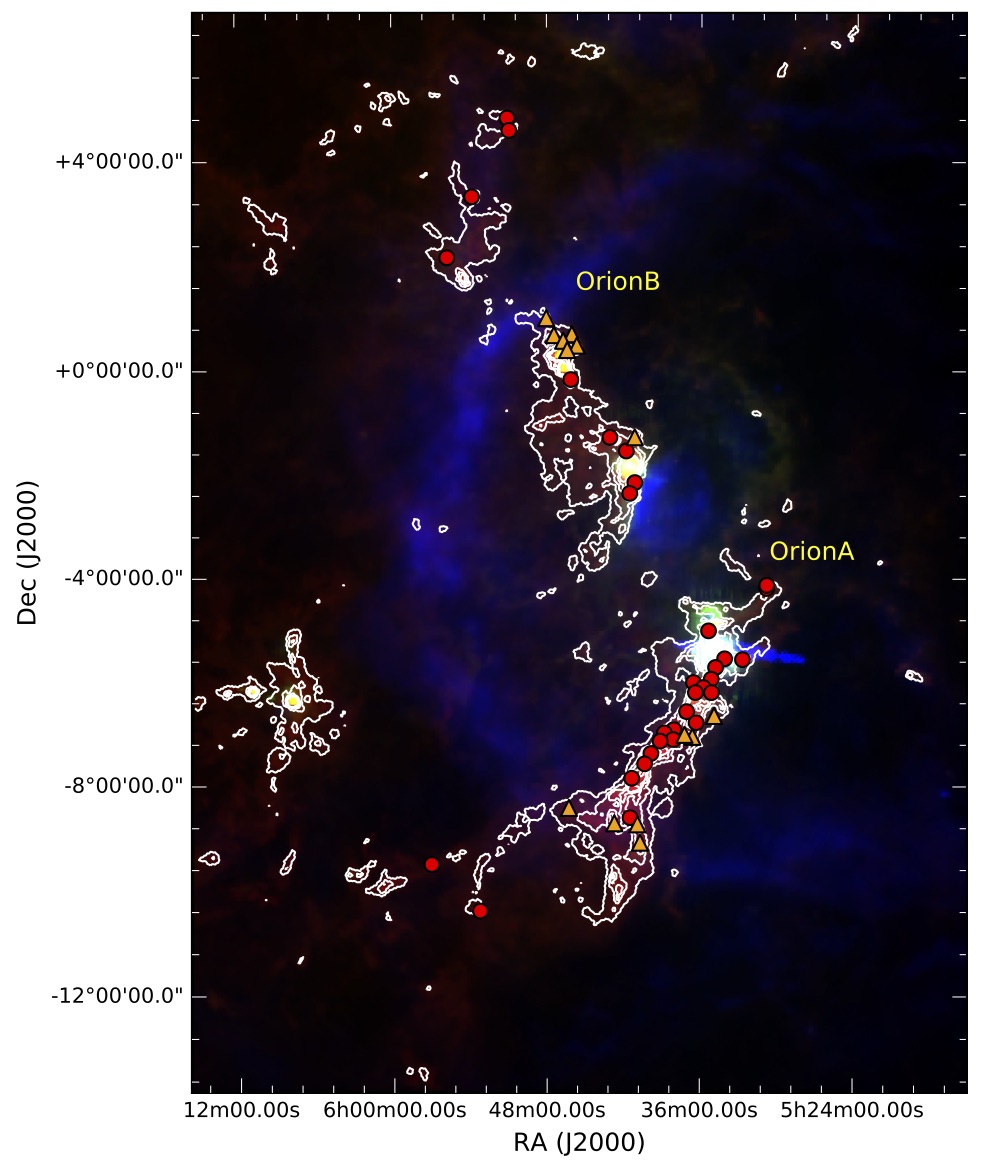}
\caption{\label{fig:OrionAandB_cloud} The Orion A (lower one) and Orion B (upper one) clouds with 46 PGCCs. The peak intensity is 9122.7 MJy sr$^{-1}$ and symbols are same as Figure 1.}
\end{figure}

\clearpage
\begin{figure}
\epsscale{0.8}
\plottwo{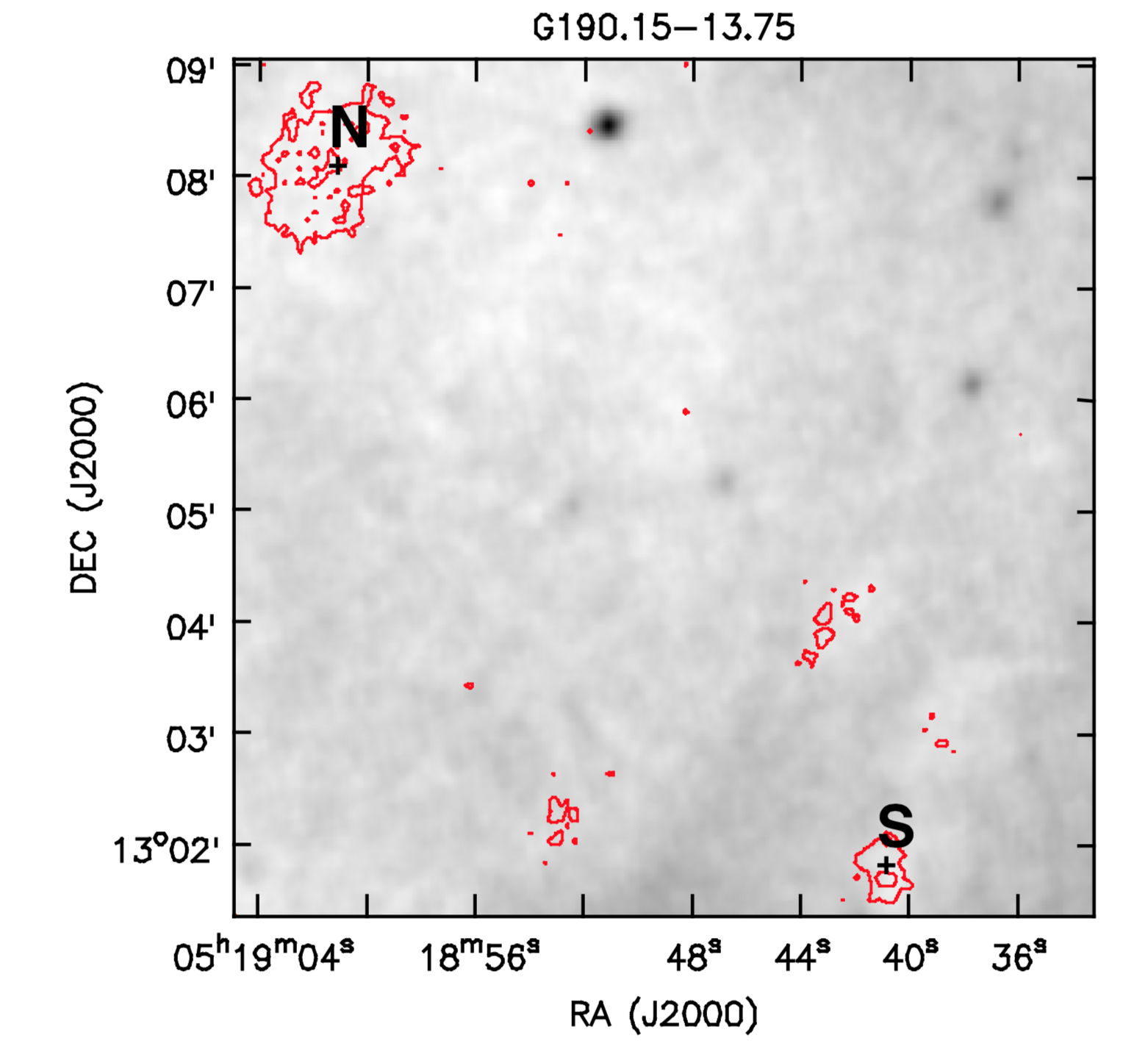}{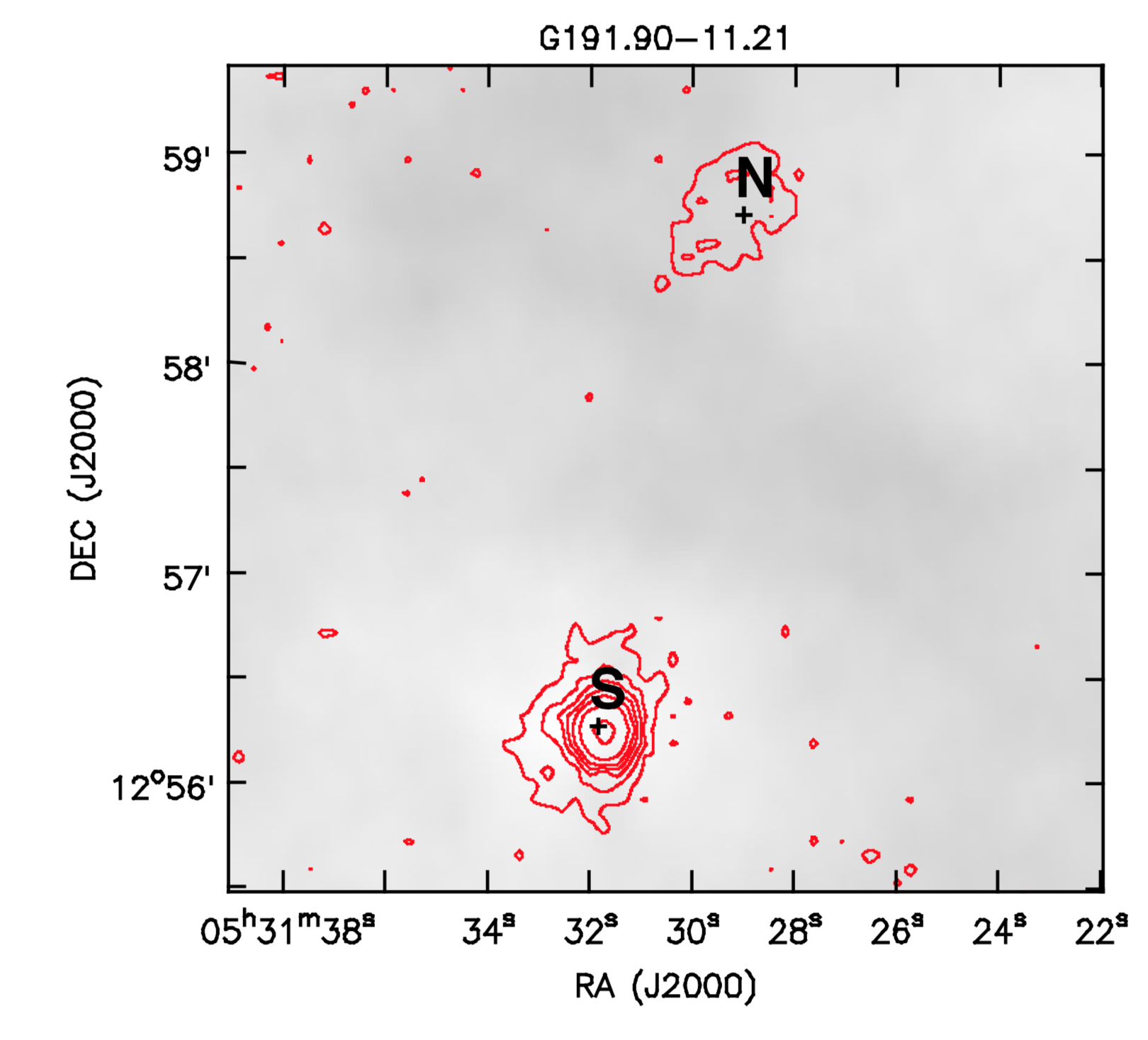}\\
\plottwo{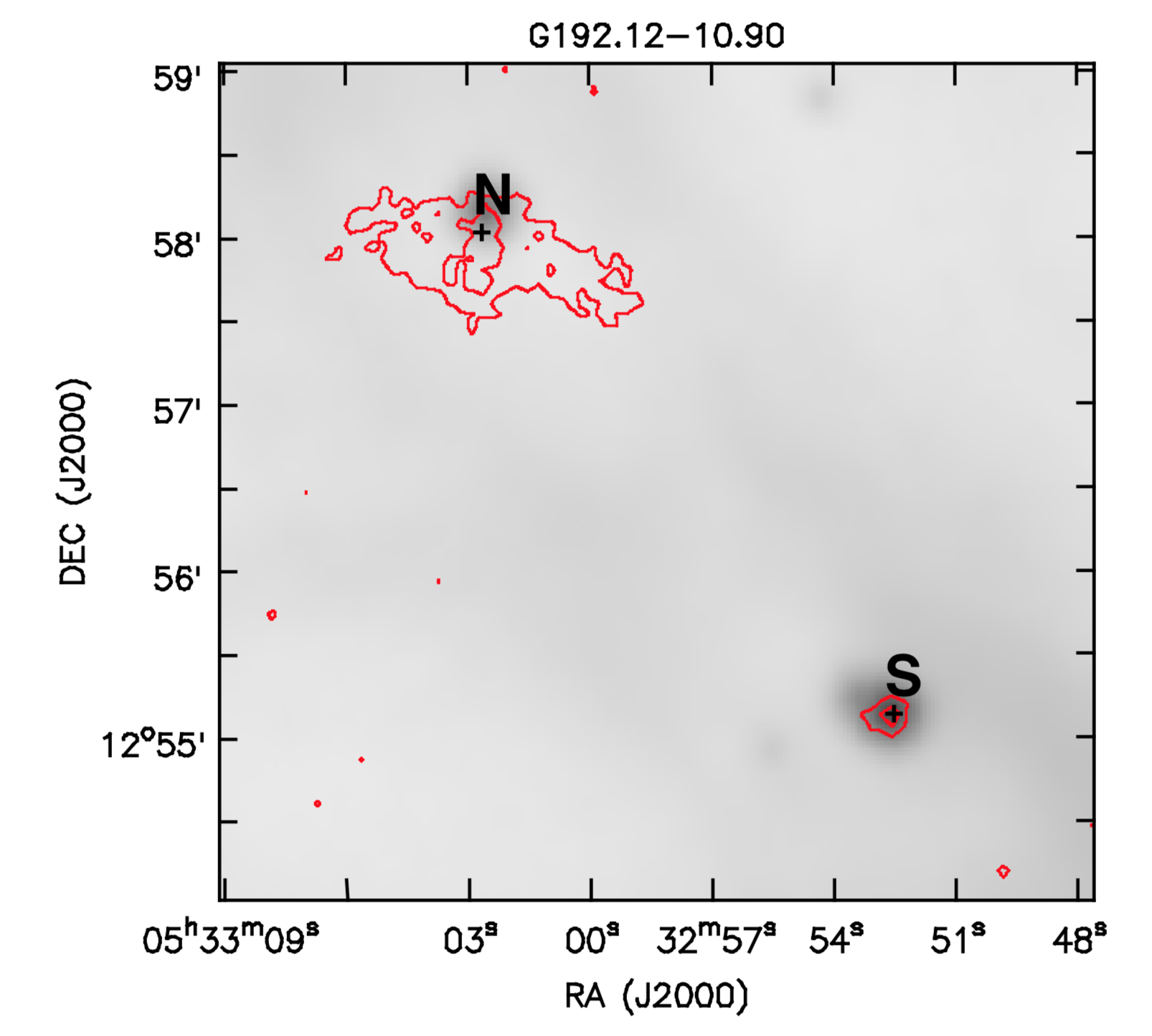}{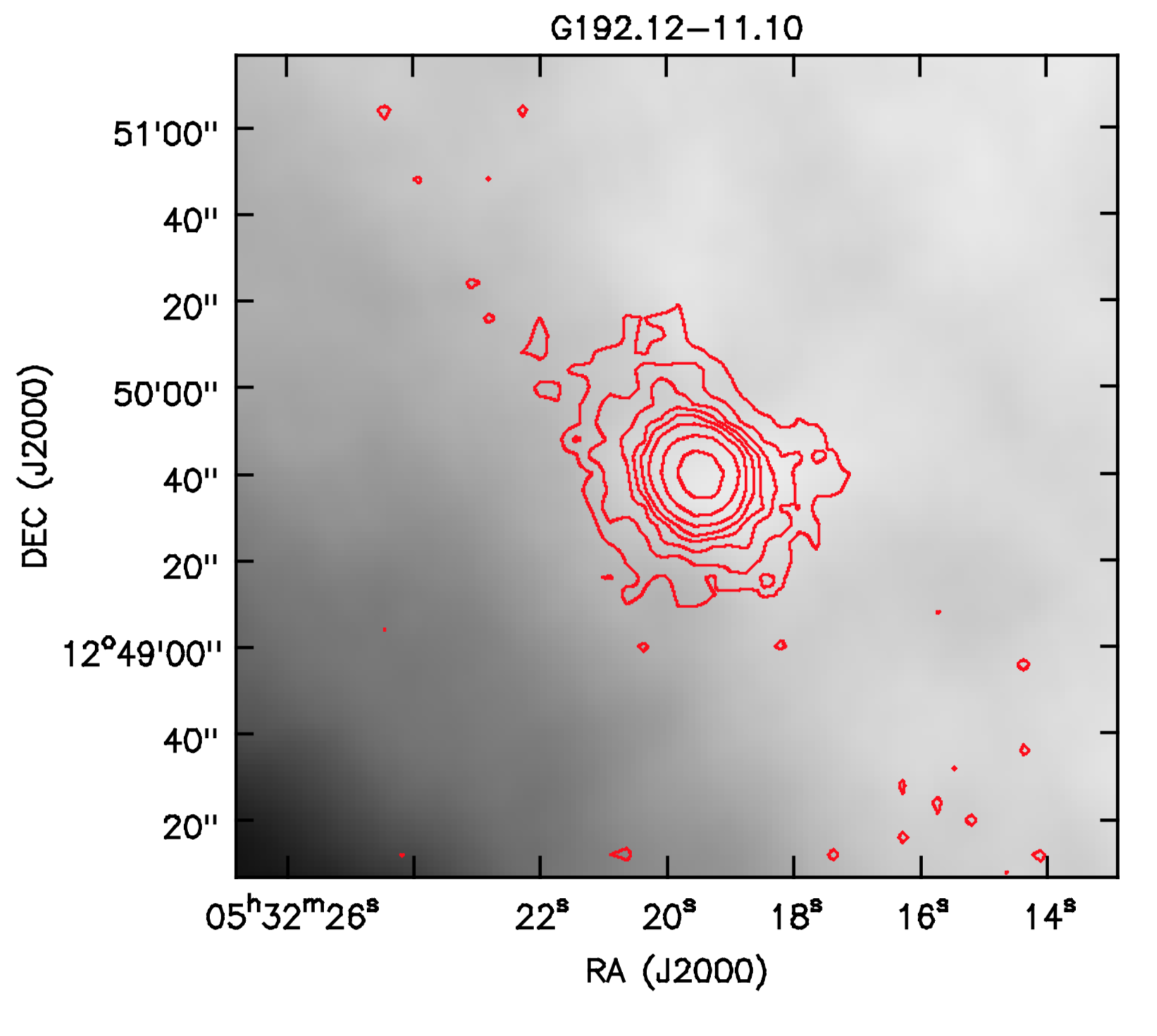}\\
\plottwo{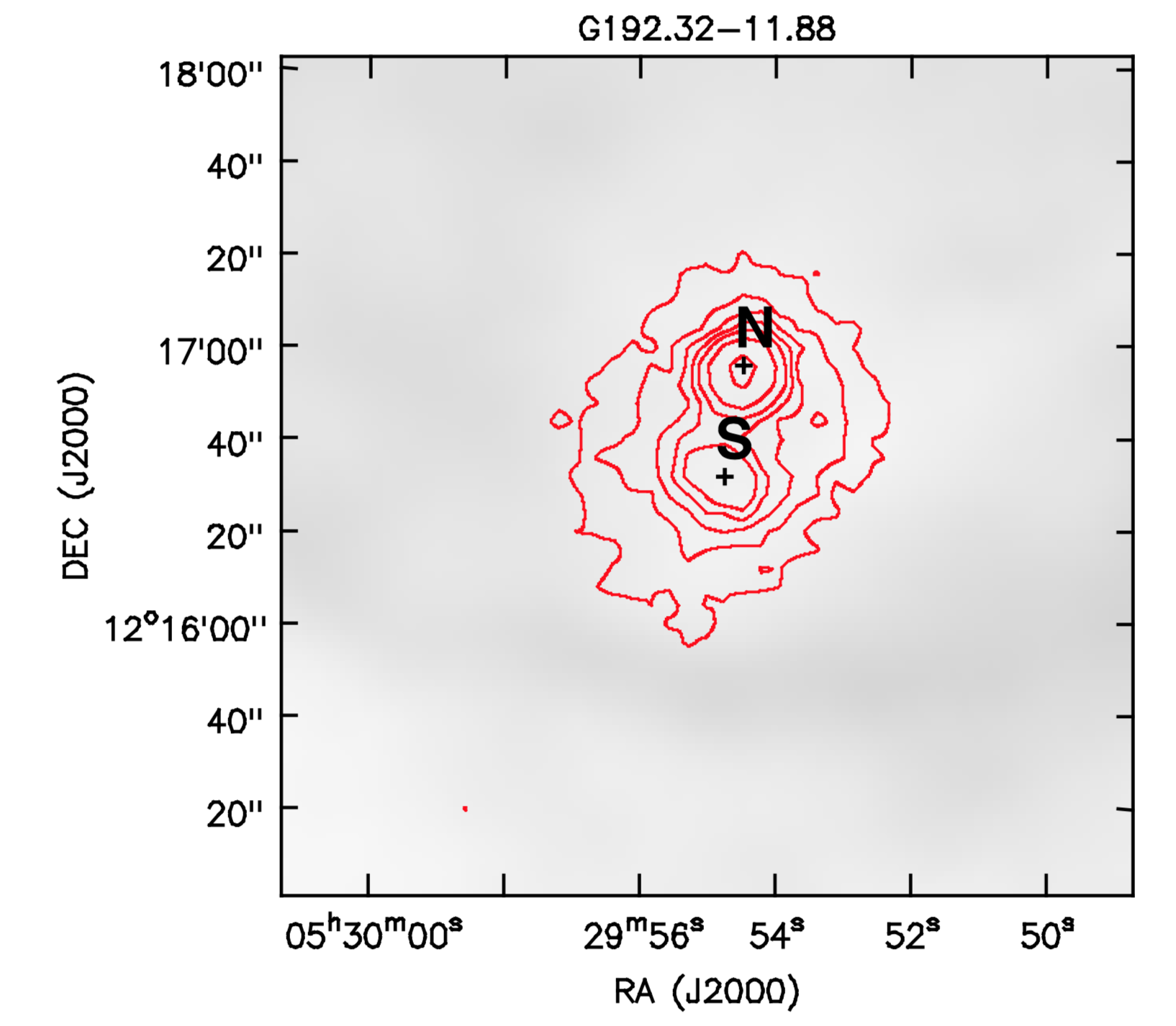}{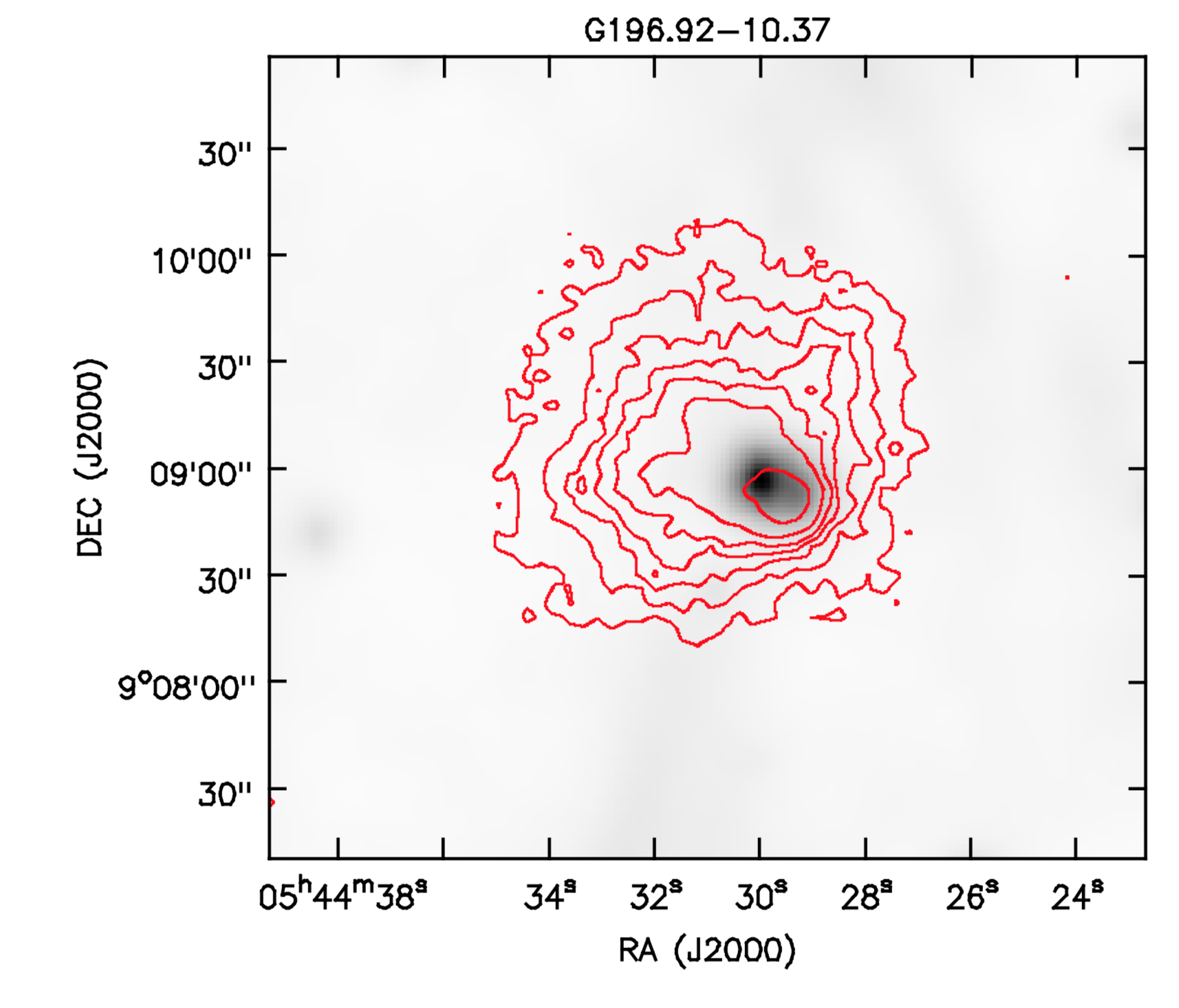}\\
\caption{ The 8 PGCCs detected at 850 $\micron$ with SCUBA-2 in the $\lambda$ Orionis cloud. The 850 $\micron$ emission distributions are overlaid on the WISE 12 $\micron$ images. The red contours represent the 3, 6, 9, 12, 15, 20, and 30$\sigma_{rms}$ levels. \label{fig:Lambda_PGCCs}}
\label{f3}
\end{figure}

\clearpage
\begin{figure}
\epsscale{0.8}
\plottwo{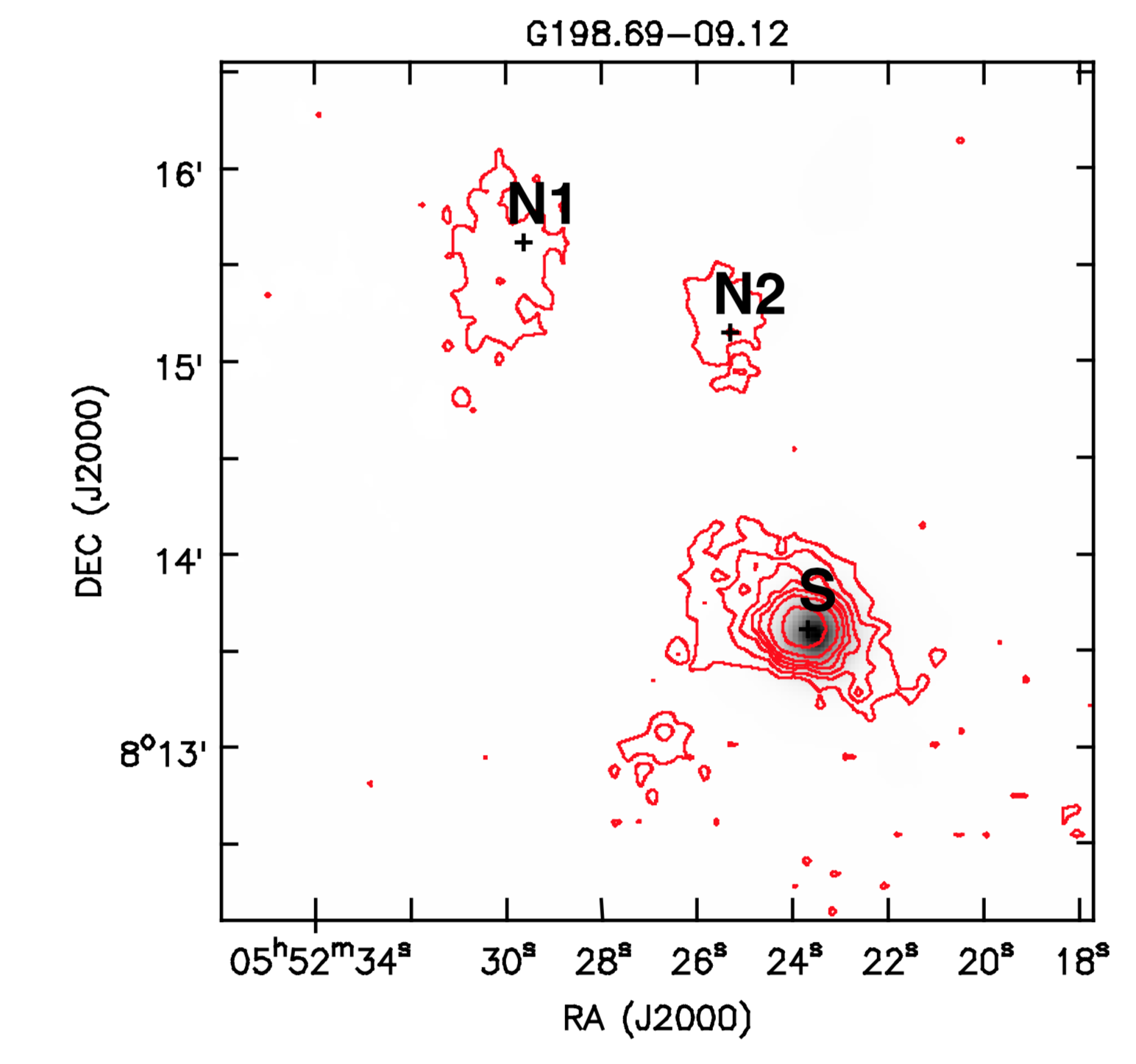}{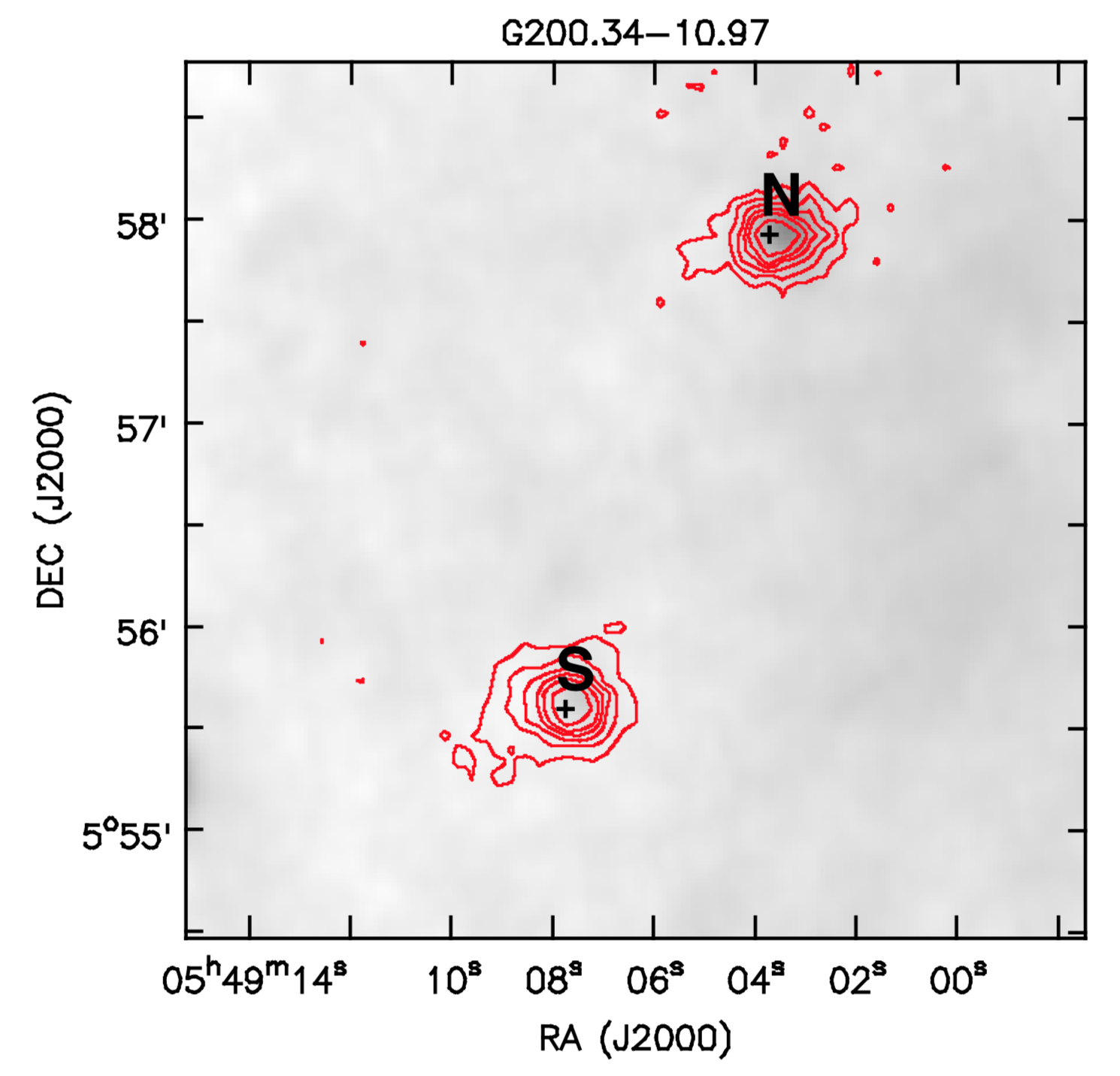}\\
\end{figure}

\clearpage
\begin{figure}
\epsscale{0.8}
\plottwo{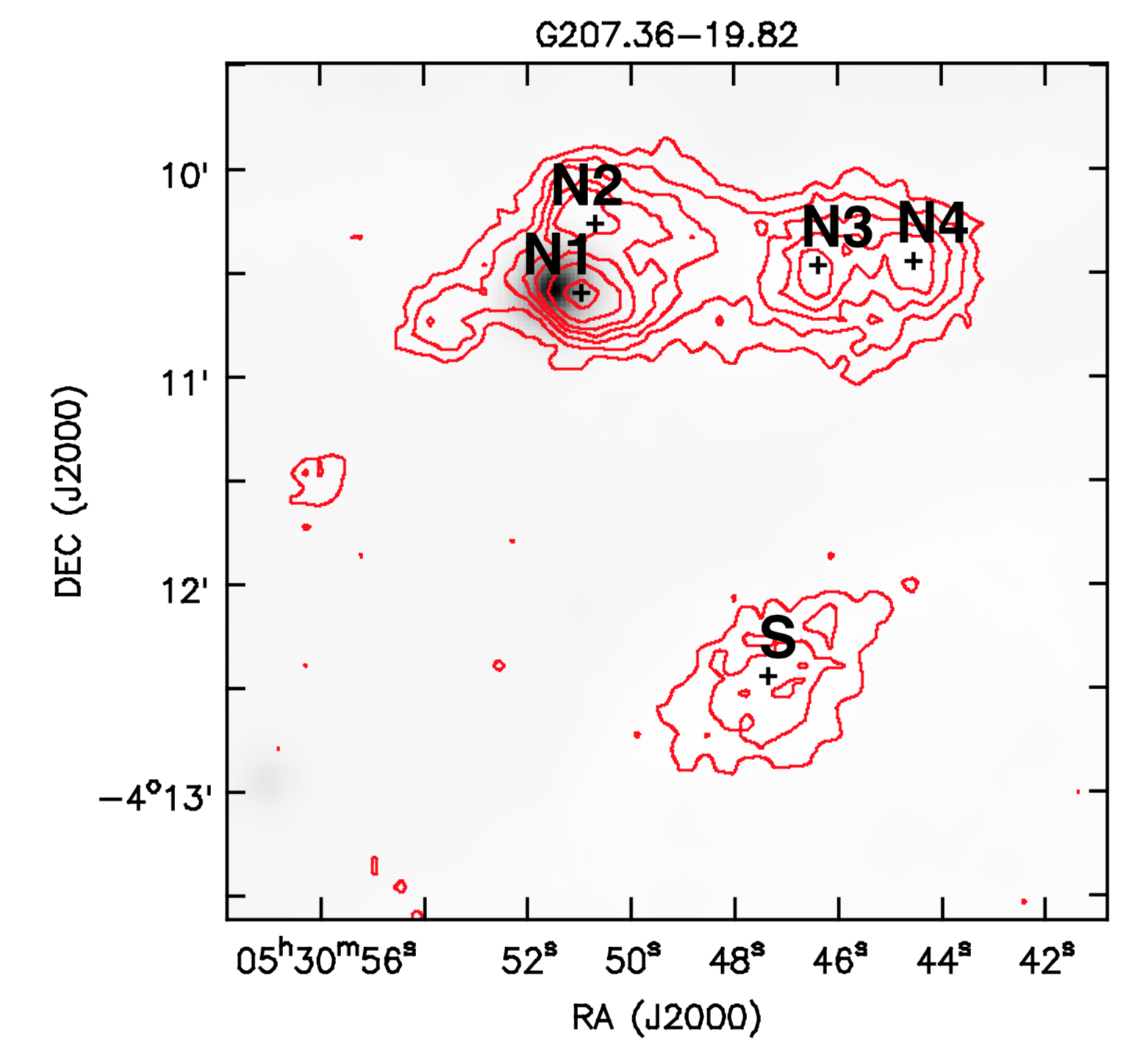}{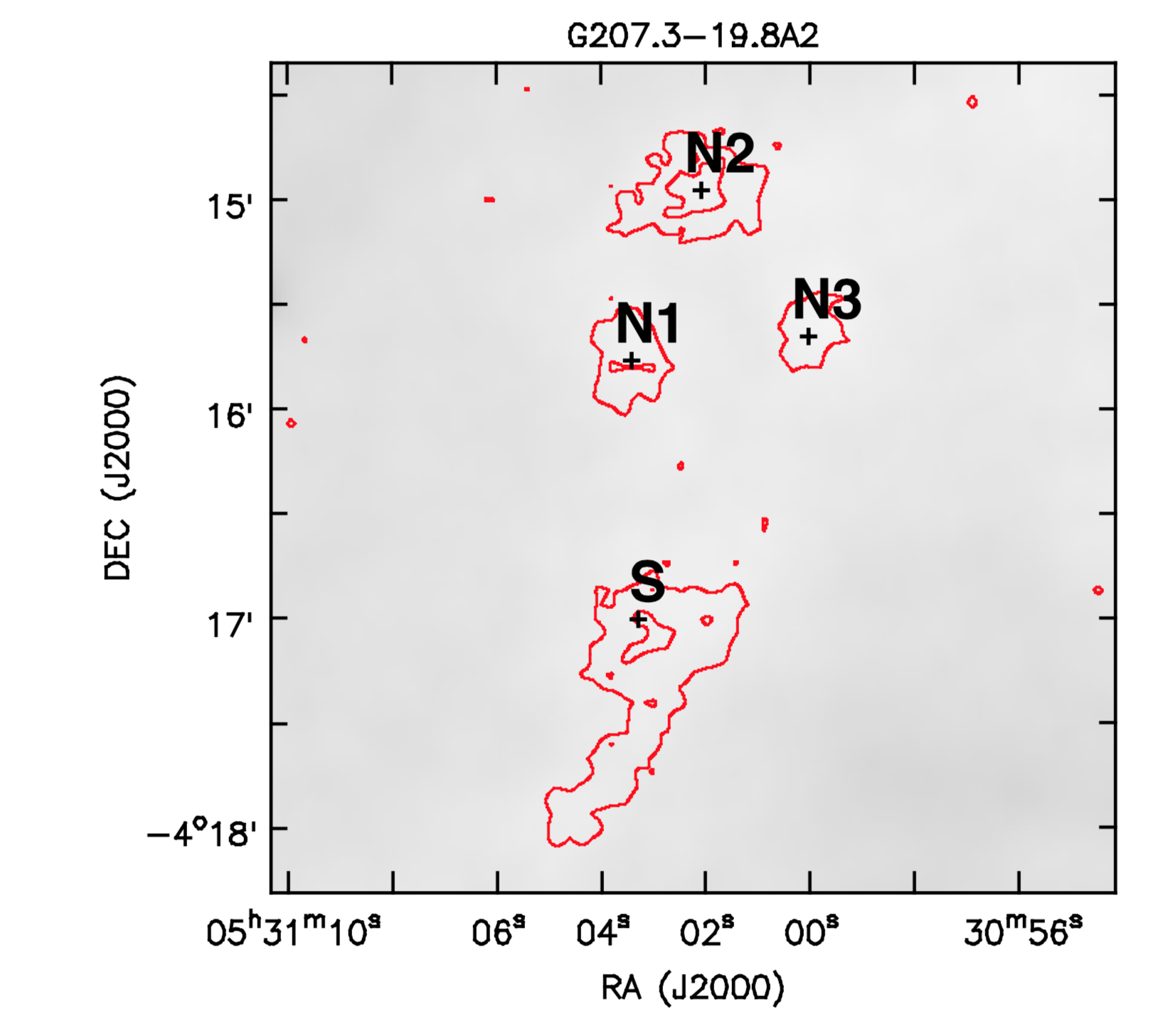}\\
\plottwo{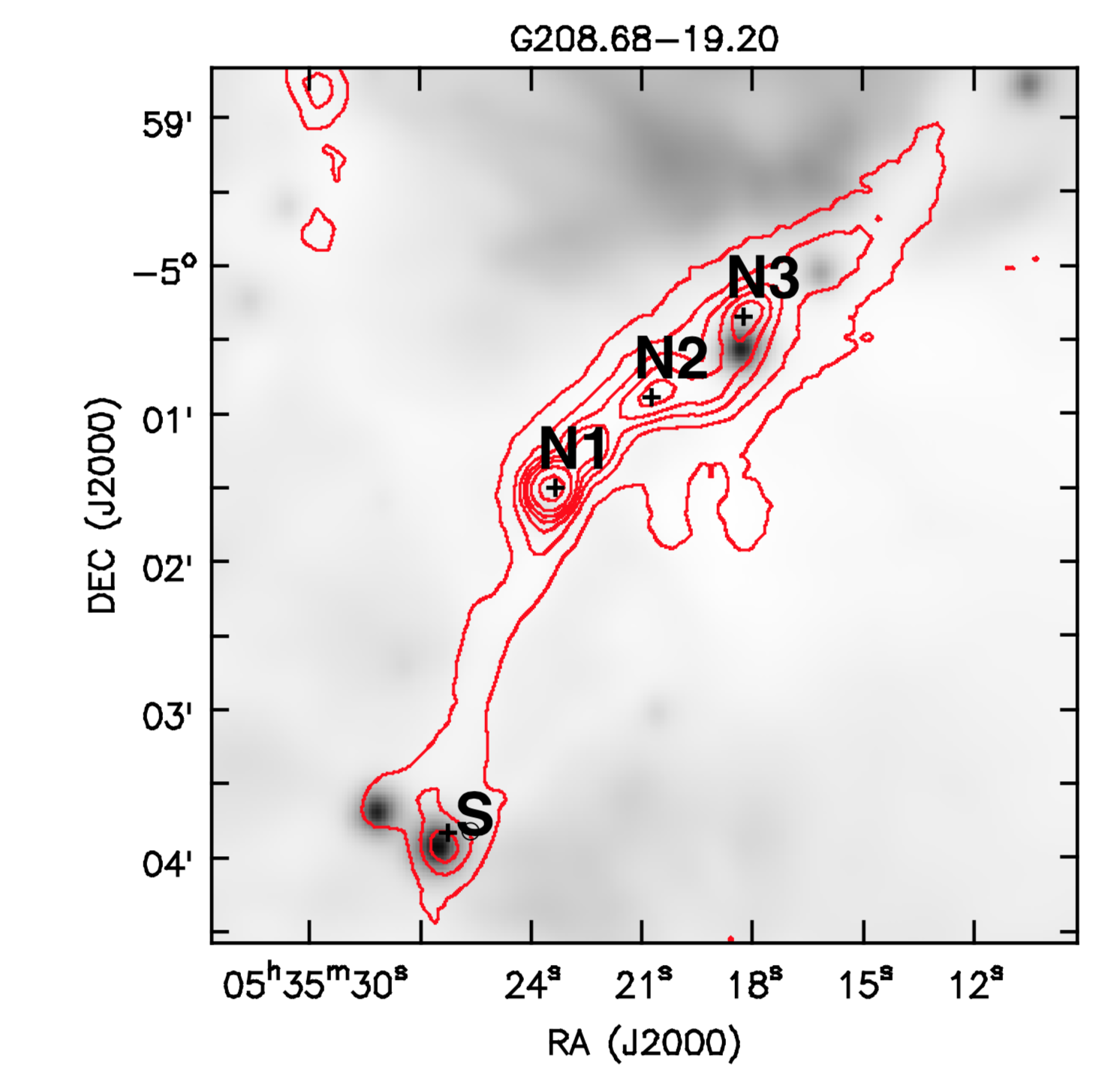}{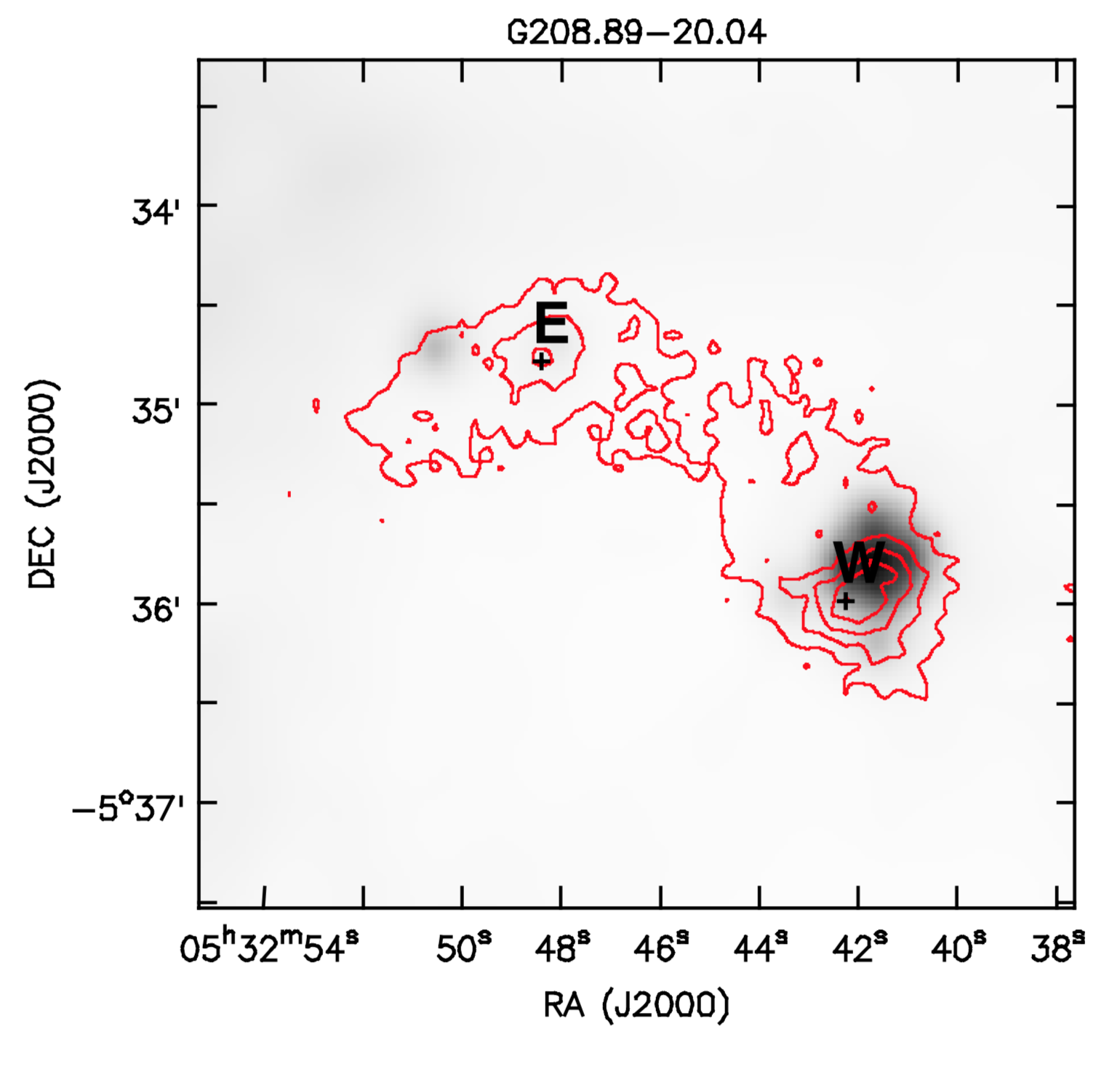}\\
\plottwo{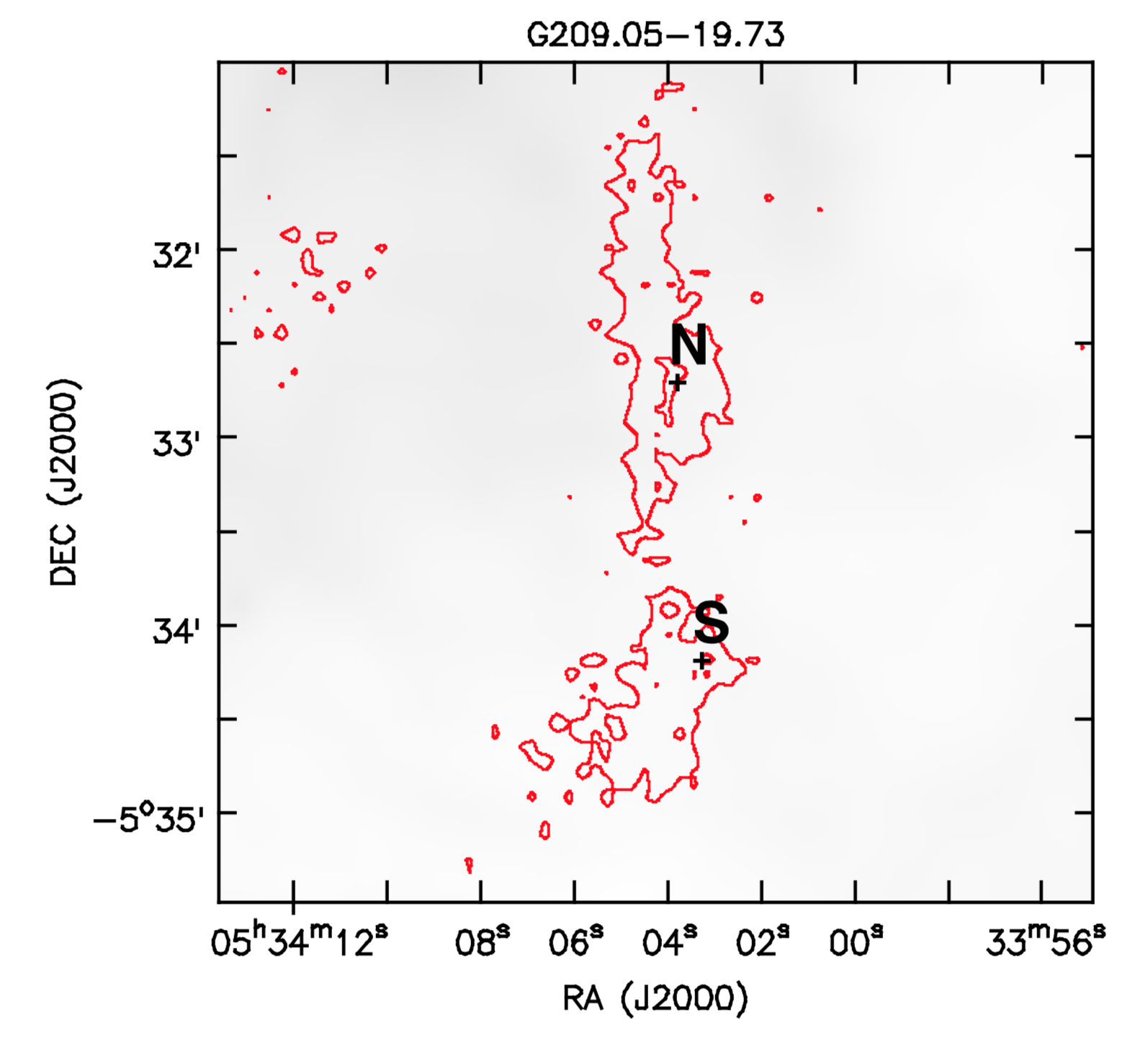}{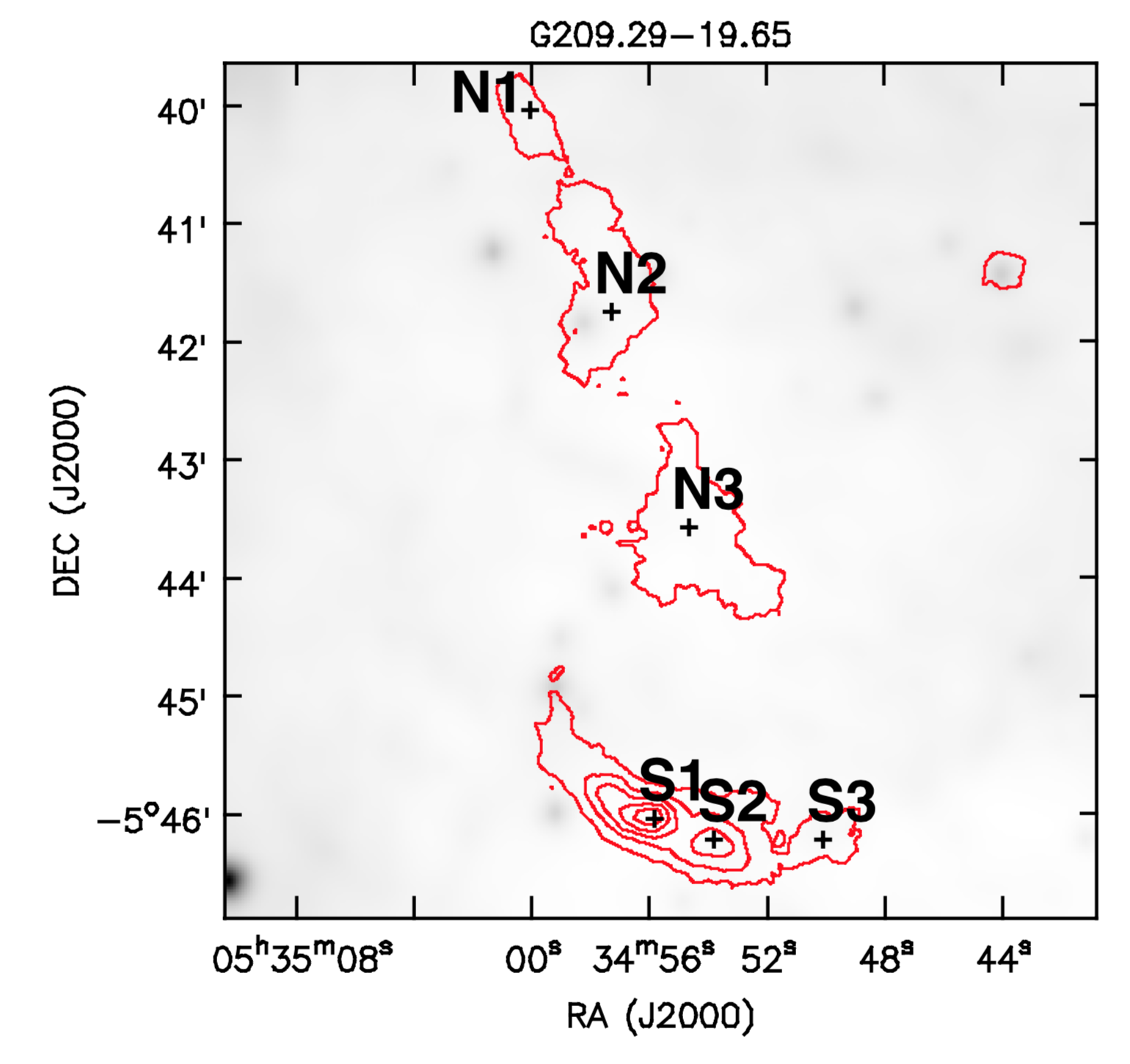}\\
\caption{ Same as in Figure 3 except for the 23 PGCCs detected at 850 $\micron$ with SCUBA-2 in the Orion A cloud. \label{fig:OrionA_PGCCs}}
\end{figure}

\clearpage
\begin{figure}
\epsscale{0.8}
\plottwo{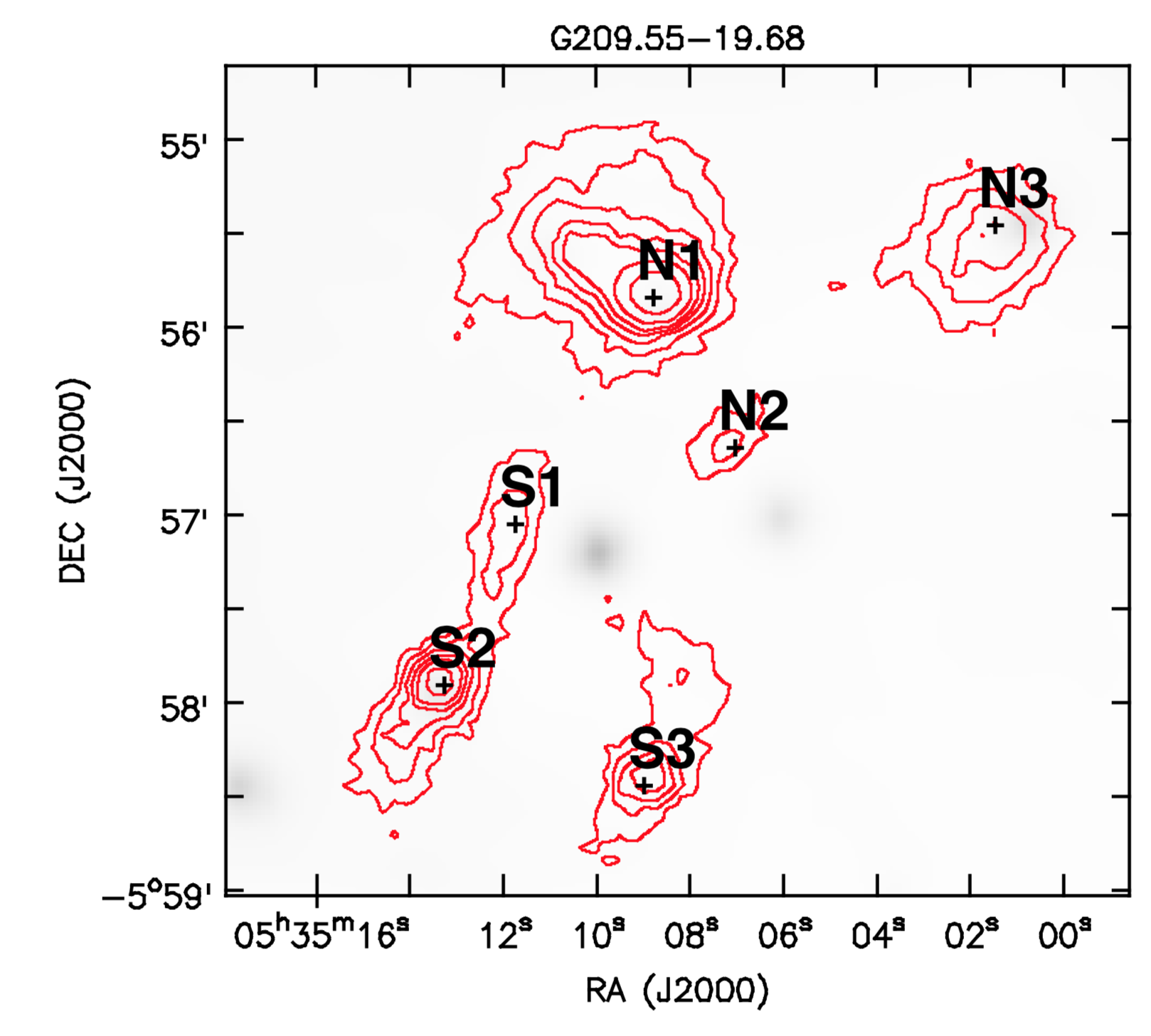}{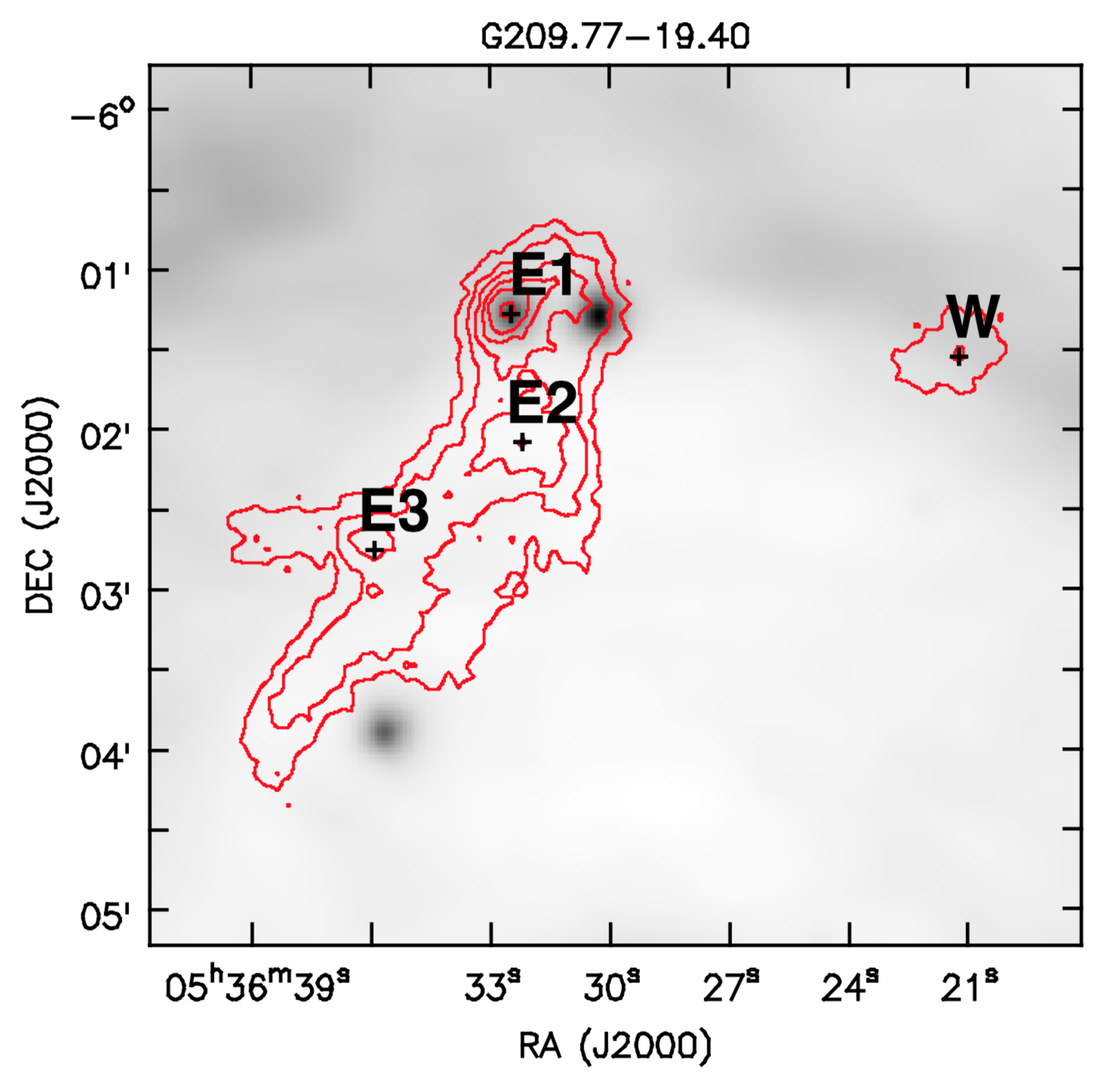}\\
\plottwo{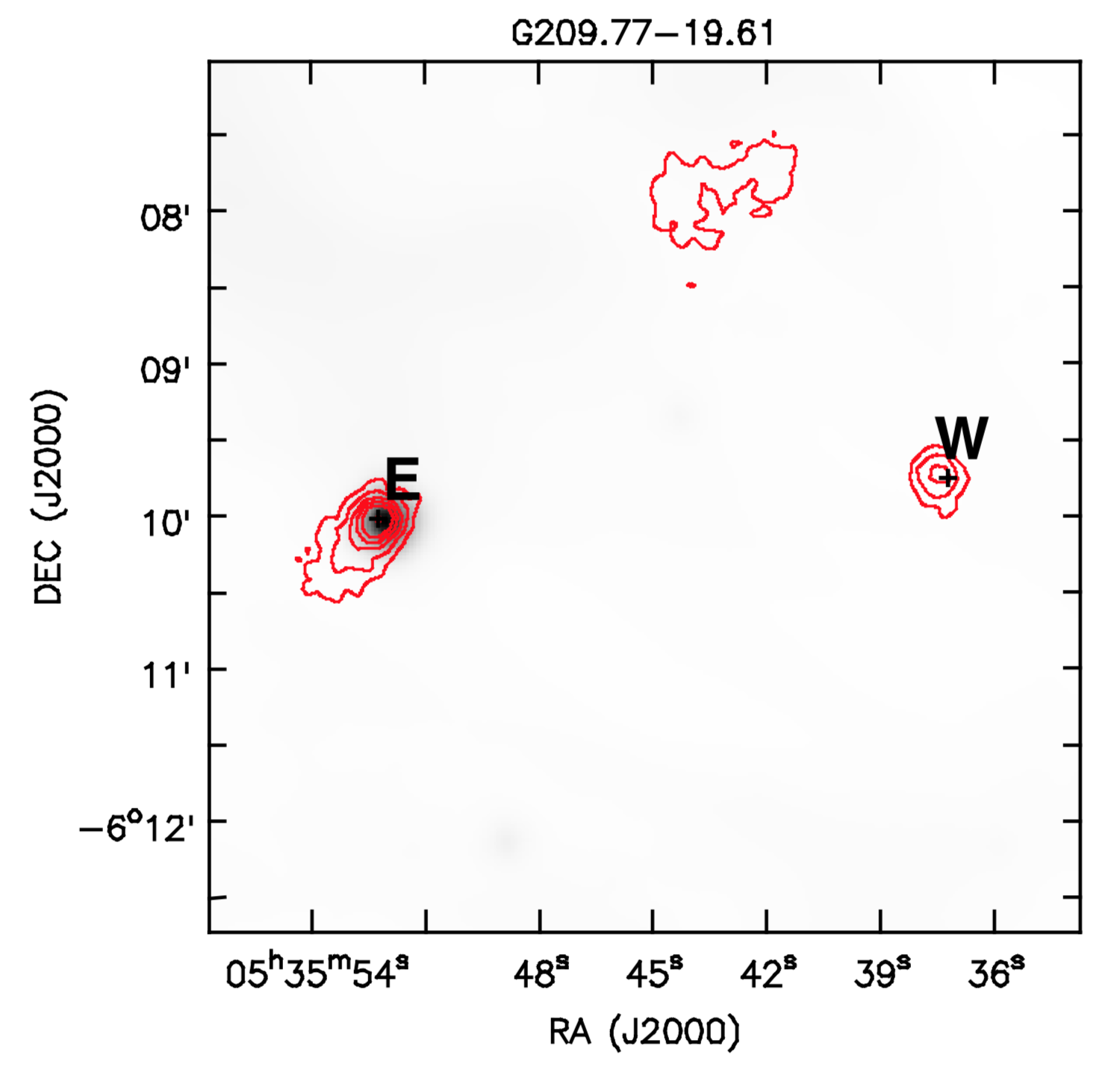}{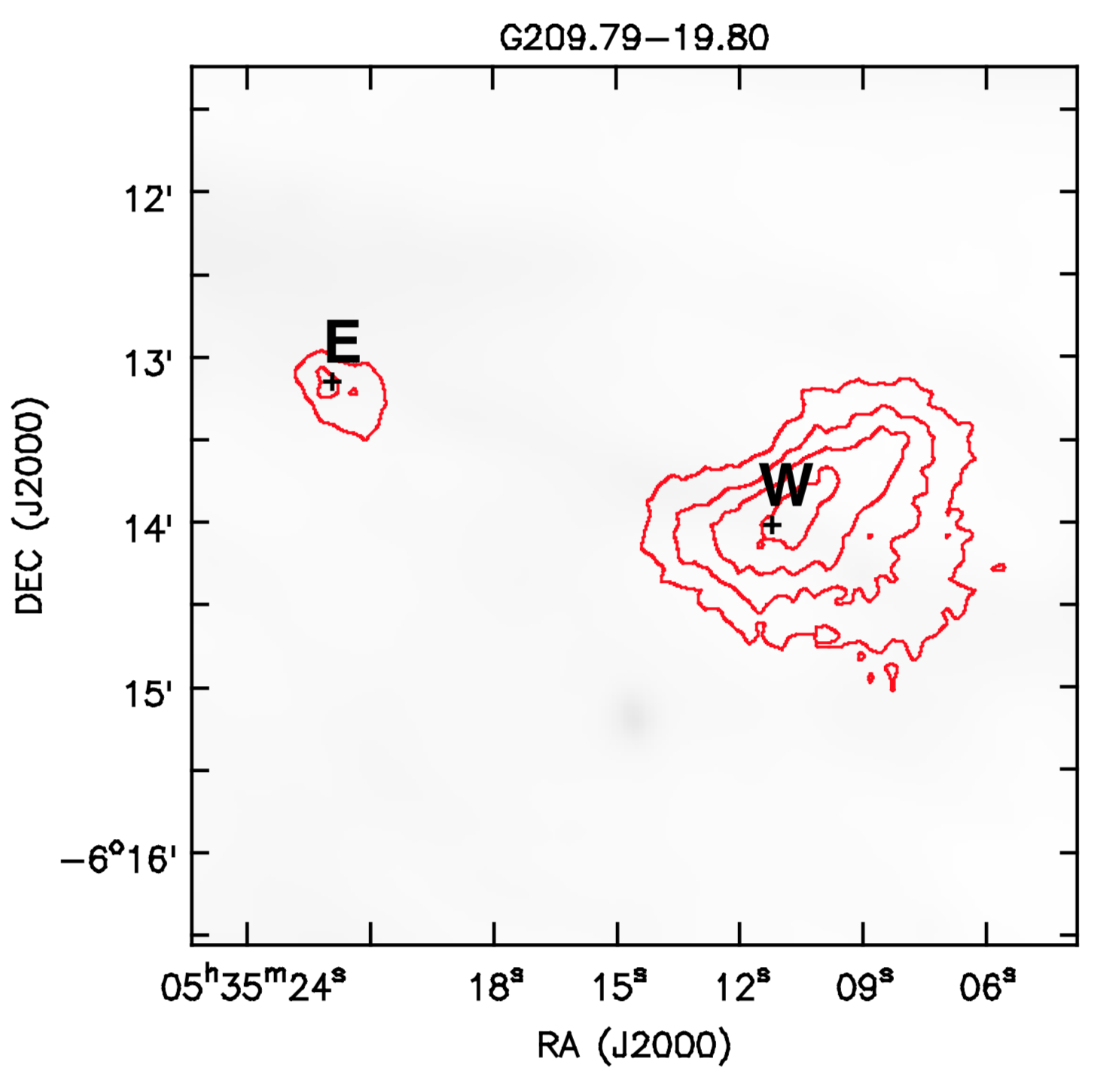}\\
\plottwo{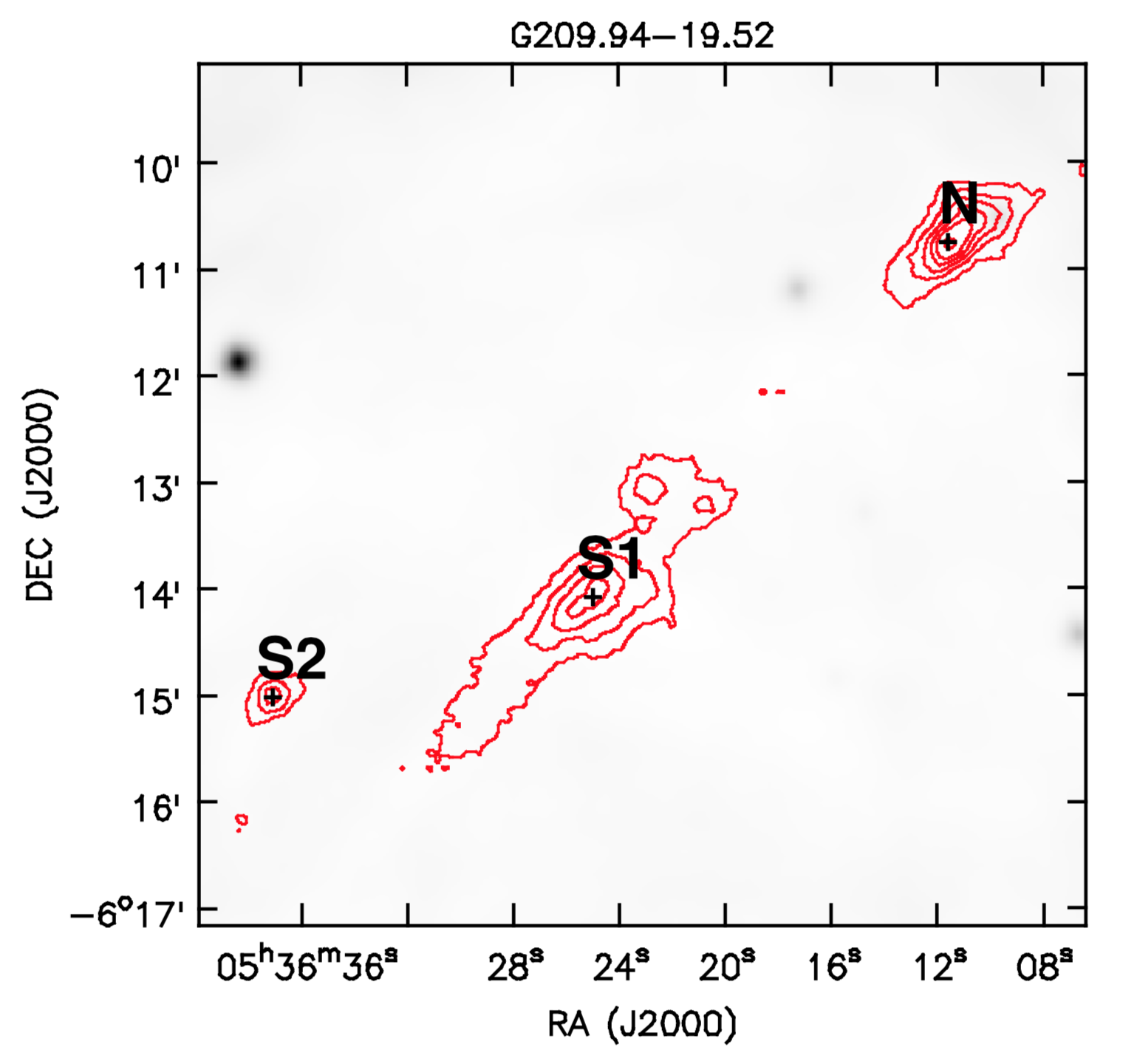}{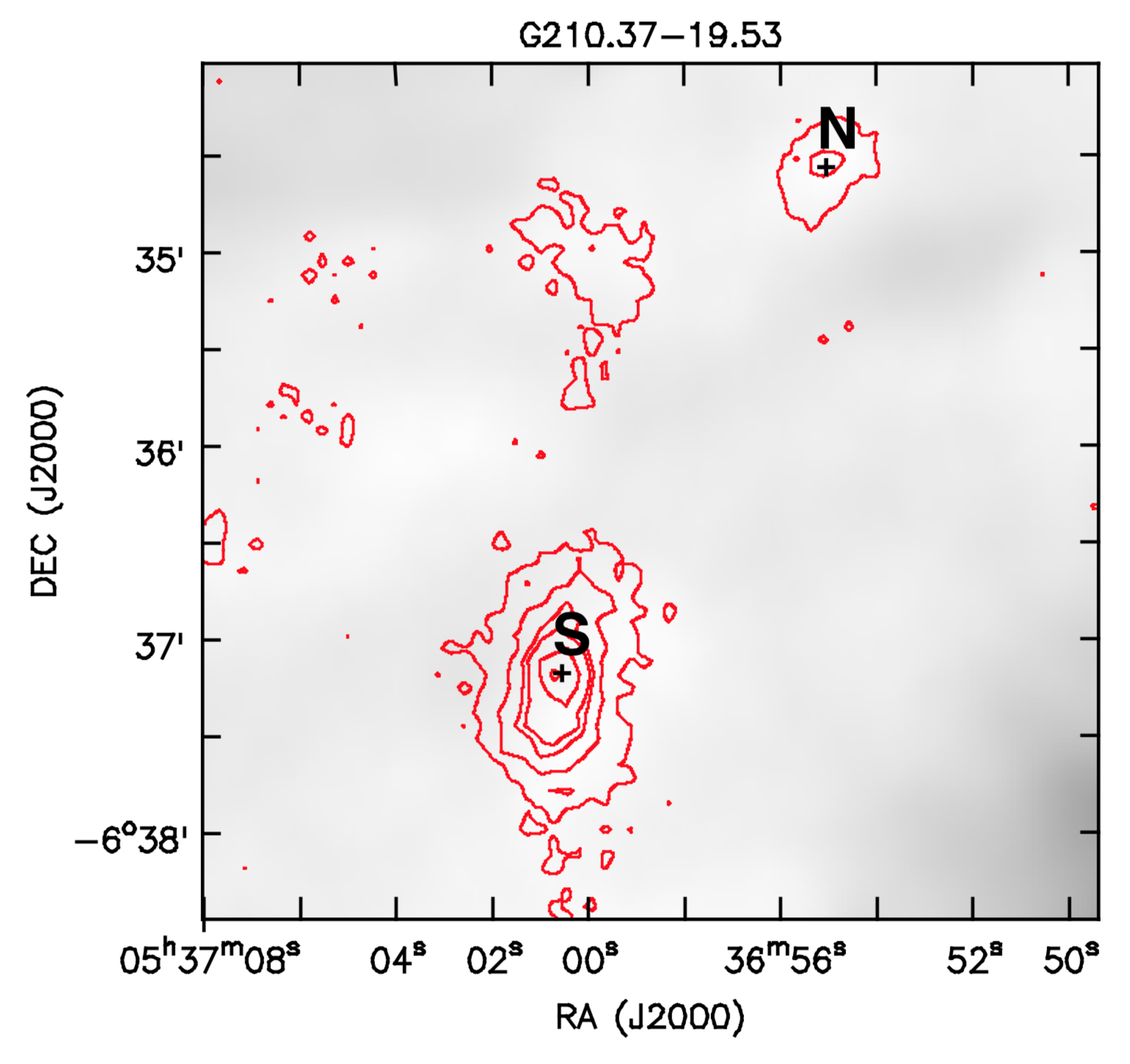}\\
\end{figure}

\clearpage
\begin{figure}
\epsscale{0.8}
\plottwo{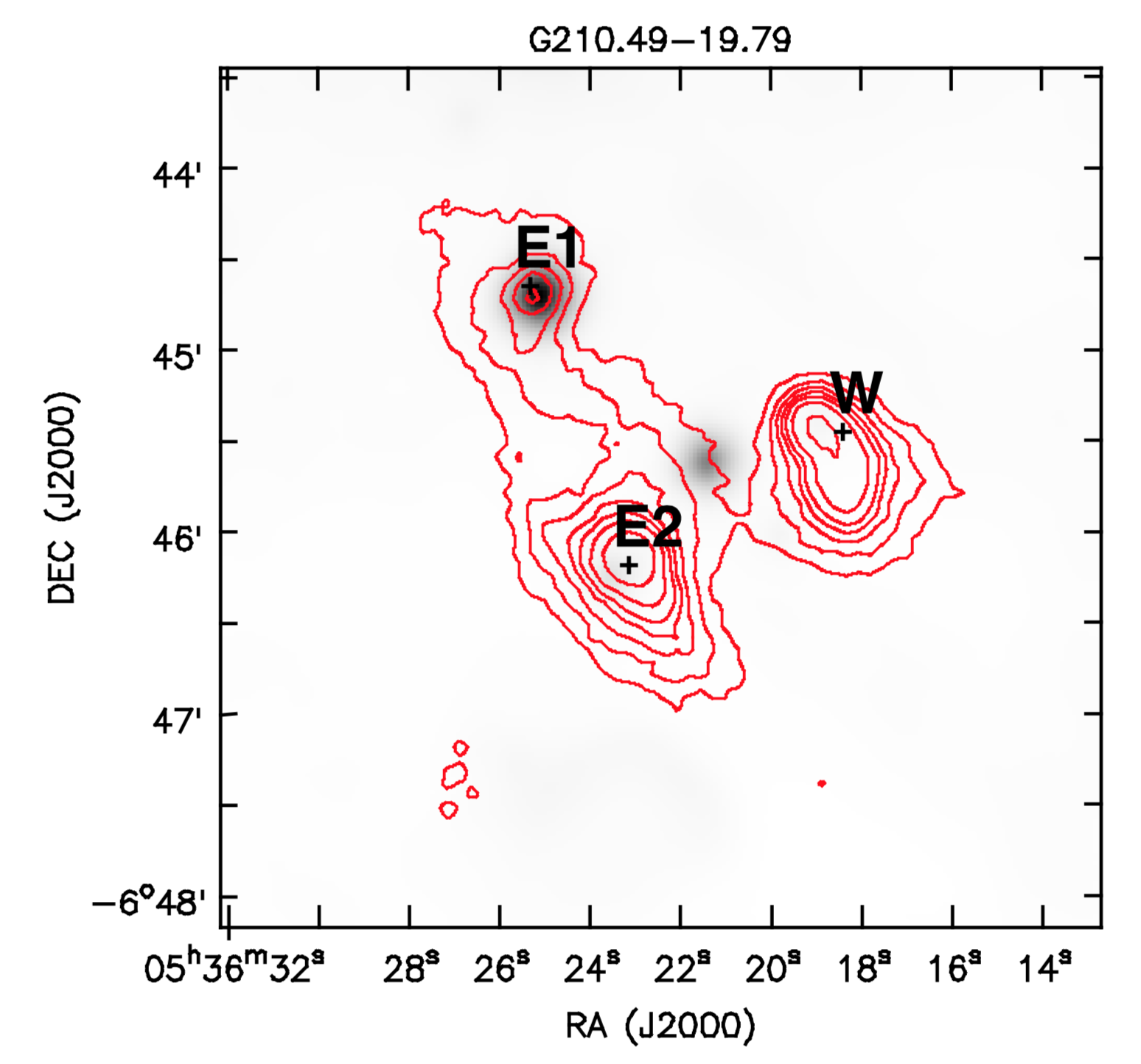}{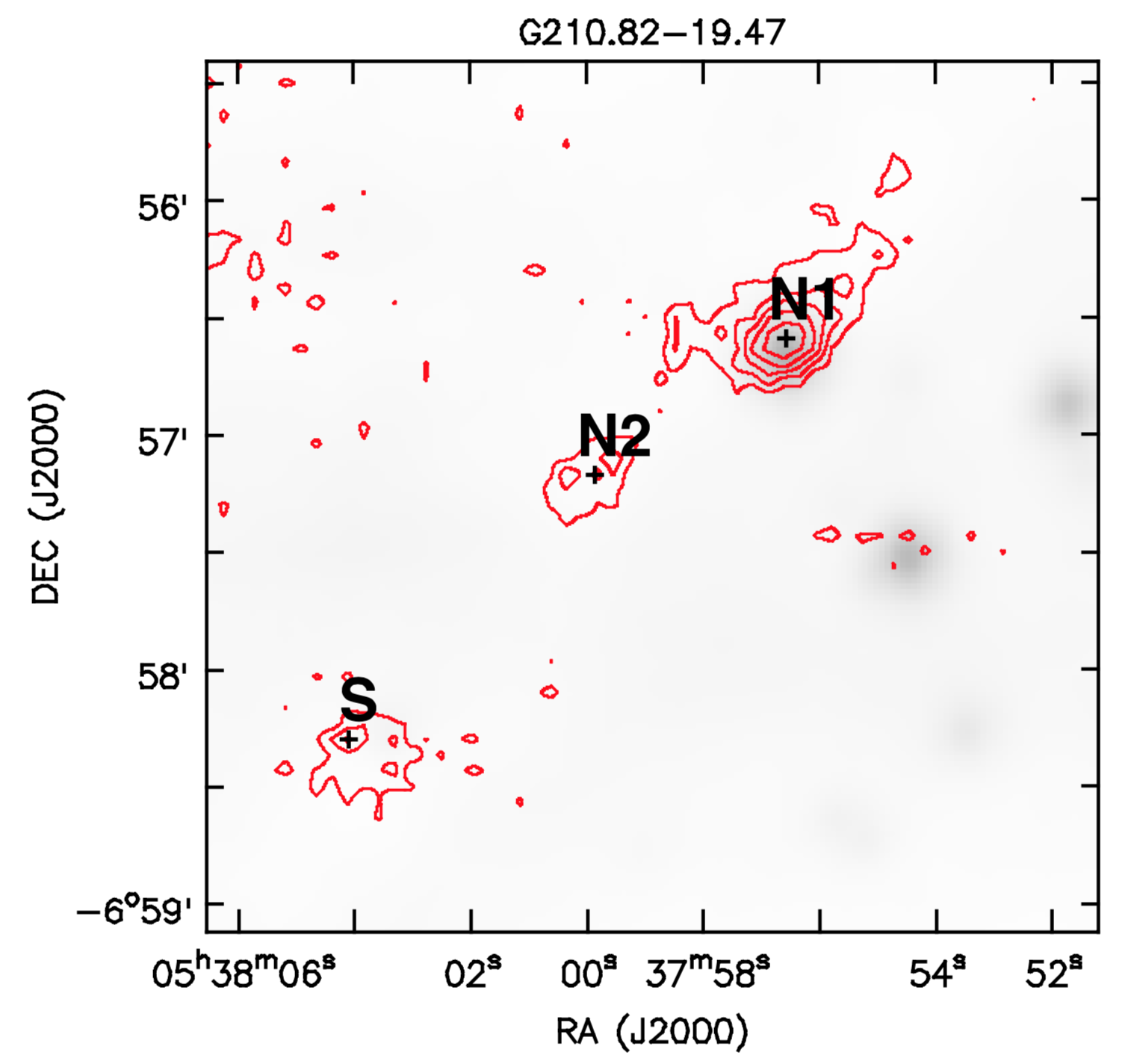}\\
\plottwo{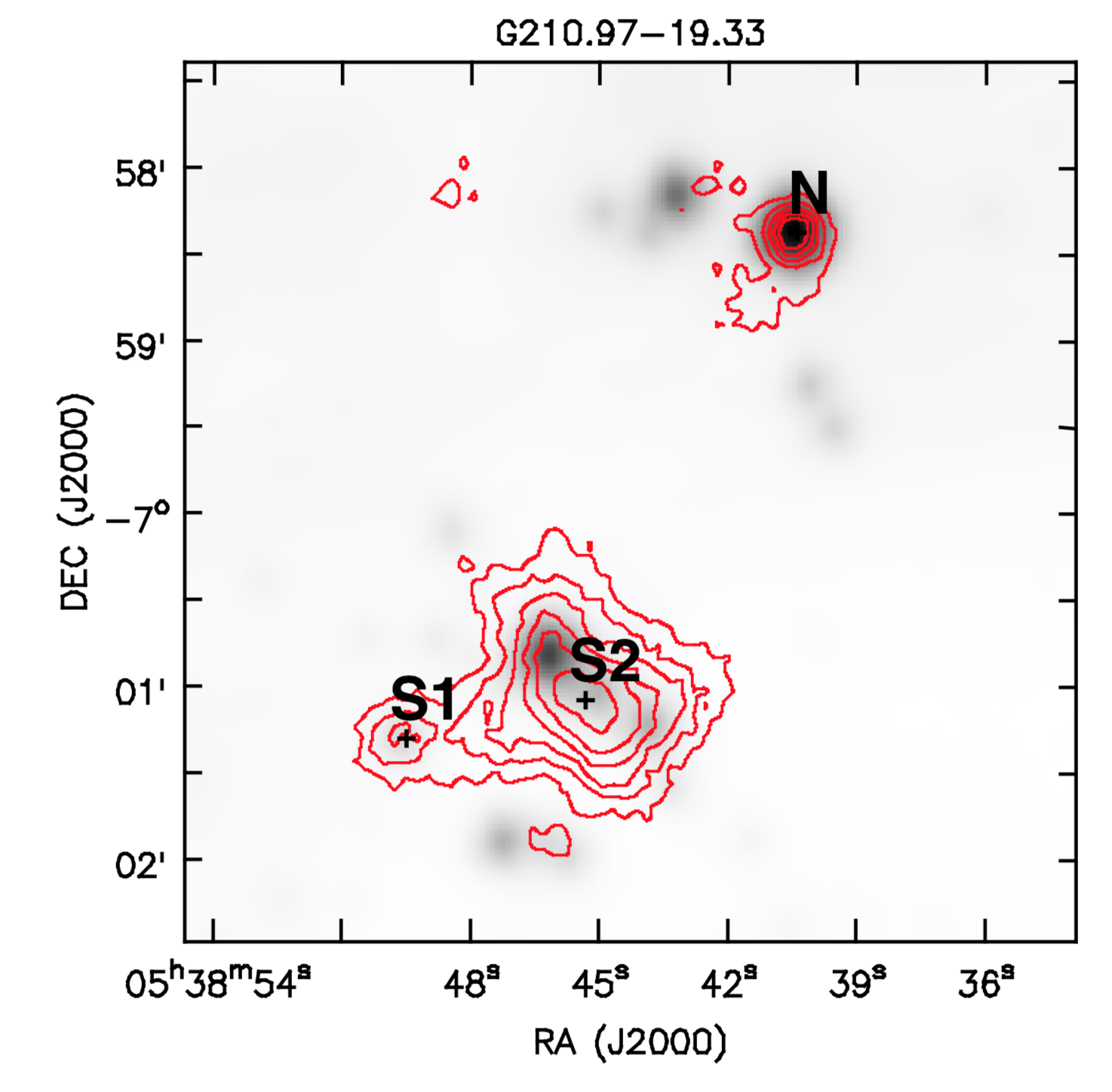}{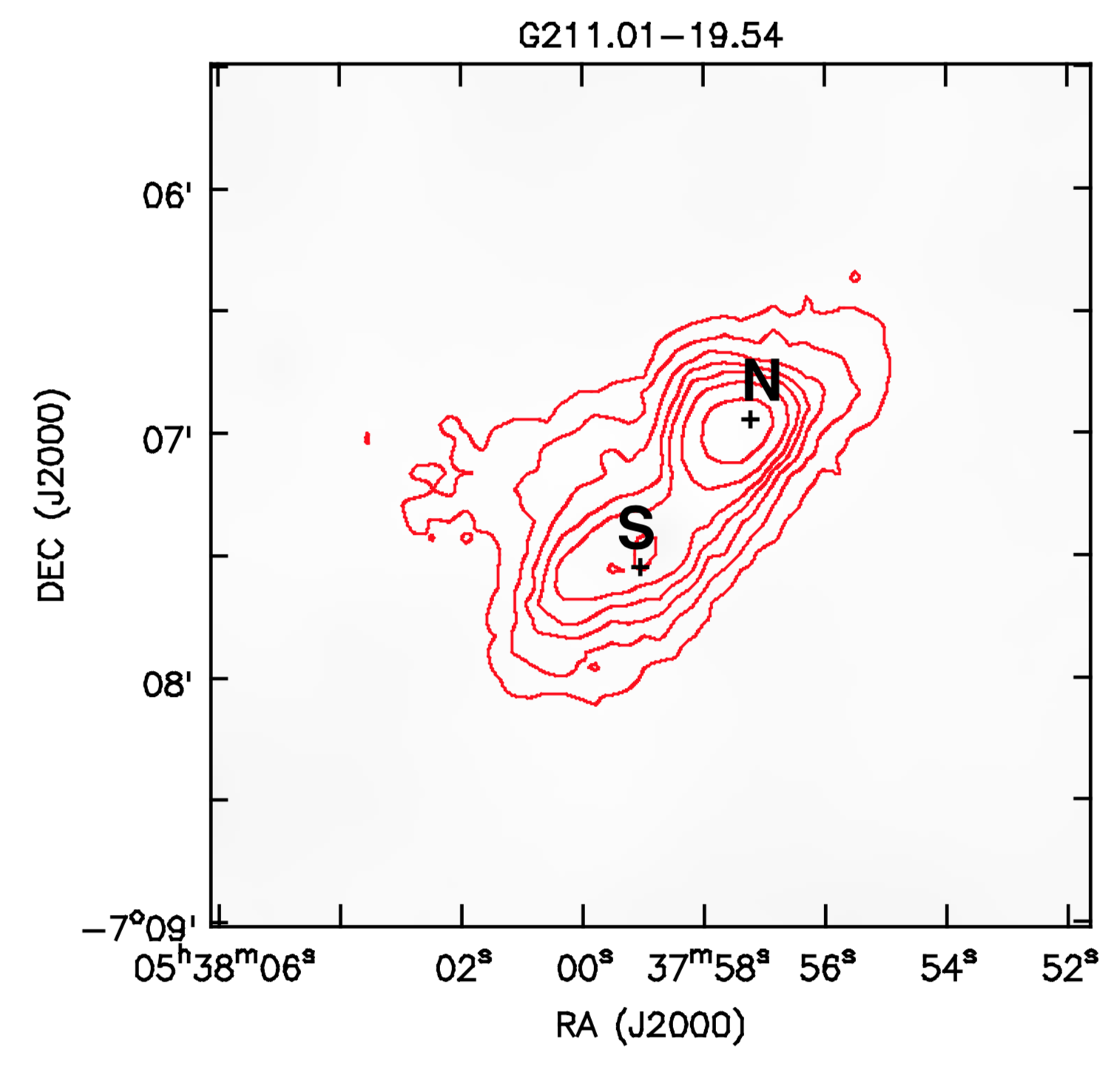}\\
\plottwo{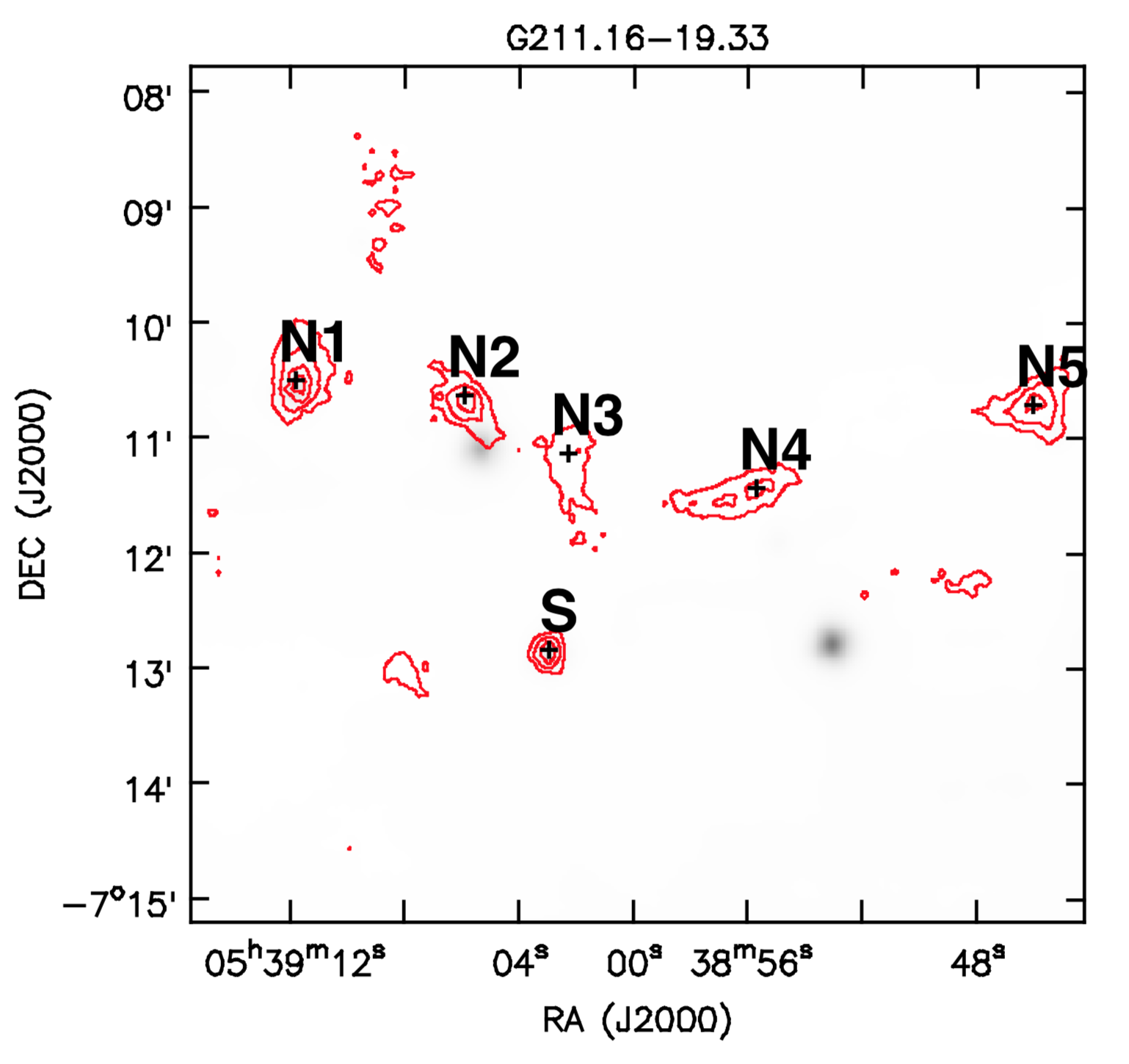}{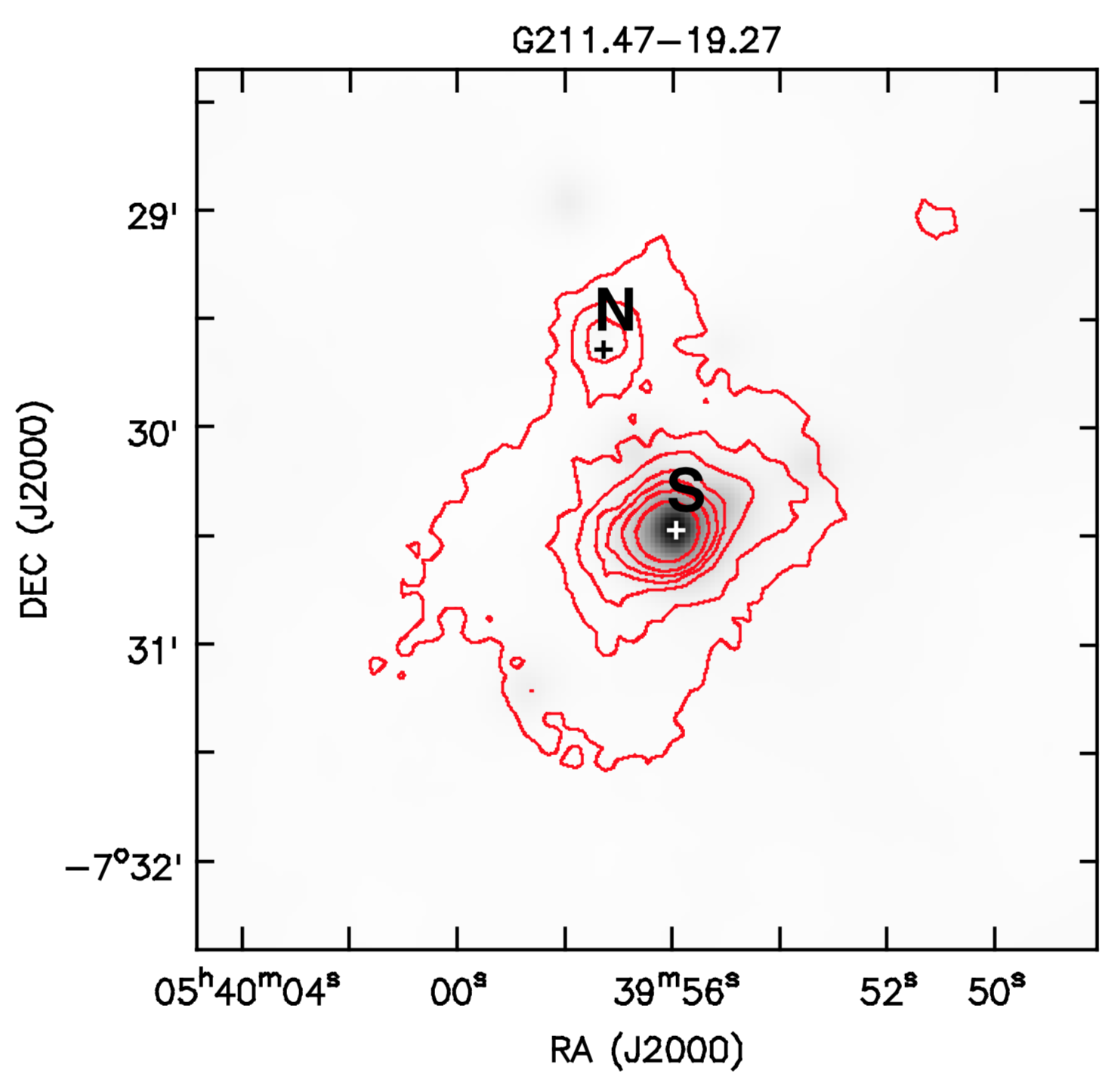}\\

\end{figure}
\clearpage
\begin{figure}
\epsscale{0.8}
\plottwo{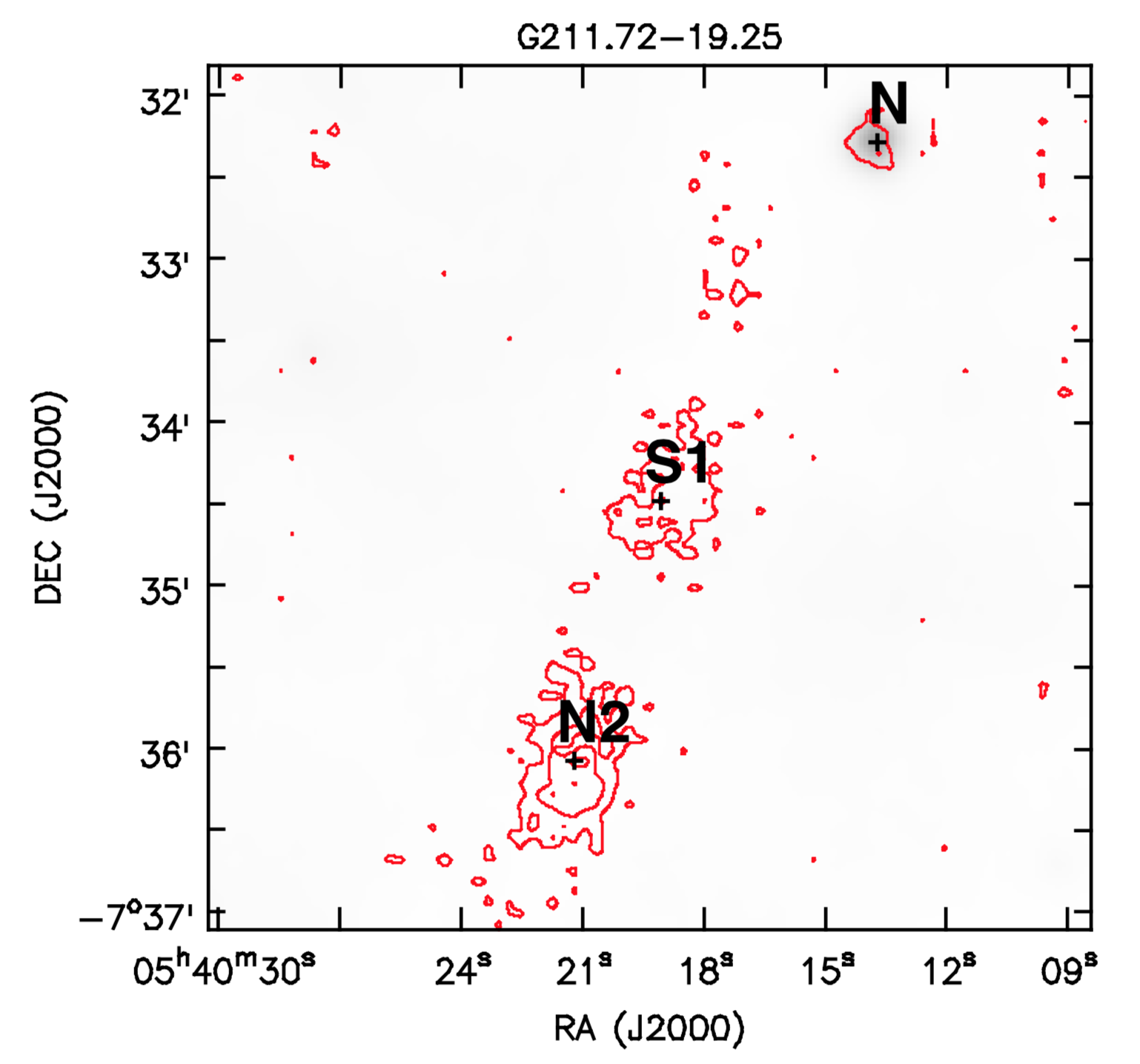}{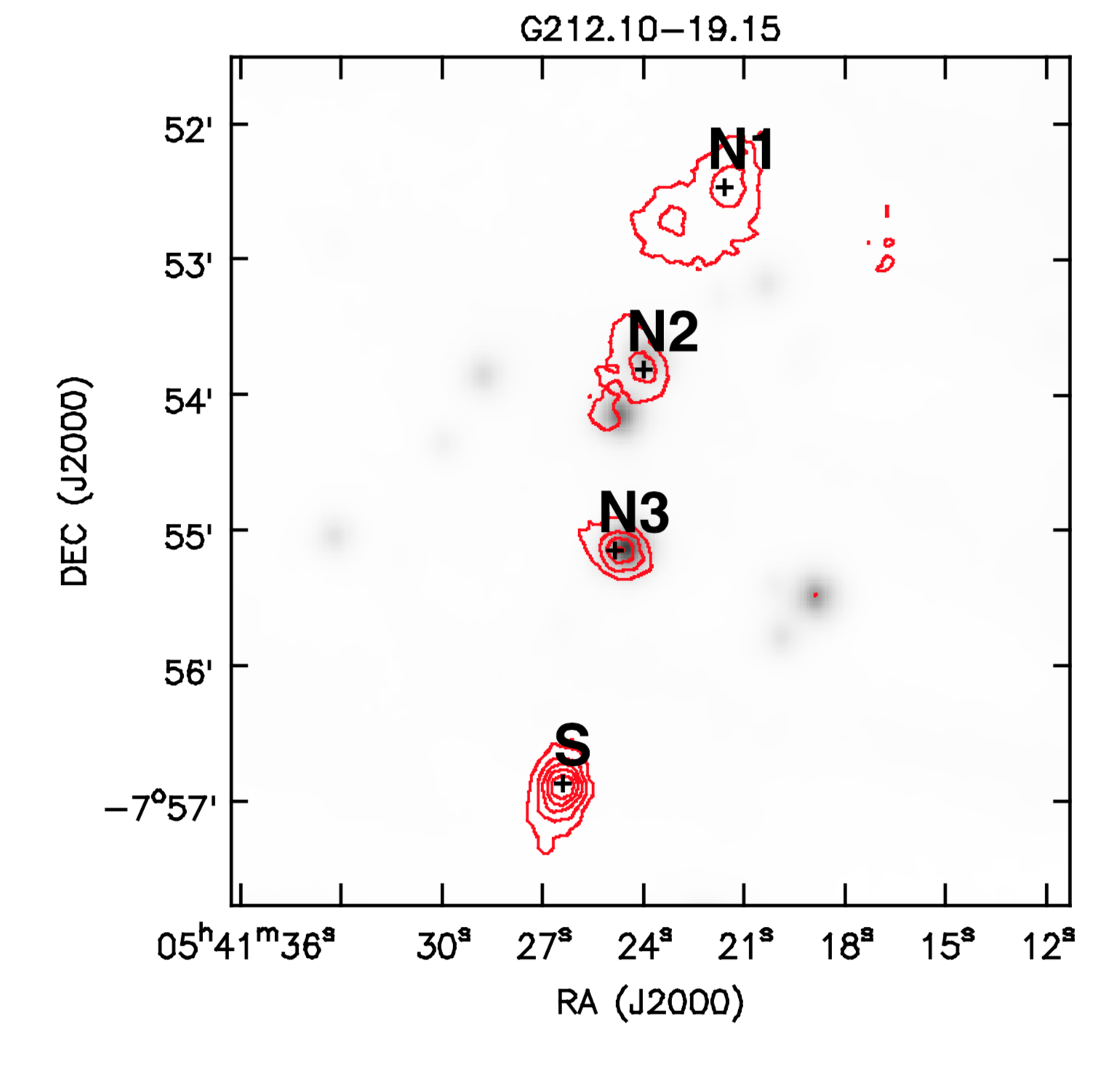}\\
\plottwo{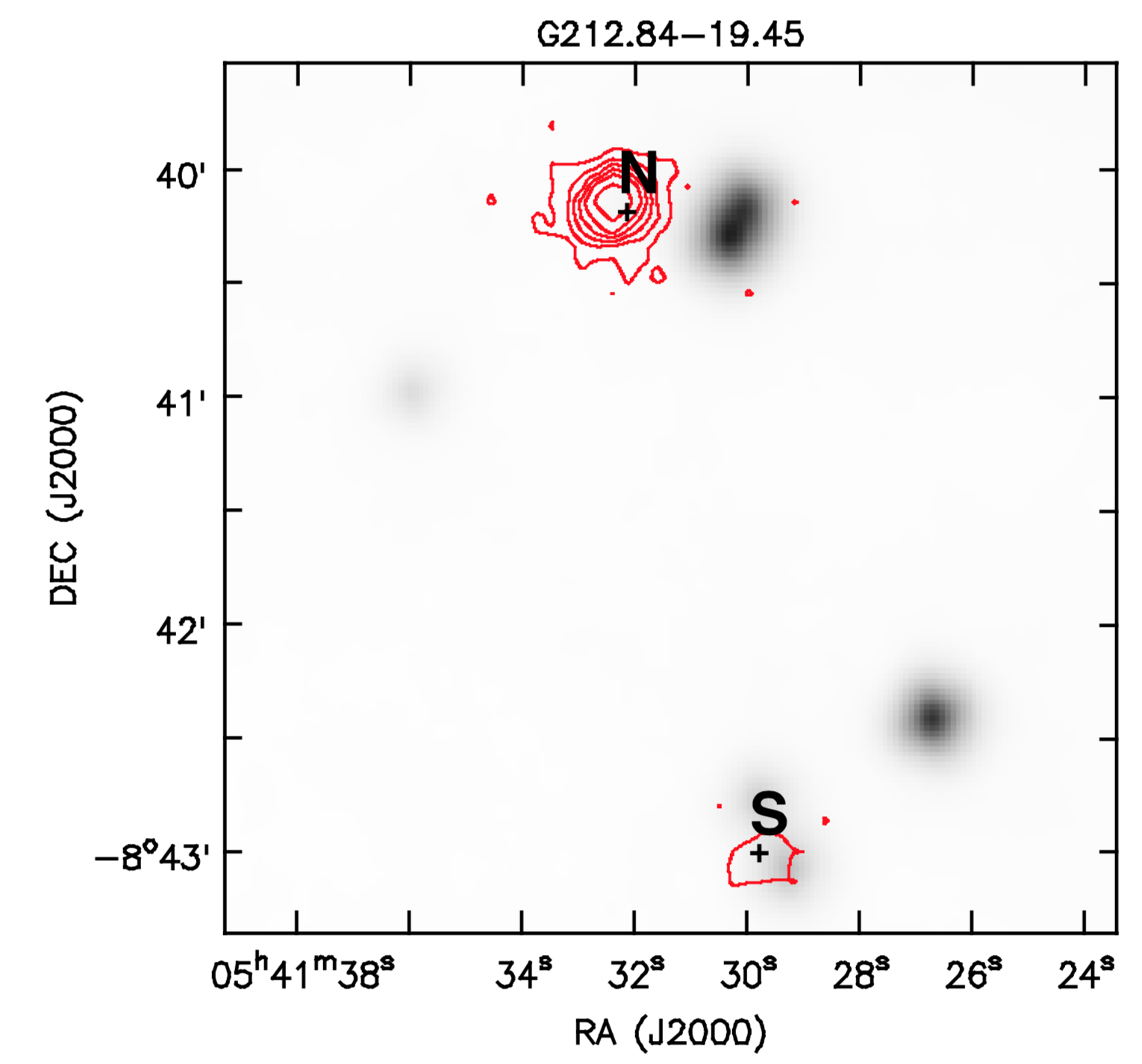}{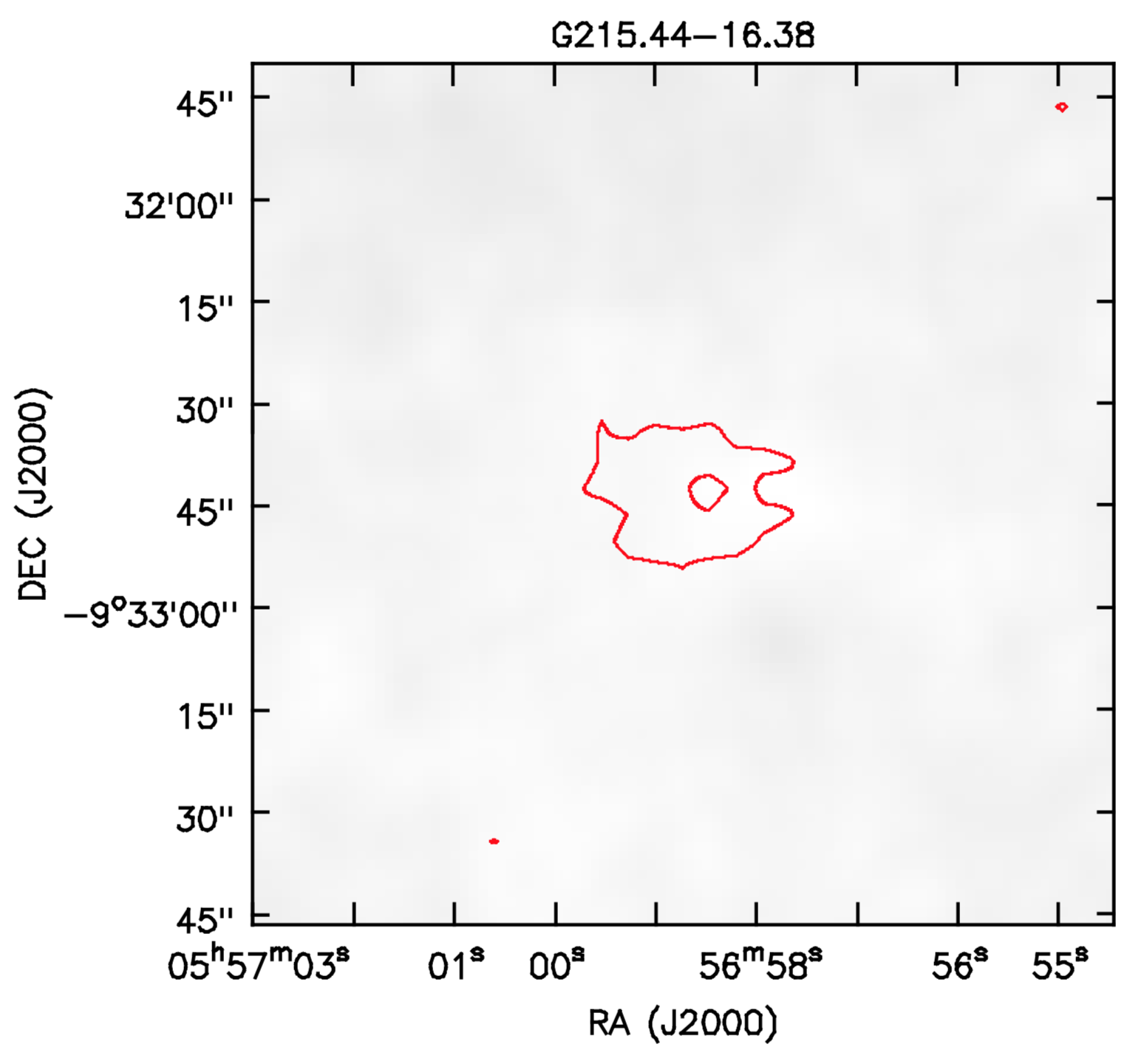}\\
\vspace*{1cm}
\hspace*{0.3cm}
 \includegraphics[width=0.35\textwidth]{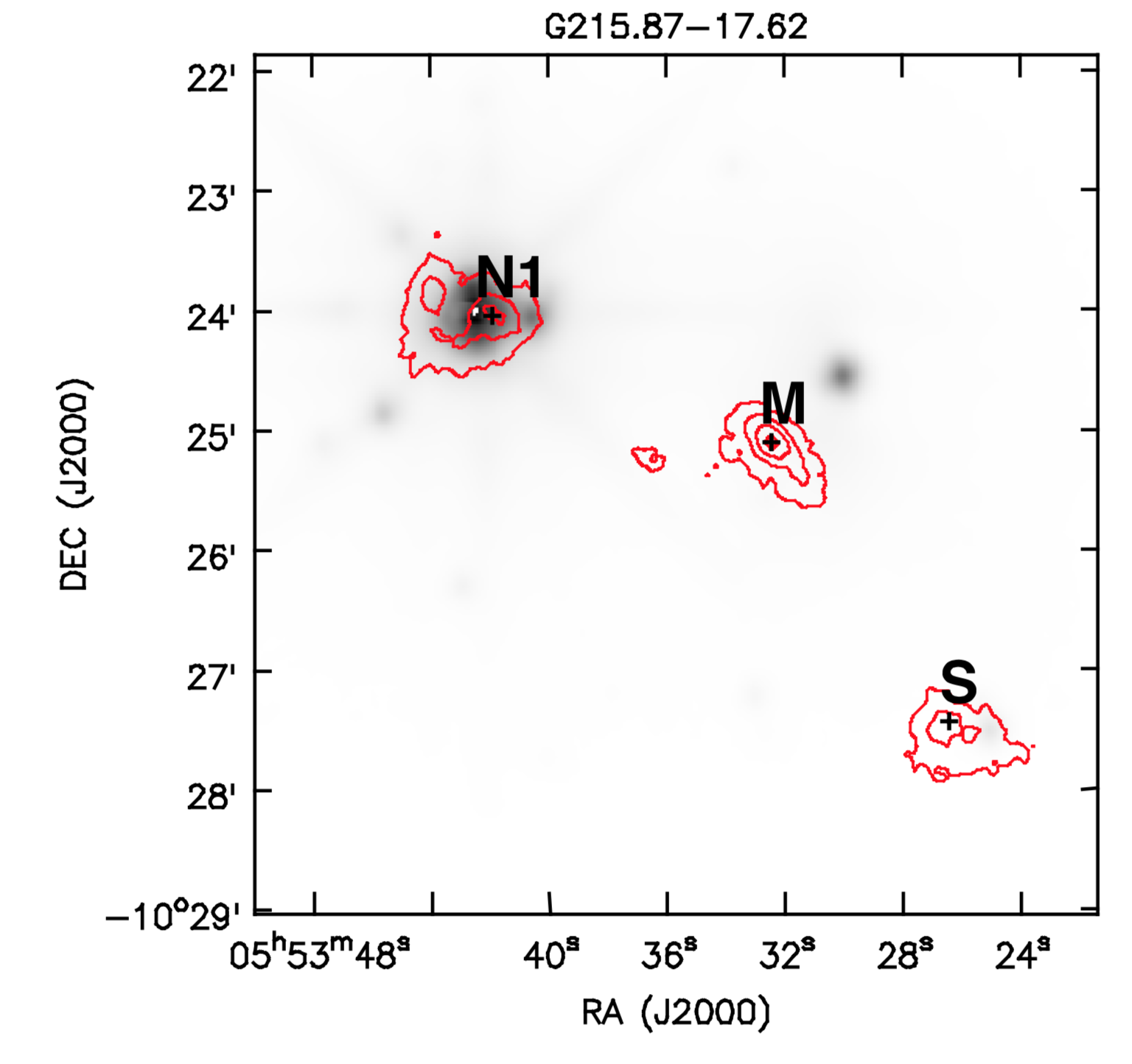}
\end{figure}

\clearpage
\begin{figure}
\epsscale{0.8}
\plottwo{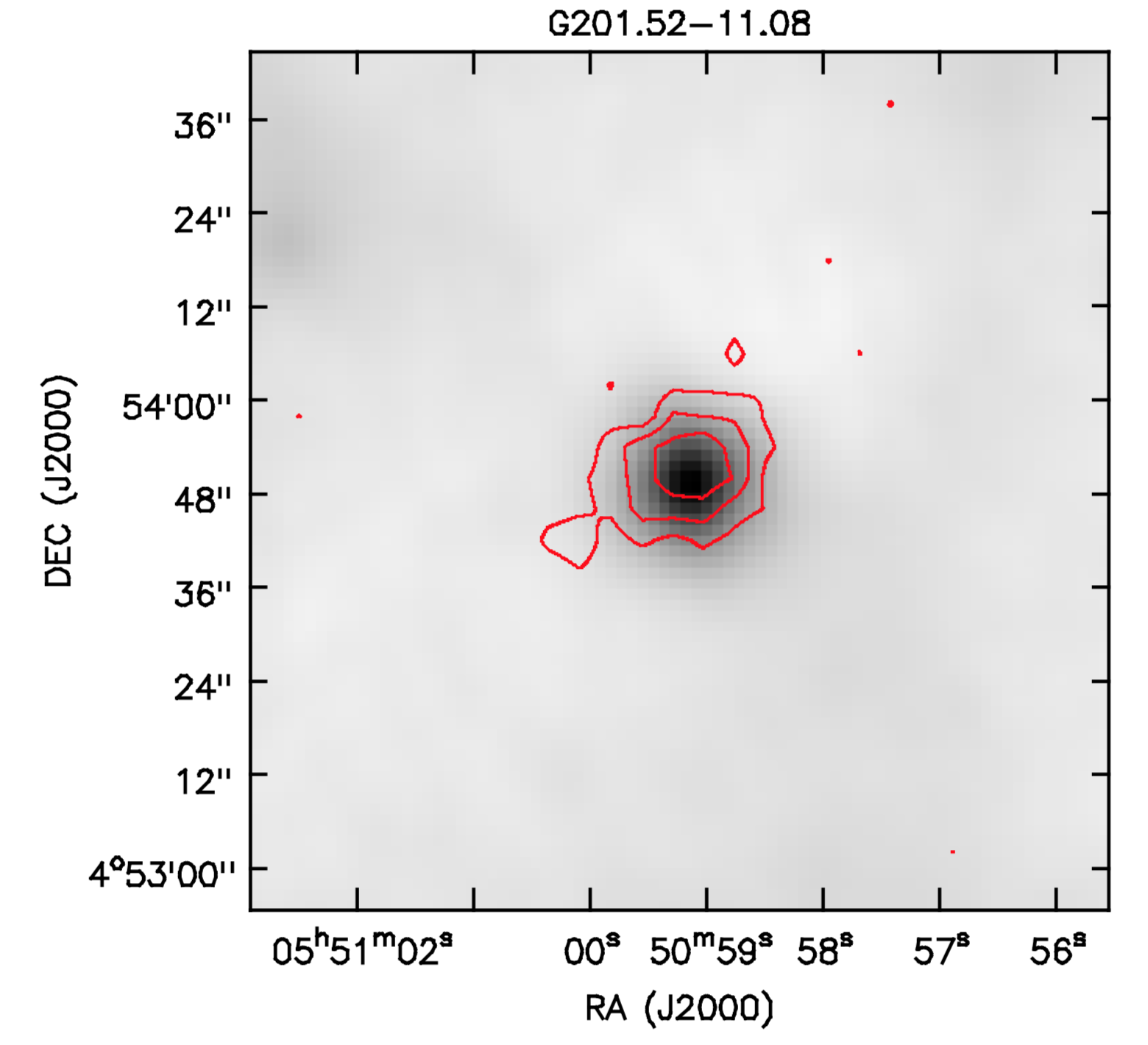}{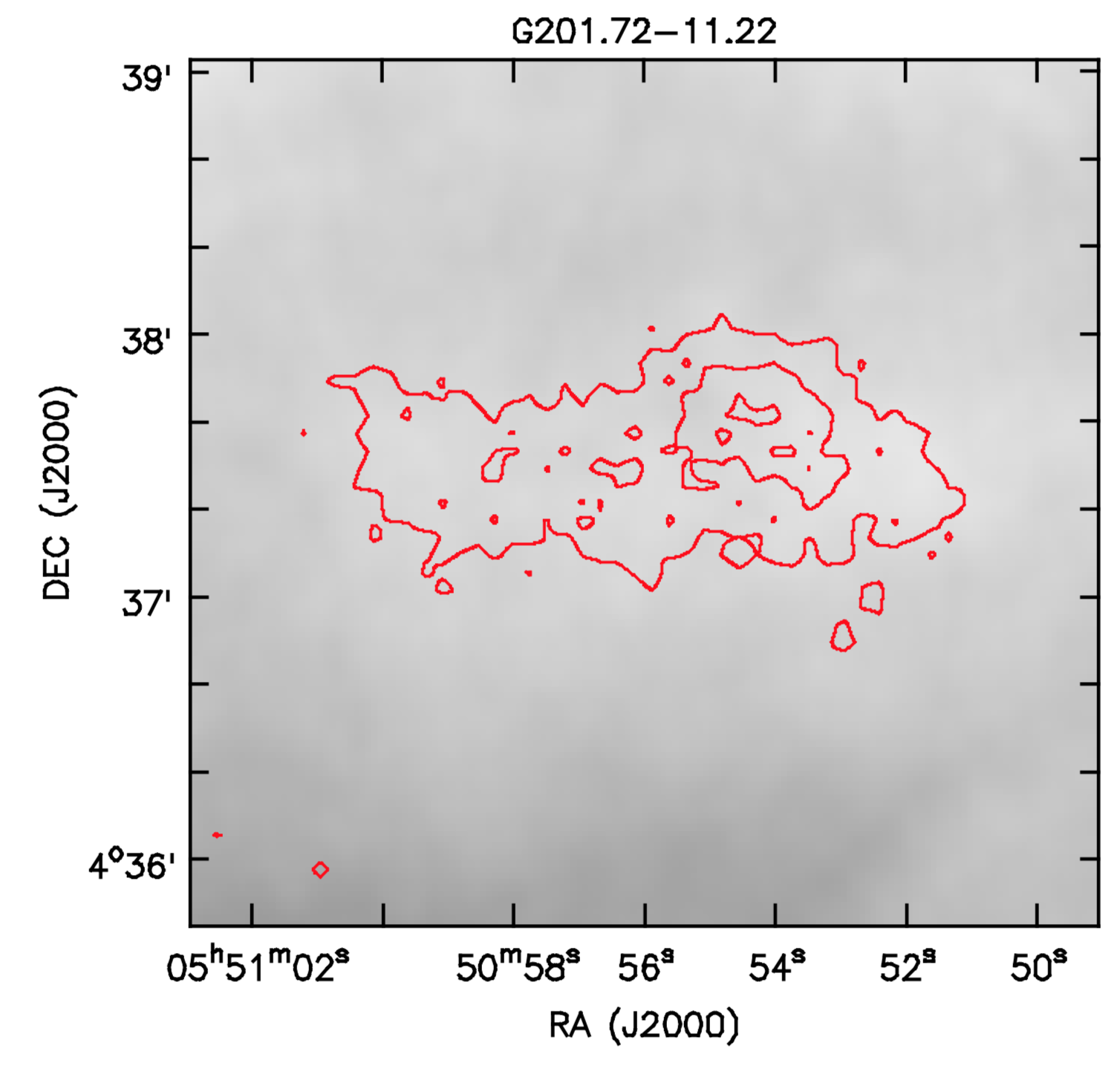}\\
\plottwo{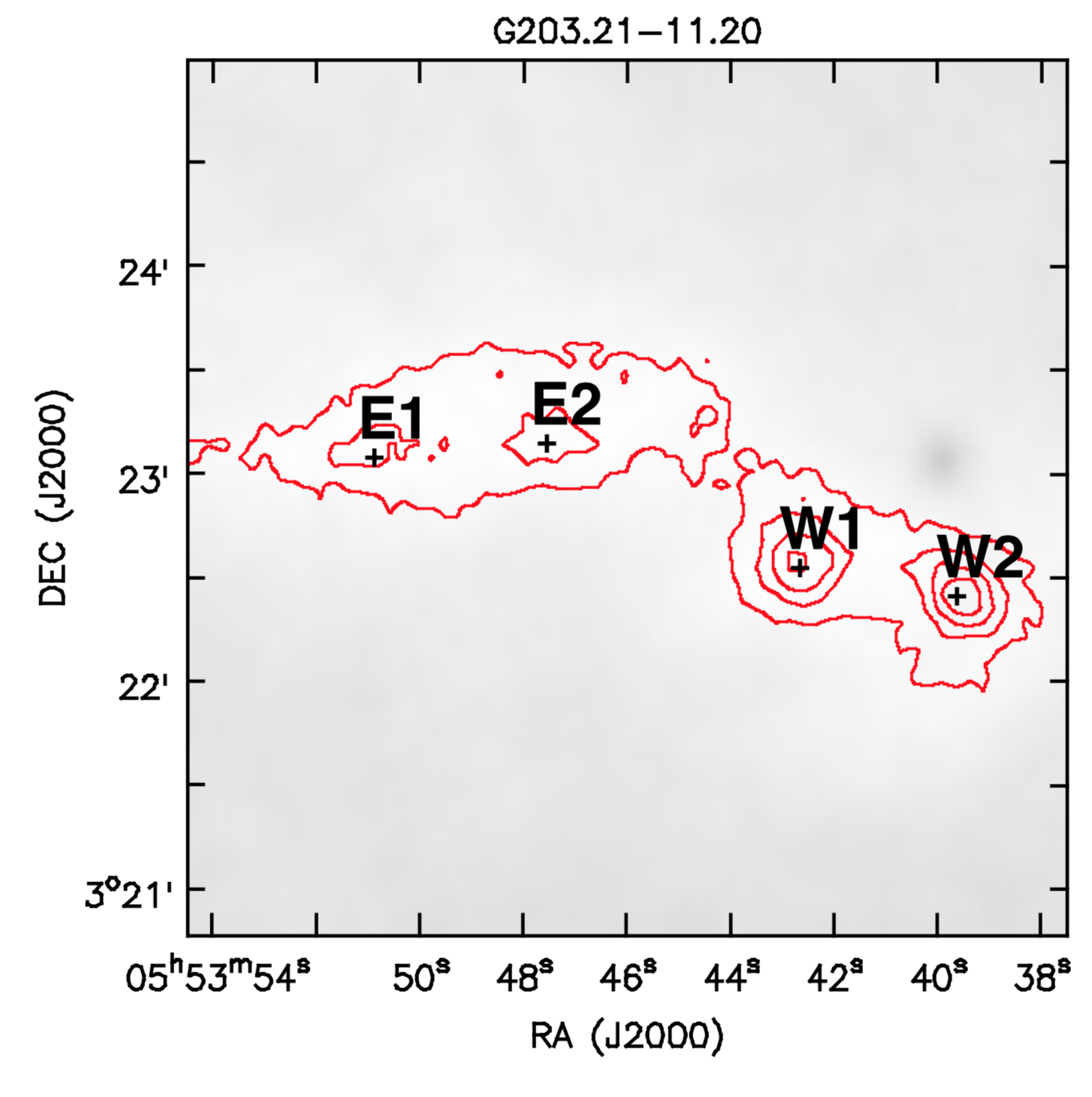}{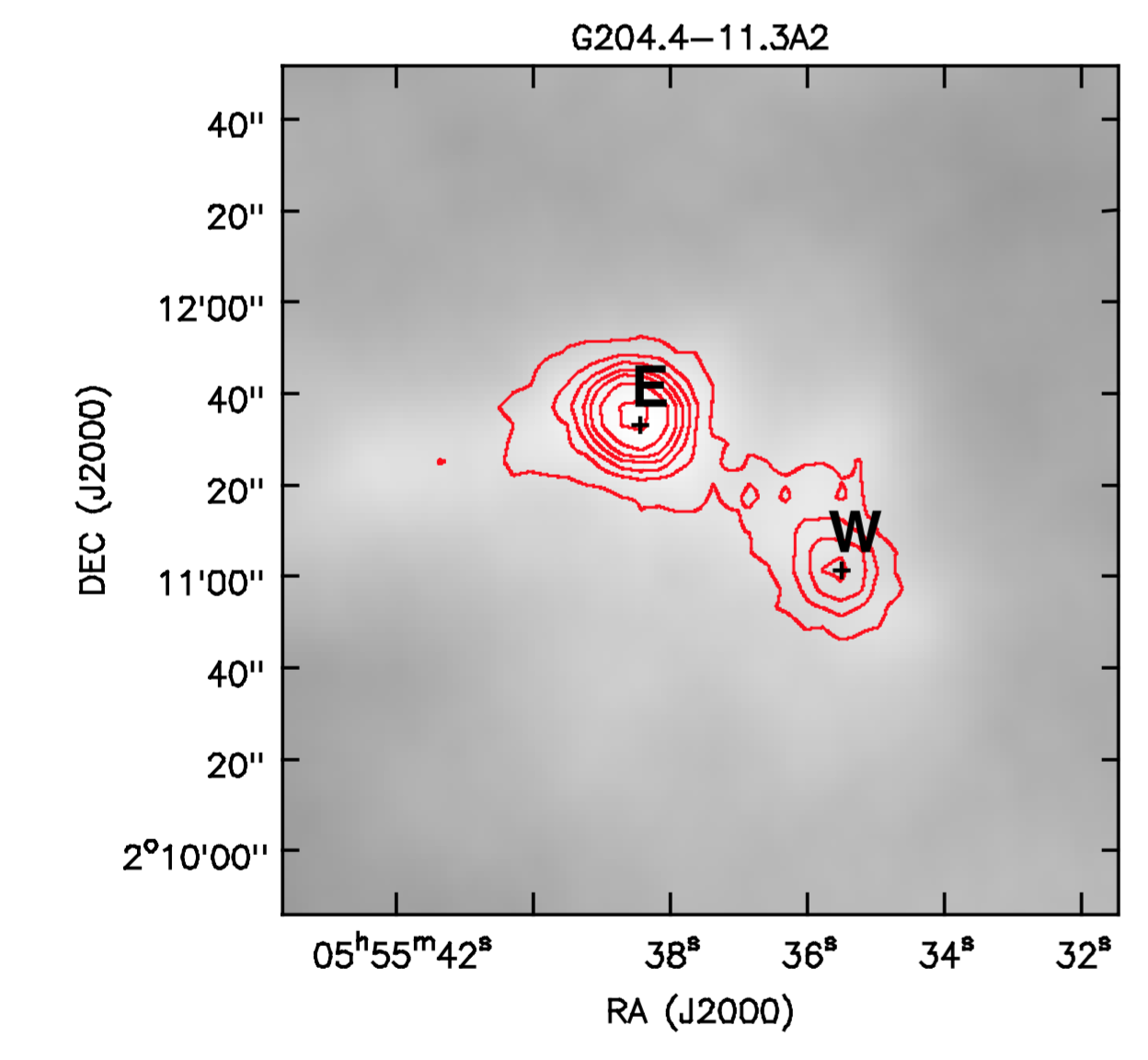}\\
\plottwo{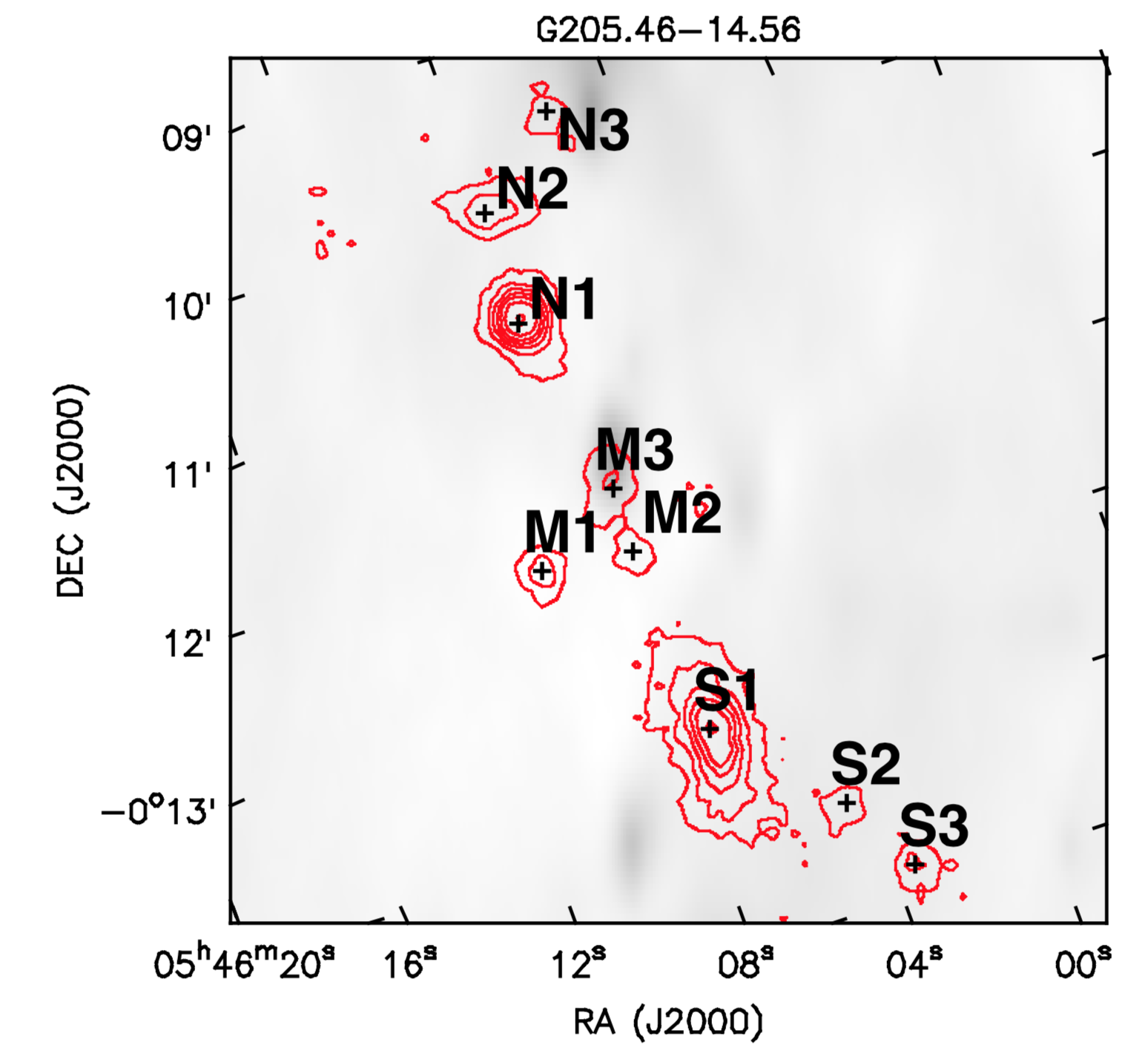}{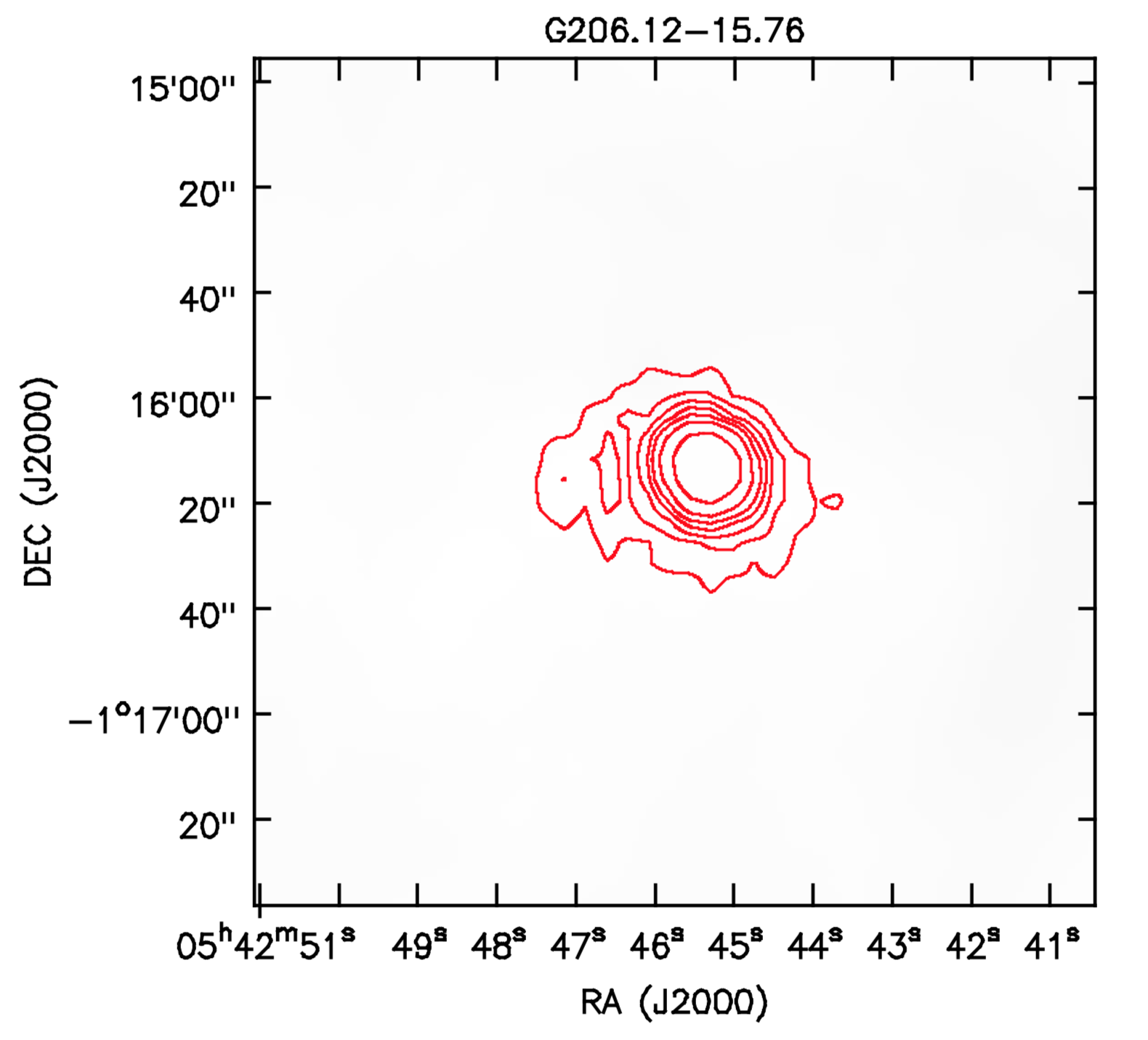}\\
\caption{ Same as in Figure 3 except for the 9 PGCCs detected at 850 $\micron$ with SCUBA-2 in the Orion B cloud. \label{fig:OrionB_PGCCs}}
\end{figure}

\clearpage
\begin{figure}
\epsscale{0.9}
\plottwo{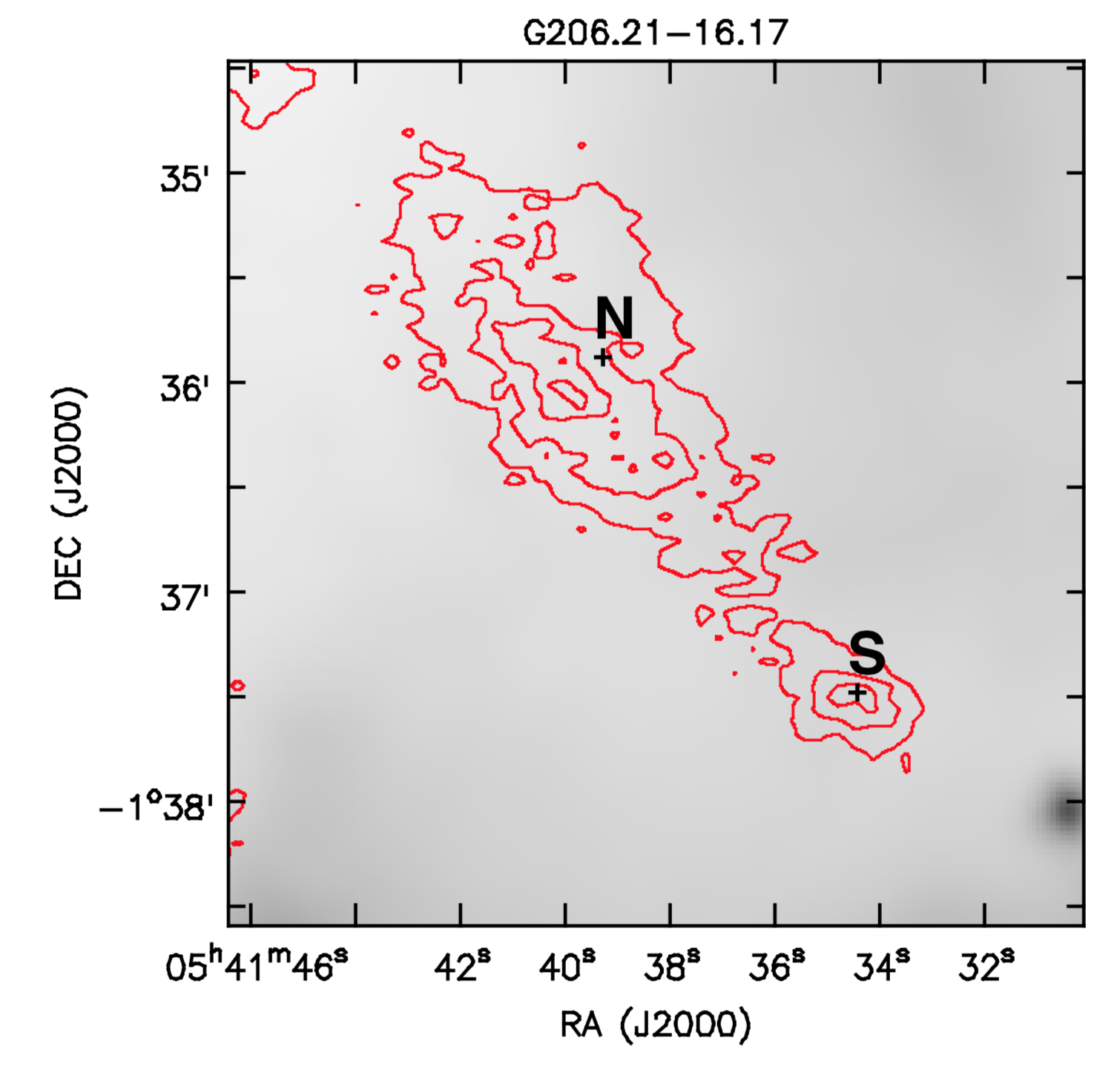}{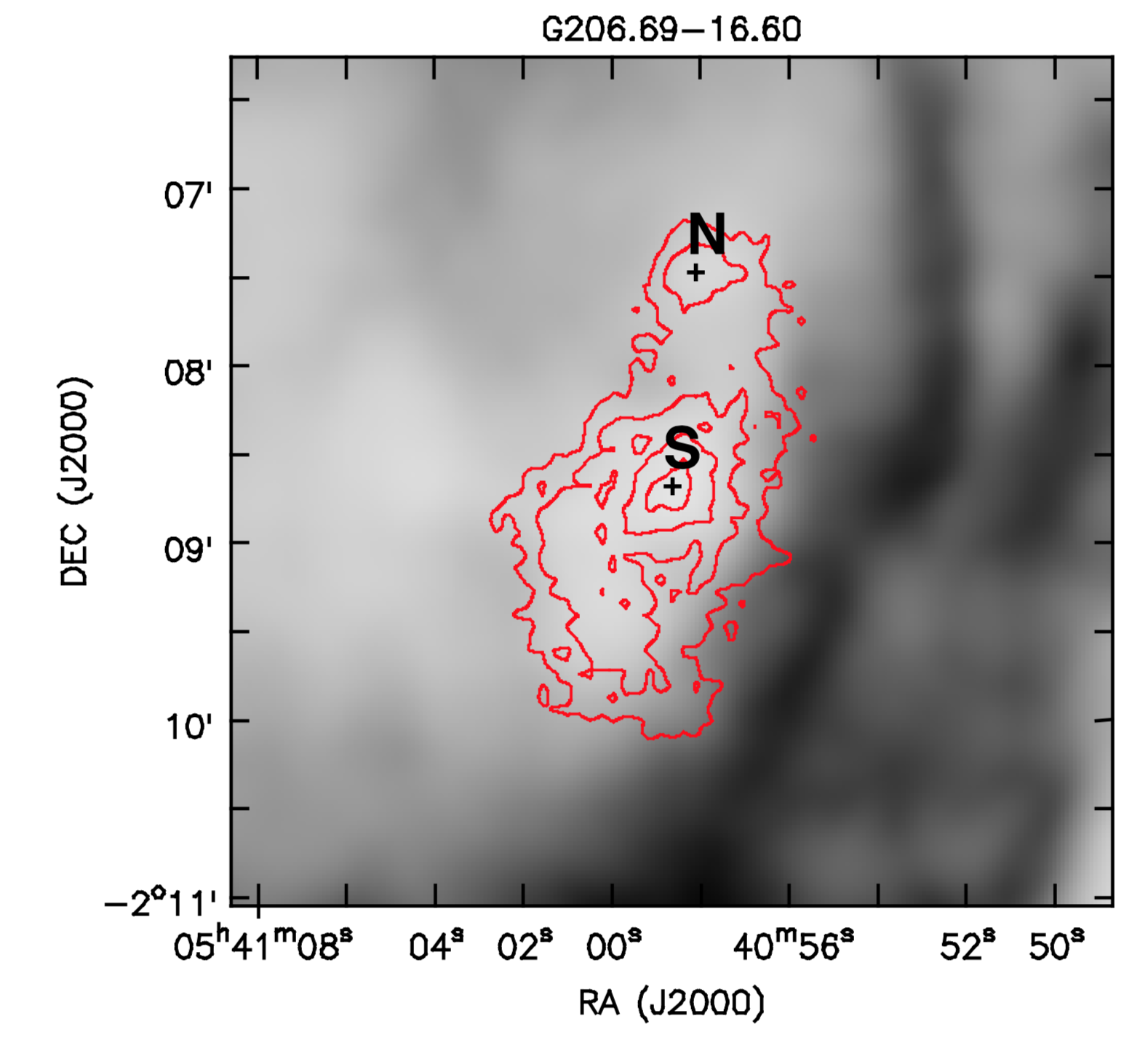}\\
\vspace*{2cm}
\hspace*{0.5cm}
\includegraphics[width=0.35\textwidth]{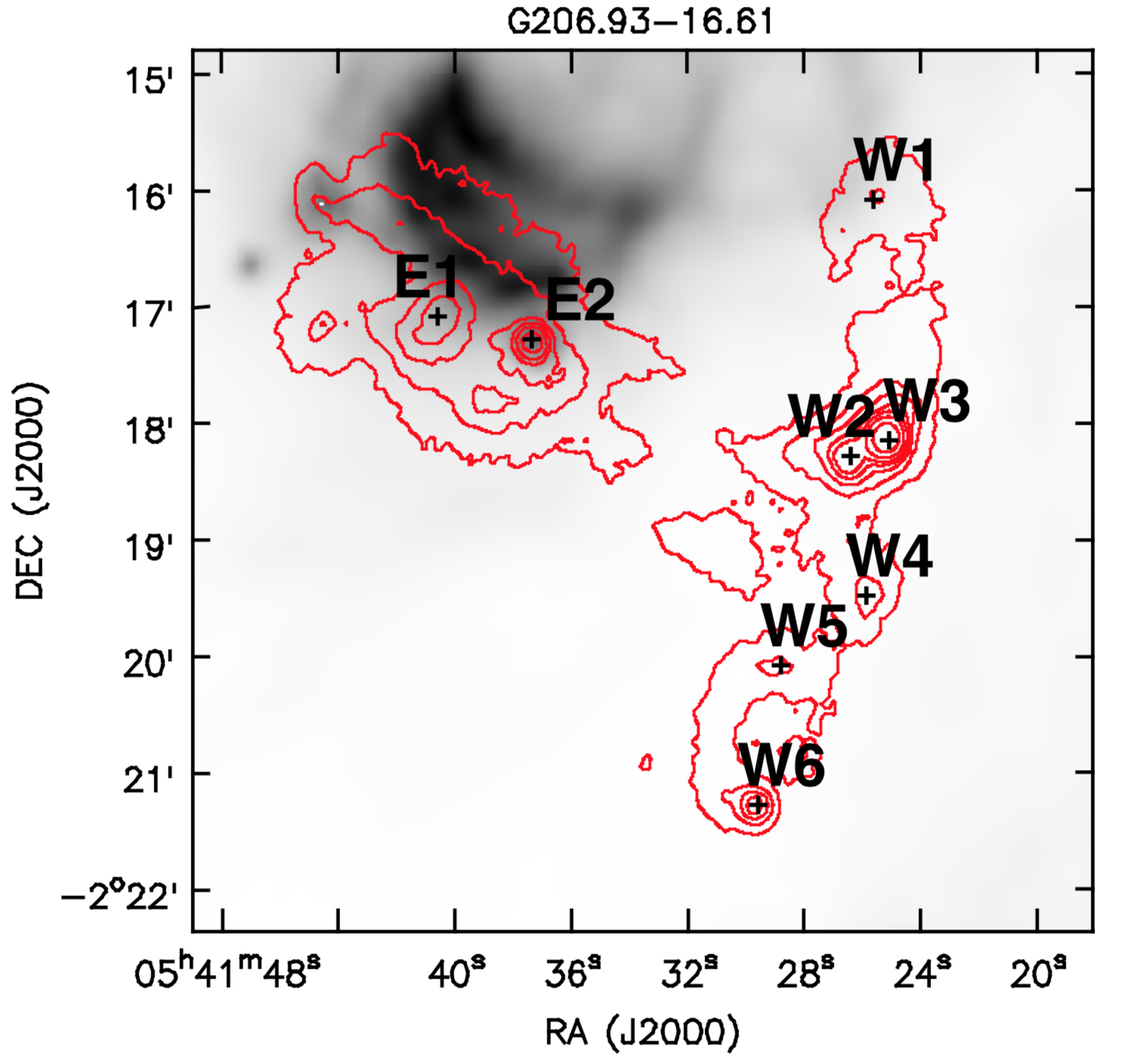}
\end{figure}

\clearpage
\begin{figure}
\epsscale{1.0}
\vspace*{1.0cm}
\plottwo{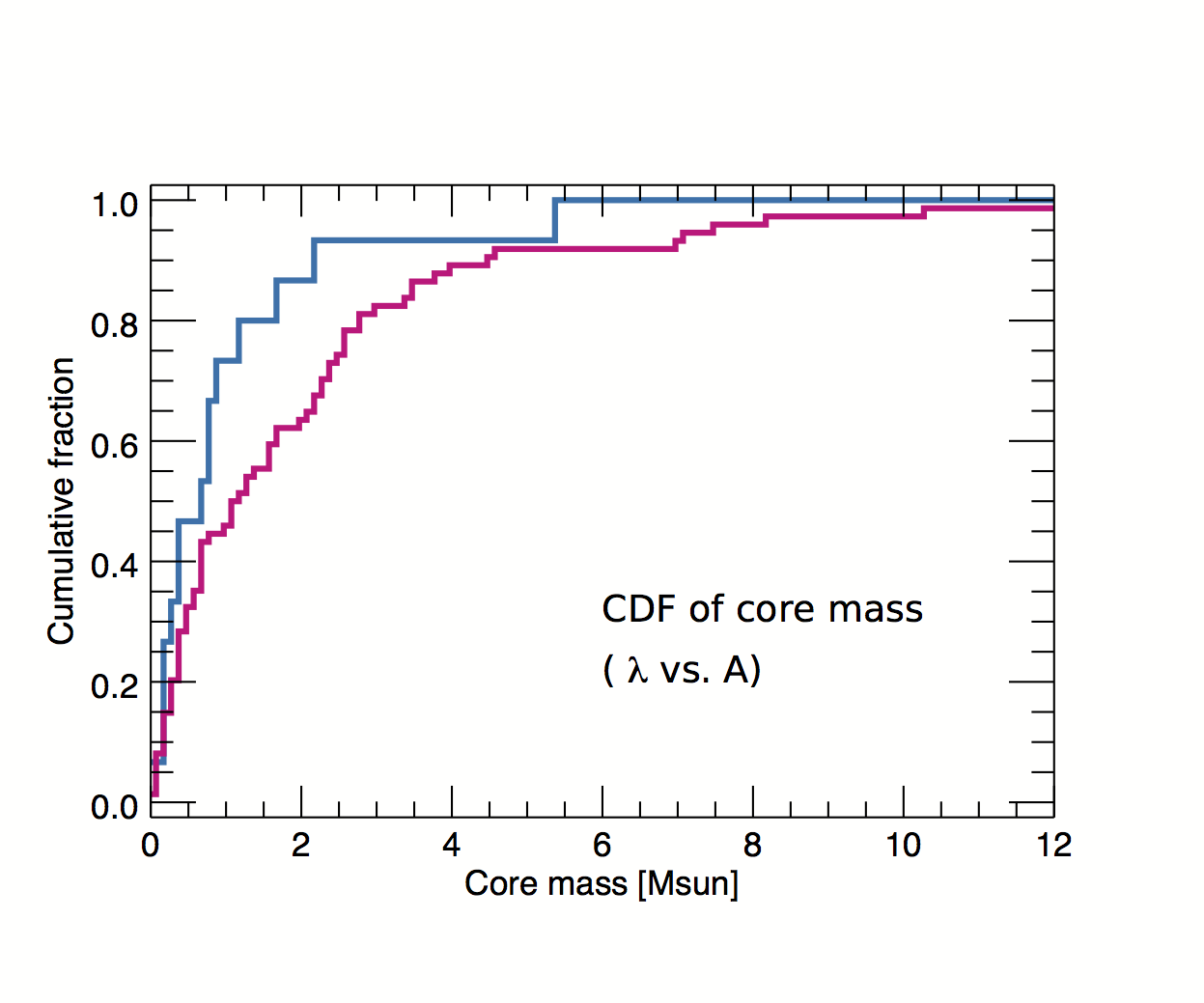}{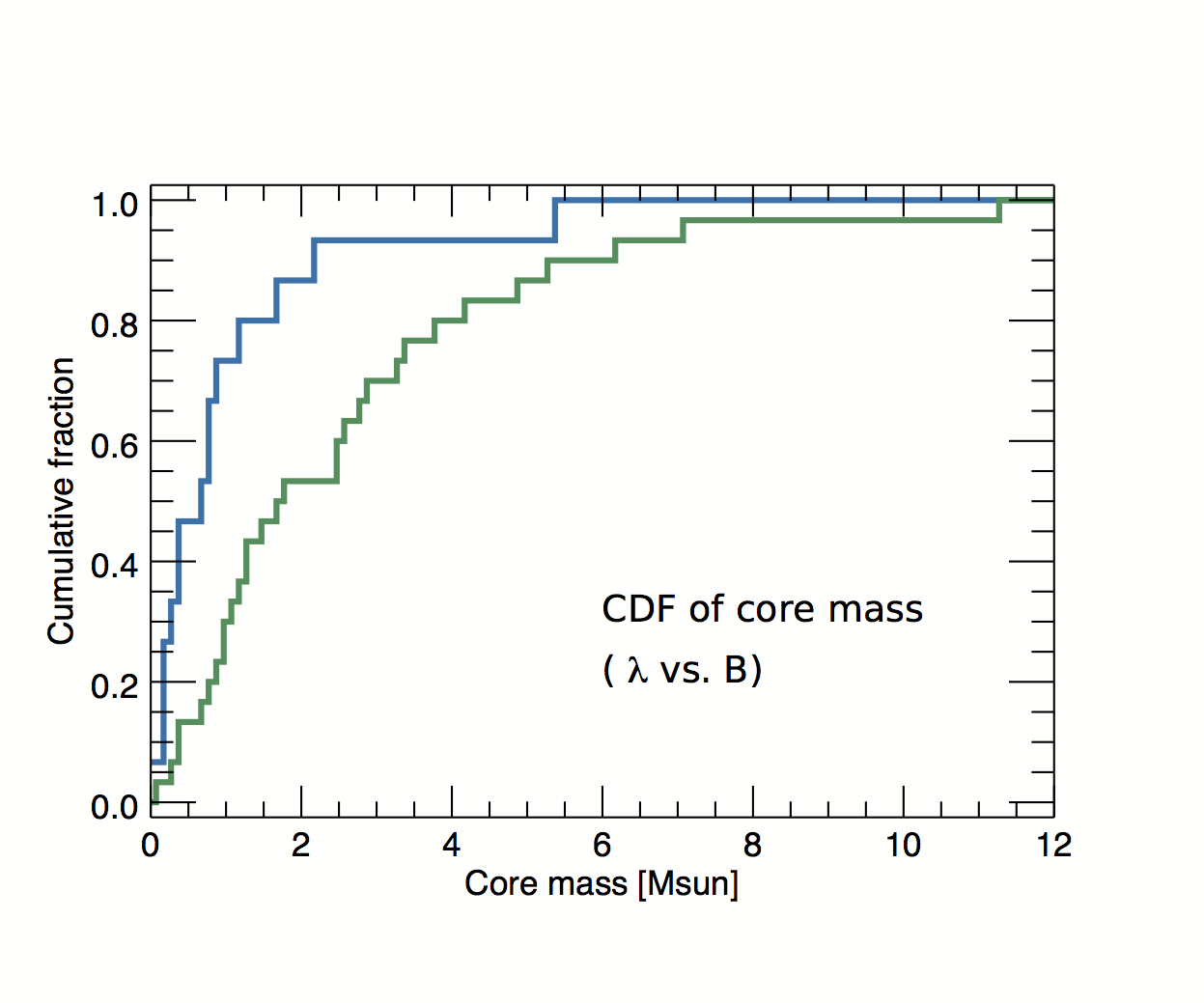}
\plottwo{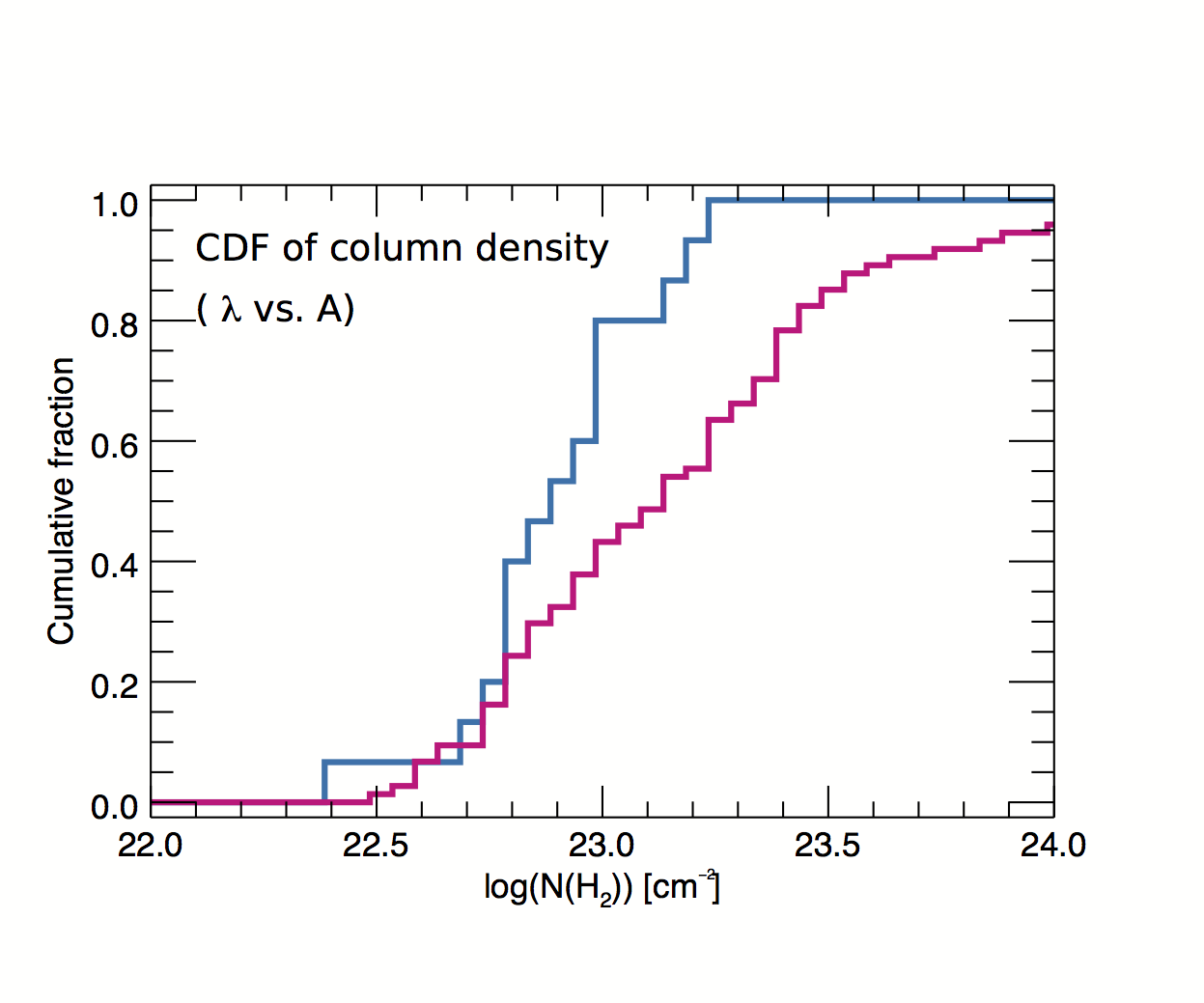}{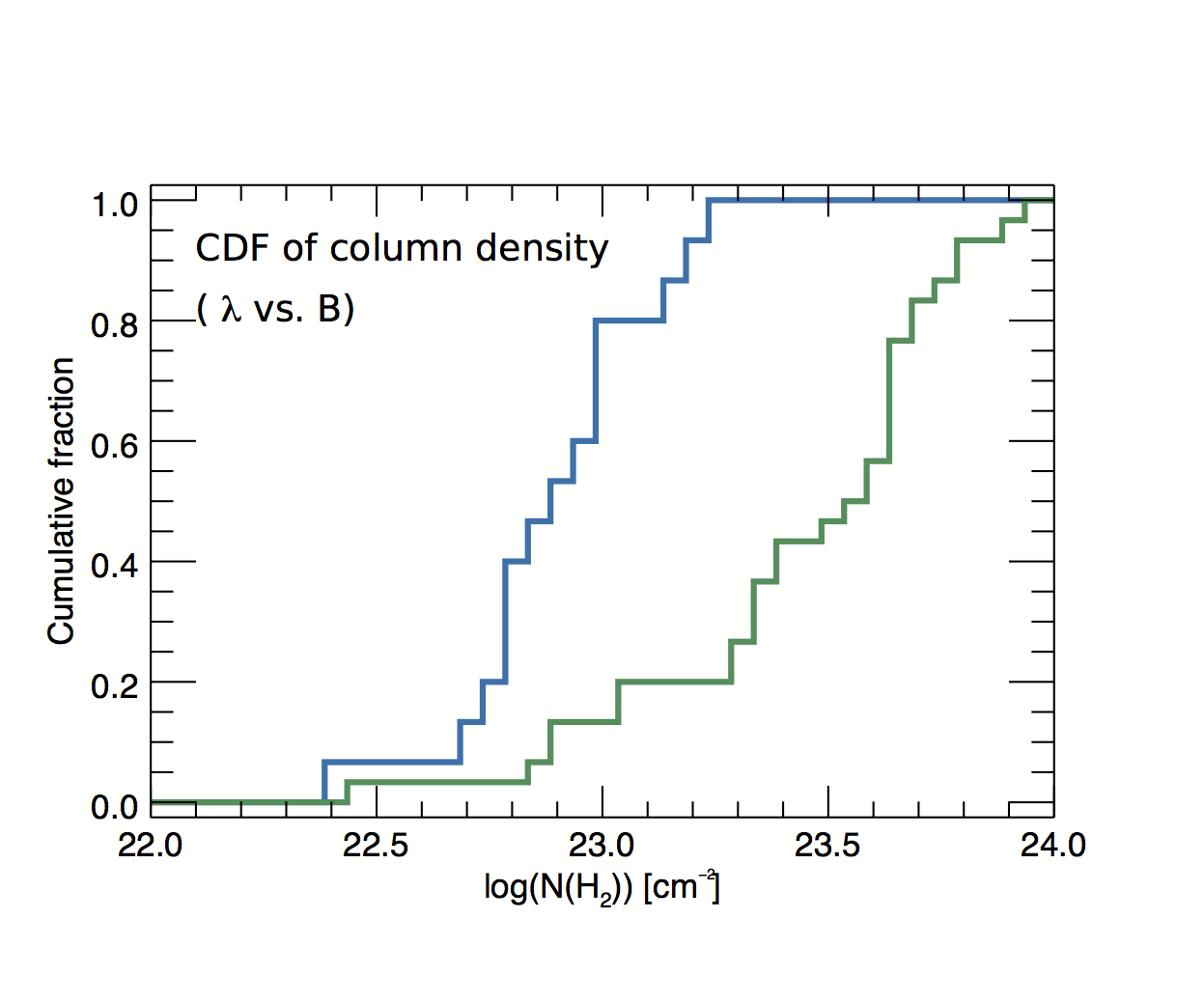}

\caption{ \label{fig:K-S}The K-S test of core masses and column densities. Upper two panels show cumulative distribution functions (CDF) of core masses. The blue solid lines show the CDF of the $\lambda$ Orionis cores and the magenta and green solid lines show the CDF of the Orion A (left) and Orion B (right) cores, respectively. 
Loewer two panels show the CDF of column densities comparing $\lambda$ Orionis cores to those of Orion A (left) and Orion B (right).  
}                            
\end{figure}

\clearpage
\begin{figure}
\epsscale{0.9}
\plotone{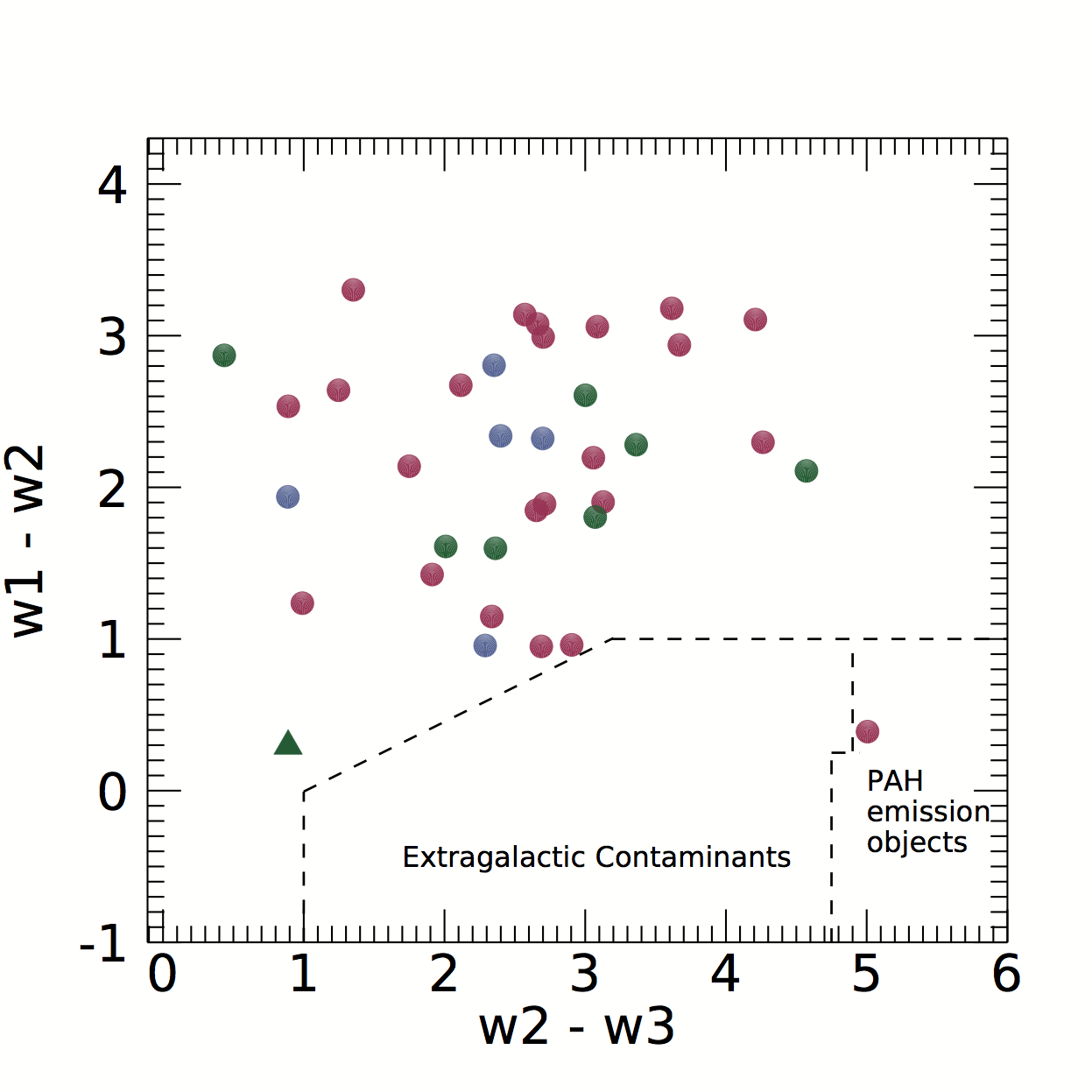}
\caption{\label{fig:color-color} A {\it WISE} color-color diagram to separating the star-forming galaxies from our YSO candidates. The dashed line shows the criterion for star-forming galaxies and PAH emission objects, and the circles represent our YSO candidates. Different colors denote different regions (blue: $\lambda$ Orionis cloud; magenta: the Orion A cloud; green: the Orion B cloud) where the YSO candidates are located.}
\end{figure}

\begin{figure}
\epsscale{0.9}
\plotone{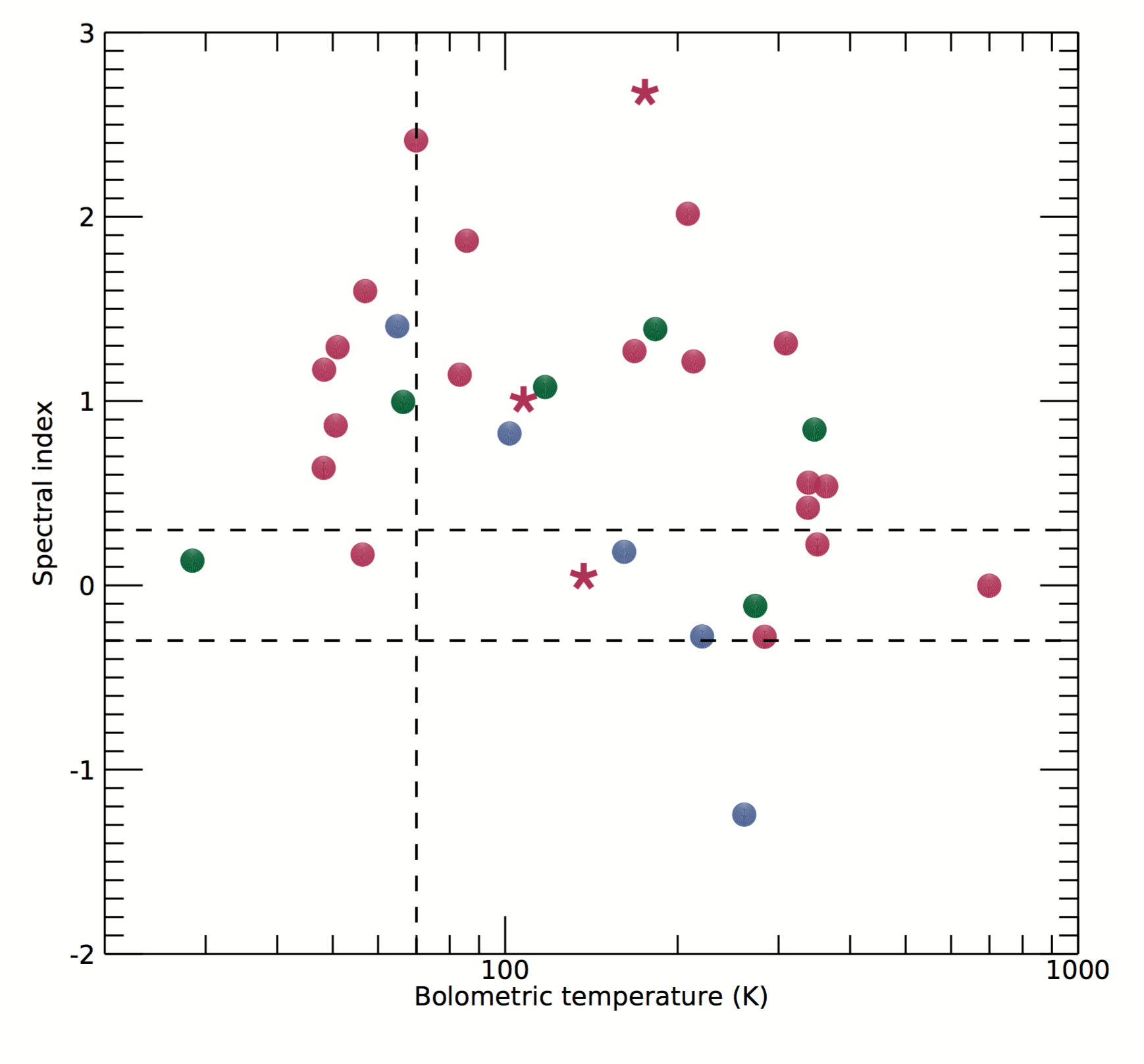}
\caption{\label{fig:classification} The 3.6 -- 22 $\micron$ spectral index $\alpha_{3.6-22}$ versus the bolometric temperature ($T_{\rm bol}$) are plotted for the 64 YSOs in our sample.  The
dashed lines delineate the regions that define the classifications (see the text for details). The colors are the same as in Figure 7. The asterisks are YSOs calculated with WISE and SCUBA-2 data, while the circles are YSOs calculated including far-infrared data.}
\end{figure}

\begin{figure}
\epsscale{0.9}
\plotone{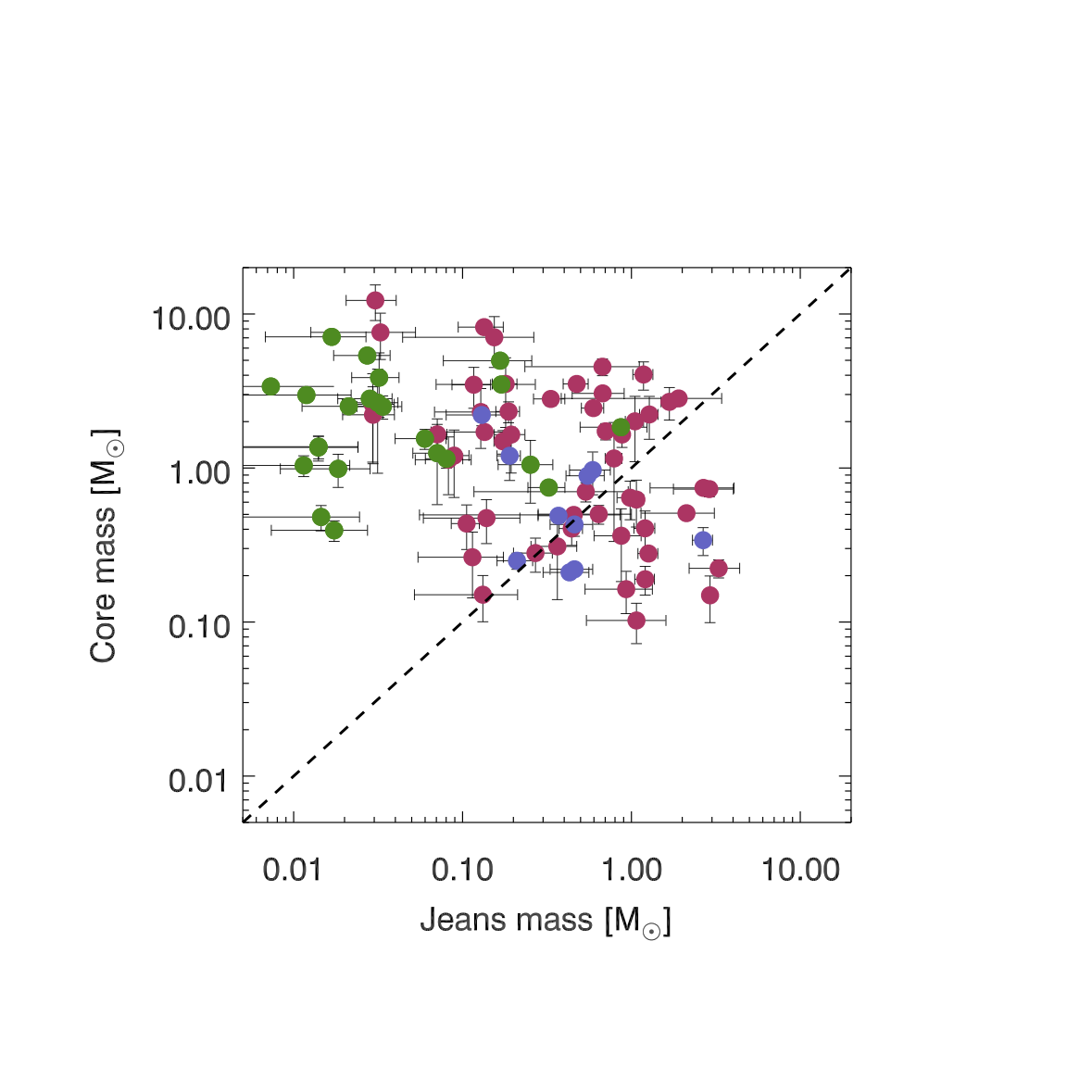}
\caption{\label{fig:jeans_mass} Jeans masses vs. core masses for 88 starless cores. The blue, magenta, and green circles represent 10 cores in the $\lambda$ Orionis cloud, 53 cores in the Orion A, and 25 cores in the Orion B clouds, respectively.}
\end{figure}

\begin{figure}
\centerline{
\includegraphics[width=9cm, height=8cm]{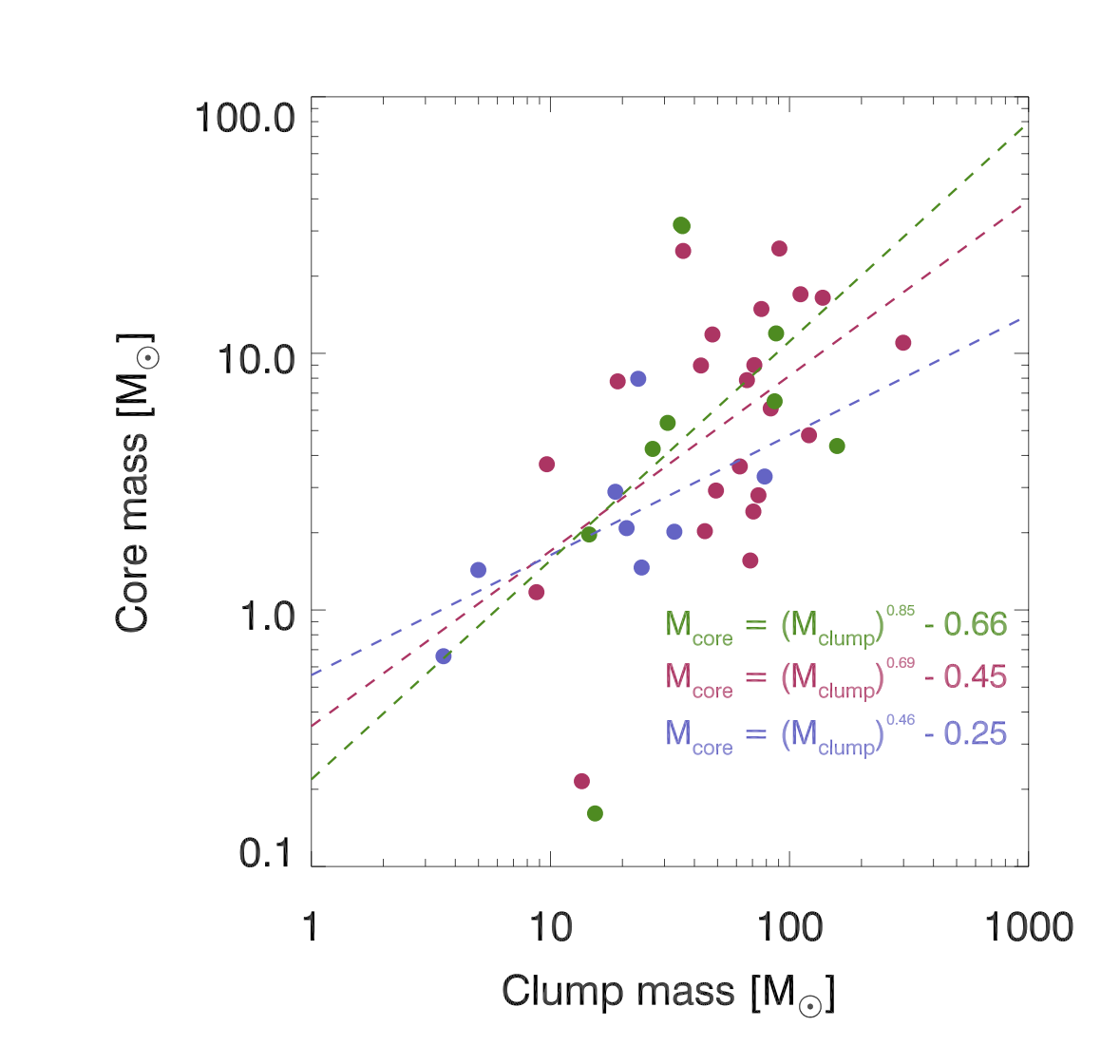}\hspace*{0.1mm}
\includegraphics[width=9cm, height=8cm]{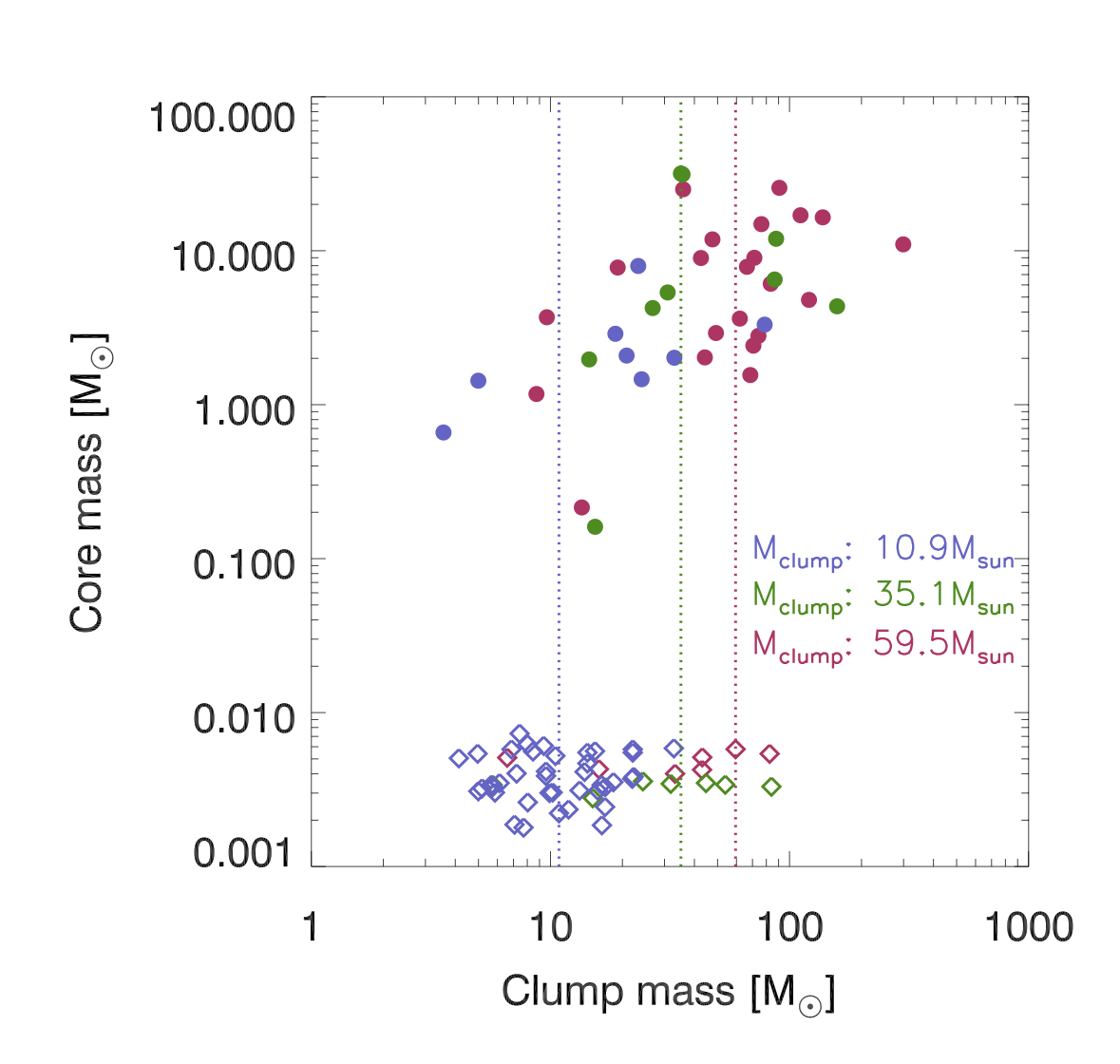}
}
 \caption{\label{fig:clump_mass} Core masses vs. clump masses. {\it Left:} The circles represent PGCCs in the $\lambda$ Orionis cloud (blue), Orion A (magenta) and Orion B (green) clouds, respectively. Each dashed line shows the correlations for each region. Functions in the right lower corner are obtained by fitting. {\it Right:} The open diamonds indicate non-detected PGCCs and the colors are the same as those in the left panel. The dotted lines represent median clump masses of $\lambda$ Orionis, Orion B, and Orion A clouds, respectively.}
\end{figure}

\clearpage
\begin{figure}
\epsscale{0.6}
\plotone{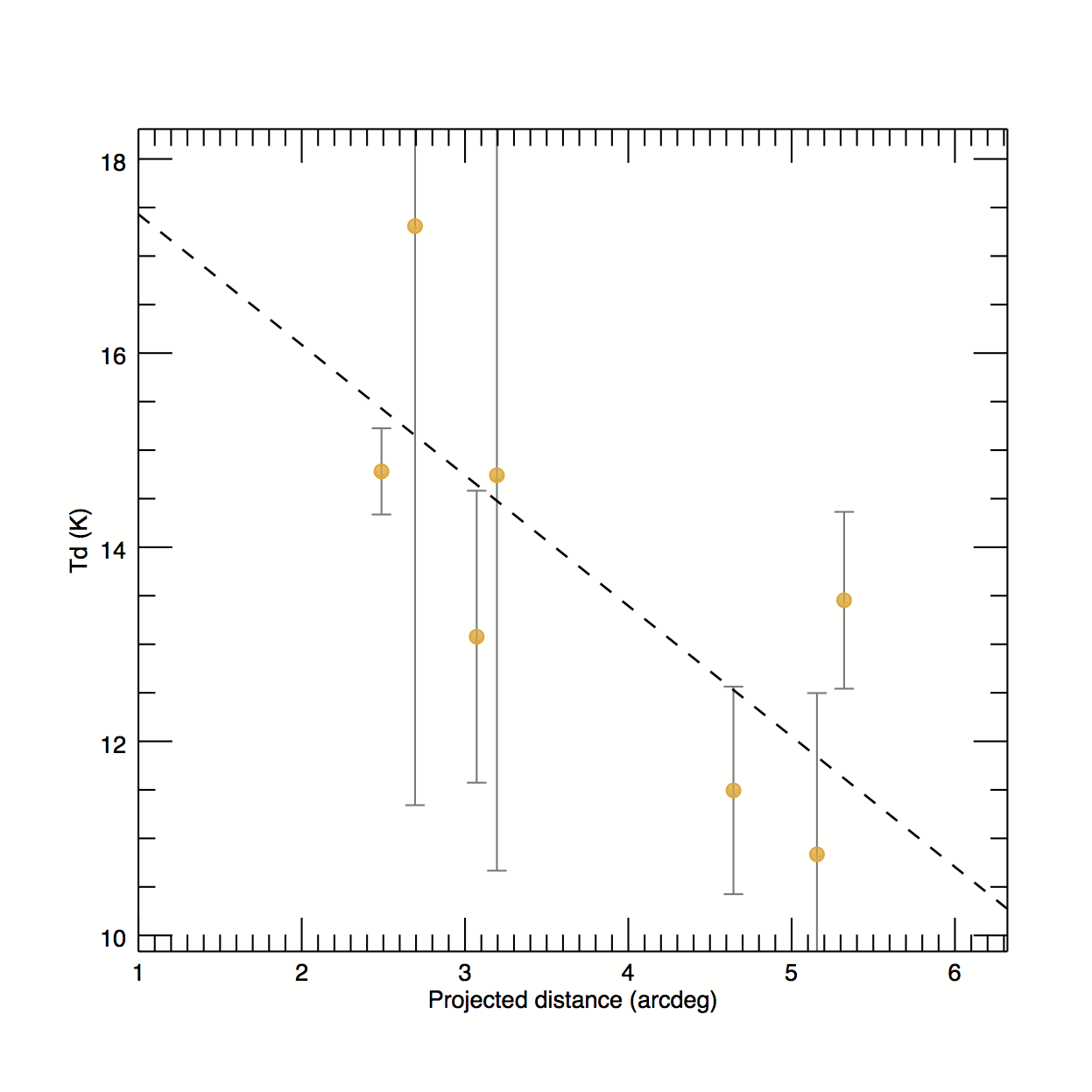}
\caption{\label{fig:projected} Dust temperature as a function of projected distance from $ ``\lambda"$ Ori. The dashed lines show least-squares fit to the data. The 8 PGCCs are detected in the $\lambda$ Orionis cloud, but only 7 PGCCs are plotted, the values for which are obtained from the PGCC catalog.}
\end{figure}

\clearpage
\begin{figure}
\begin{tabbing}
  \includegraphics[width=0.35\textwidth]{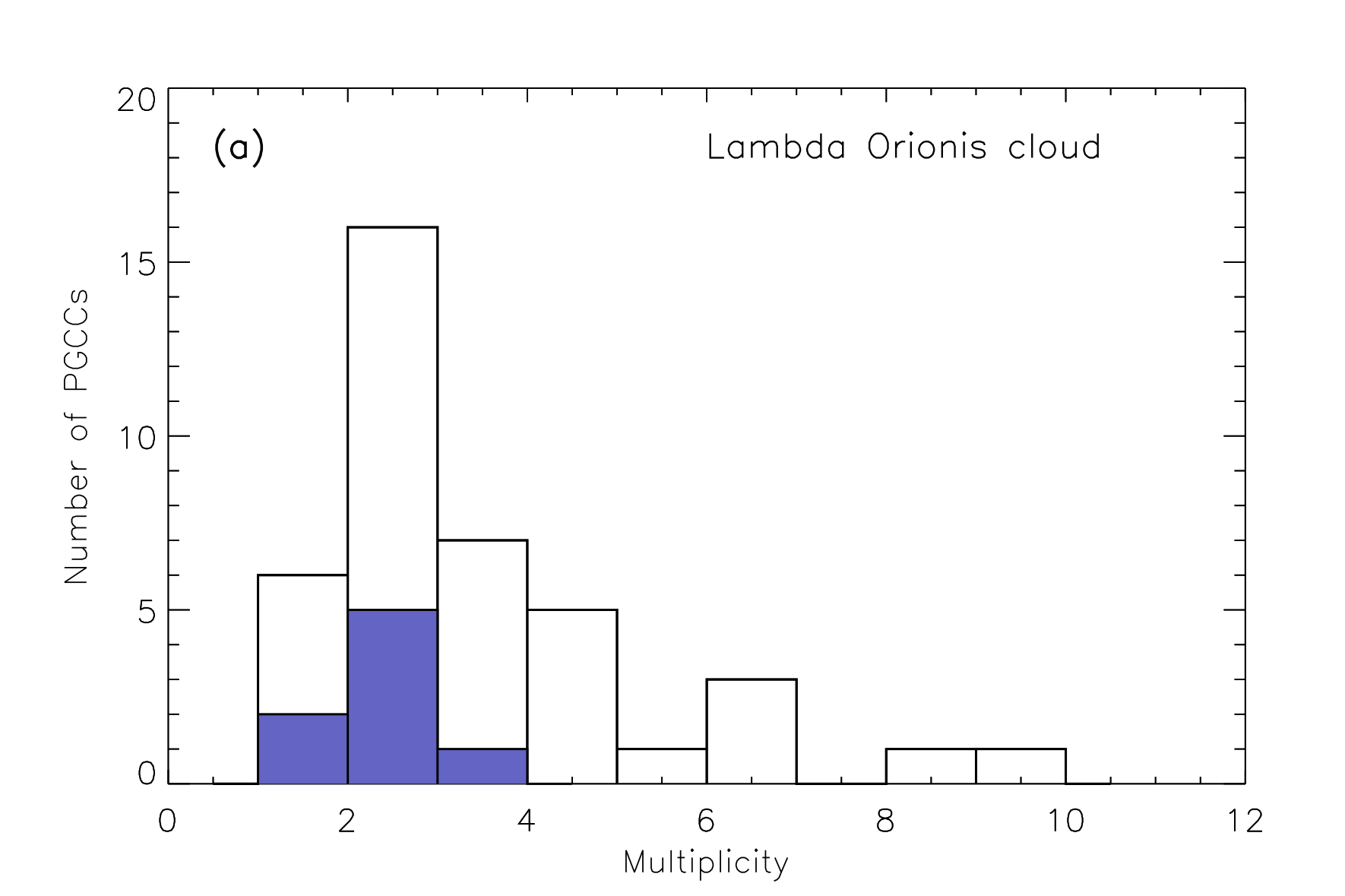}\nobreak
  \includegraphics[width=0.35\textwidth]{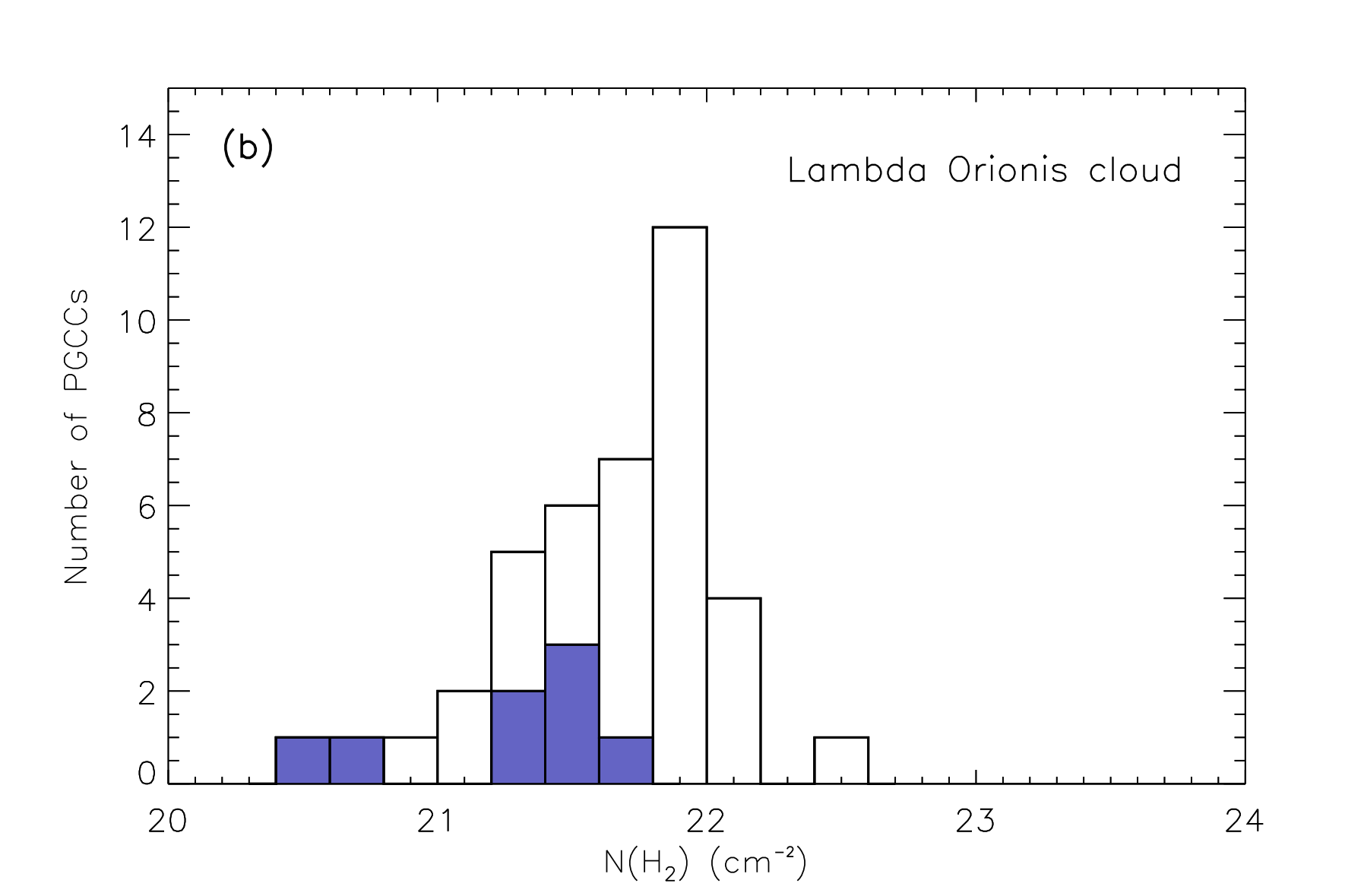}\nobreak 
   \includegraphics[width=0.35\textwidth]{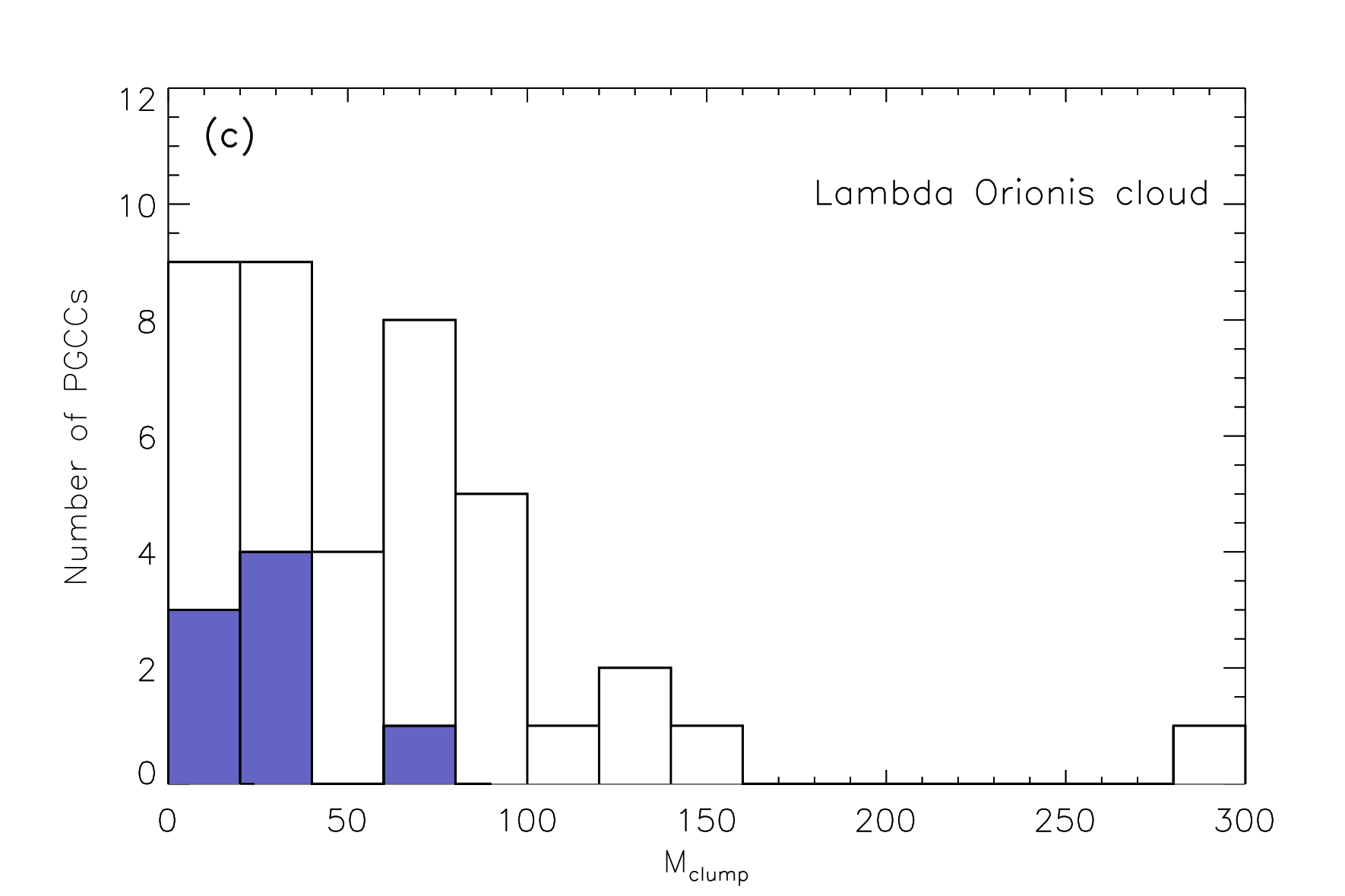}  \\
    \includegraphics[width=0.35\textwidth]{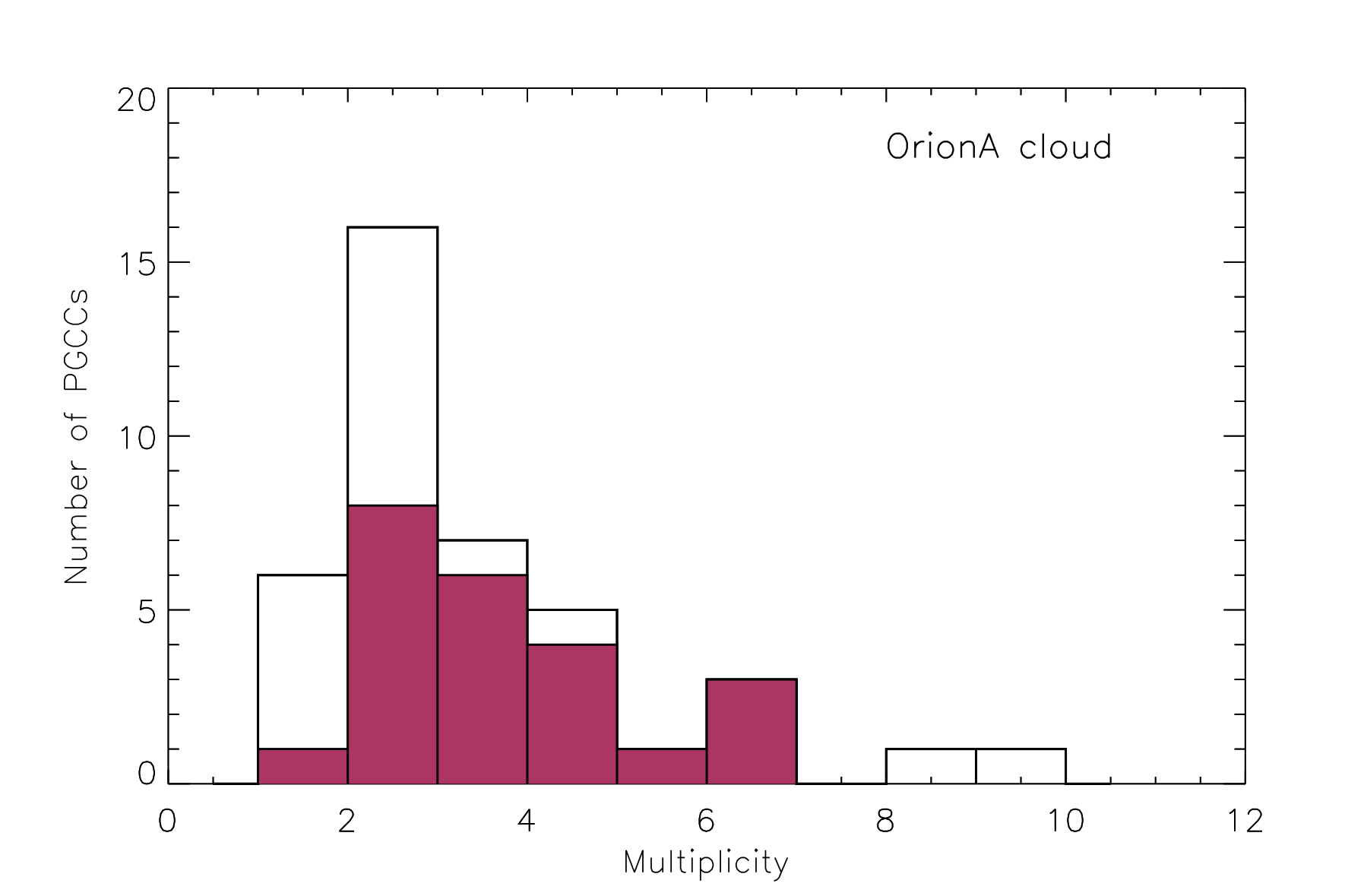}\nobreak 
  \includegraphics[width=0.35\textwidth]{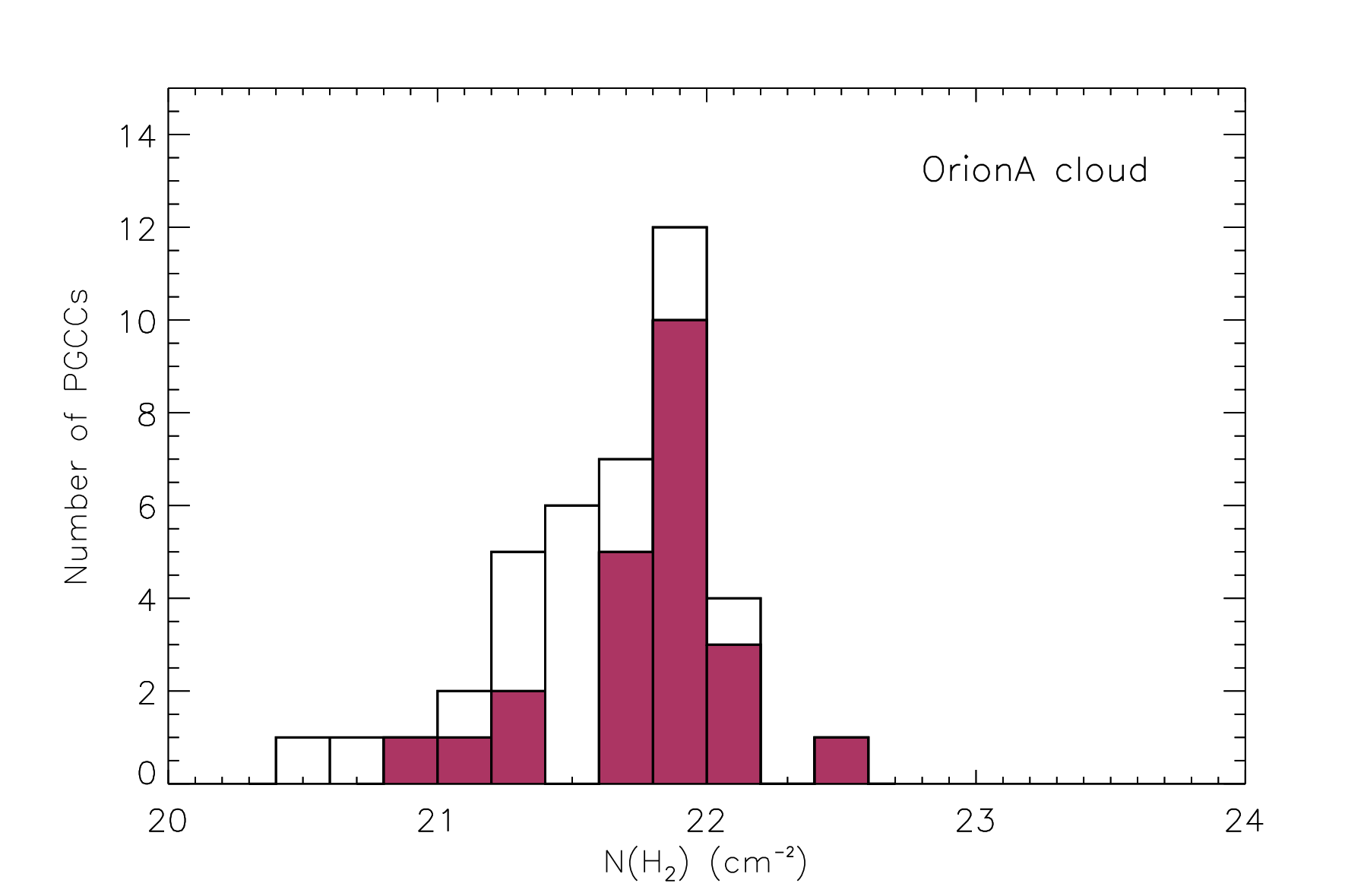} \nobreak
  \includegraphics[width=0.35\textwidth]{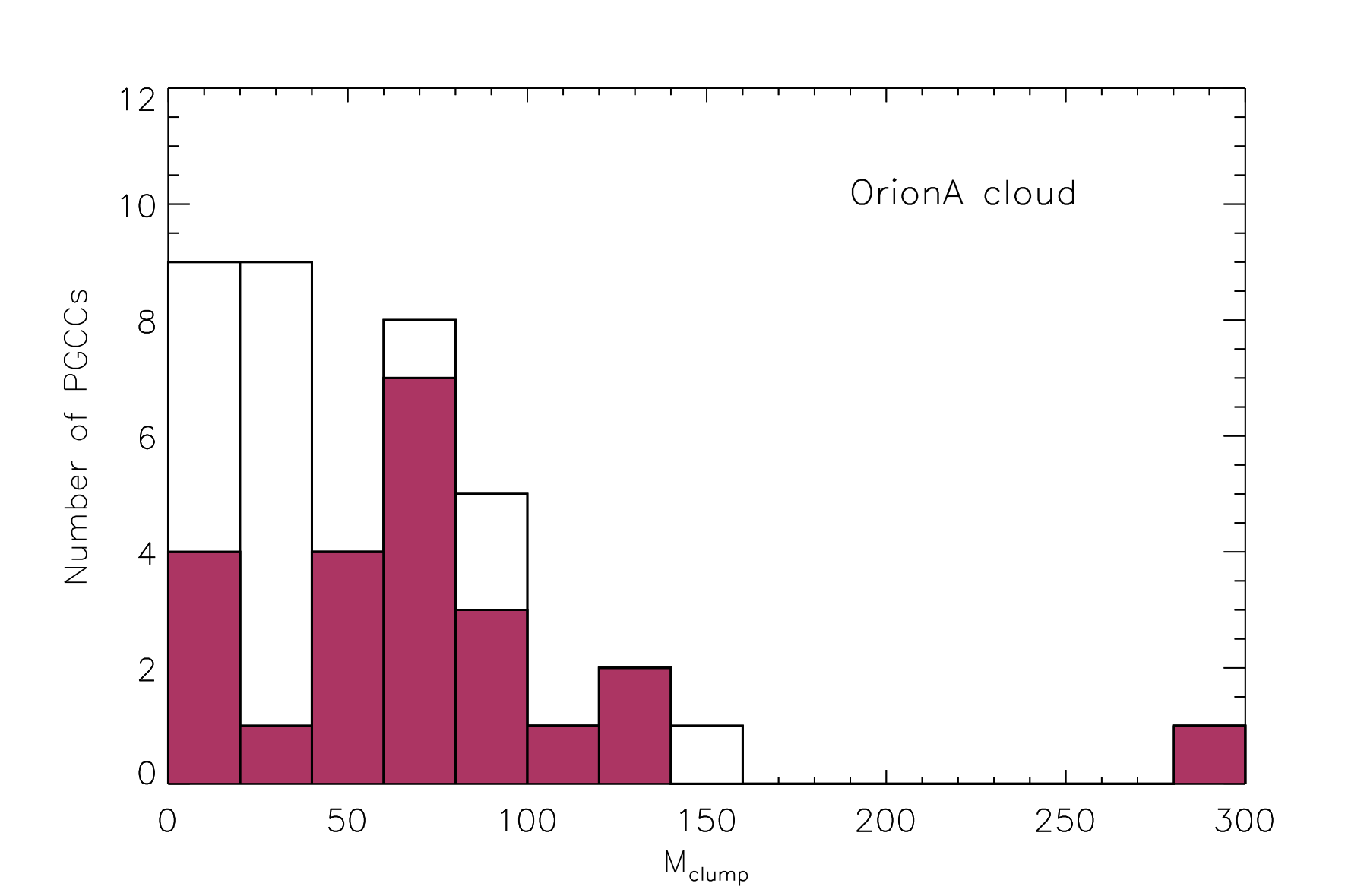}\\ 
  \includegraphics[width=0.35\textwidth]{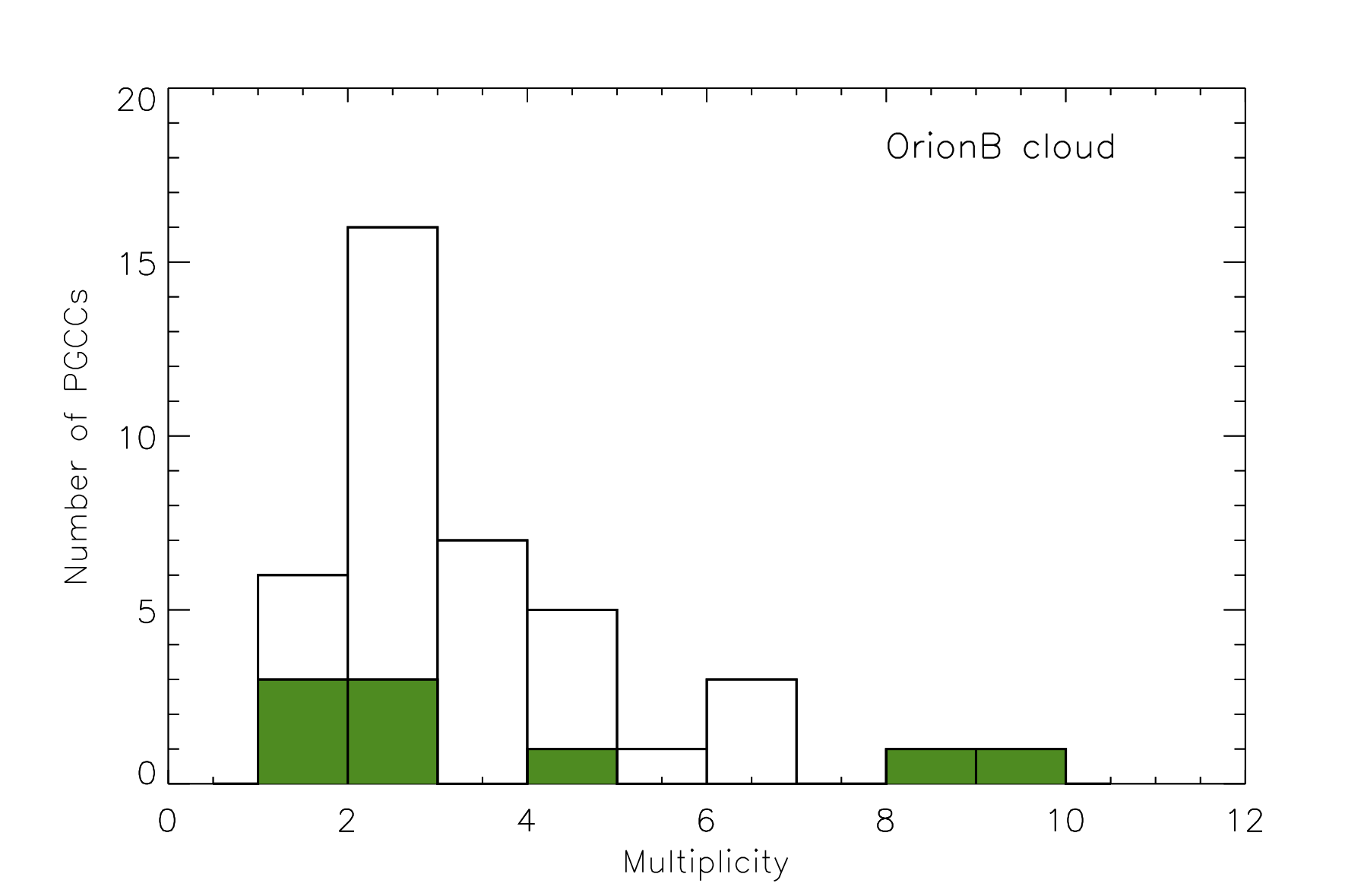}\nobreak
    \includegraphics[width=0.35\textwidth]{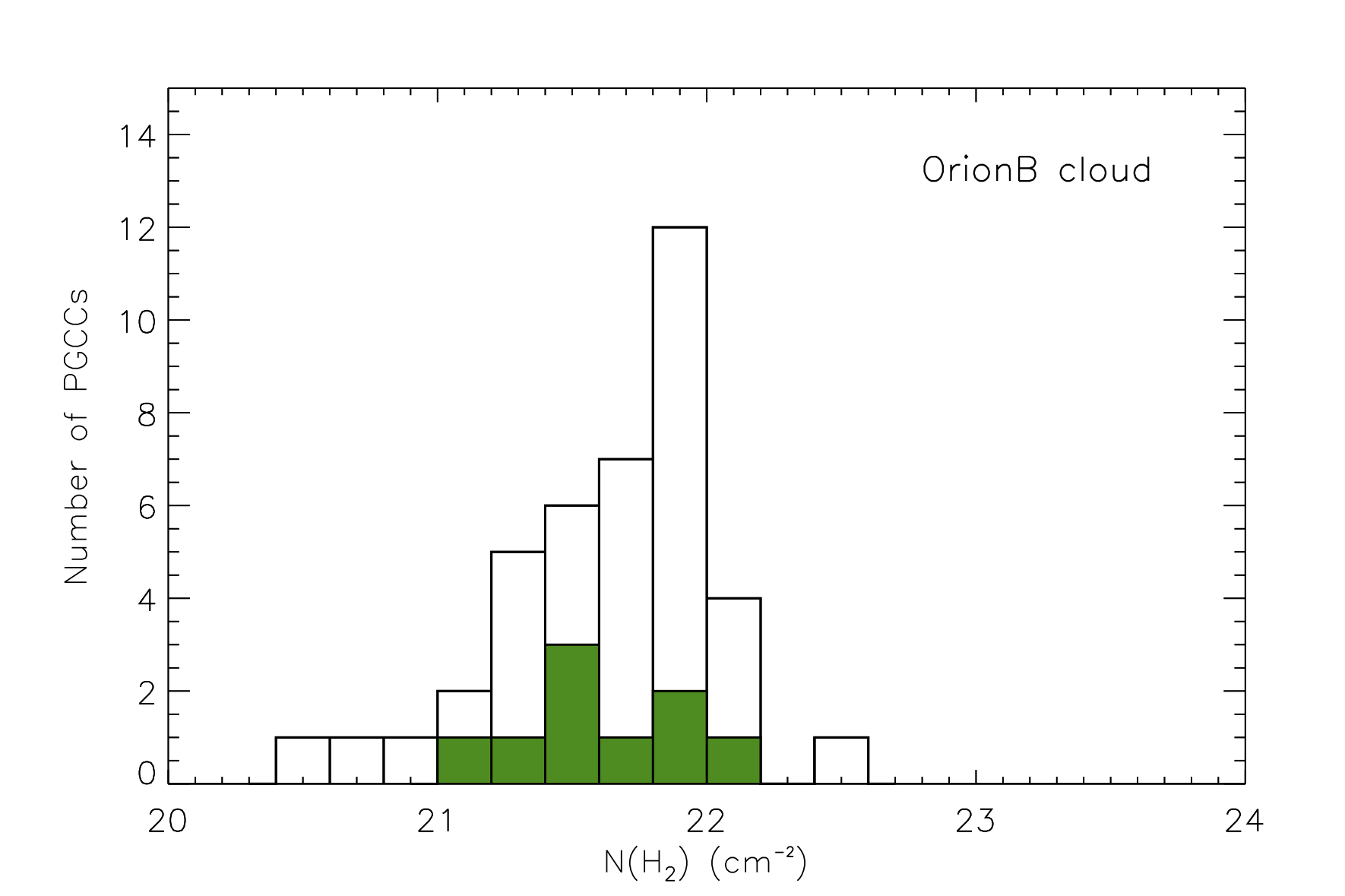} 
  \includegraphics[width=0.35\textwidth]{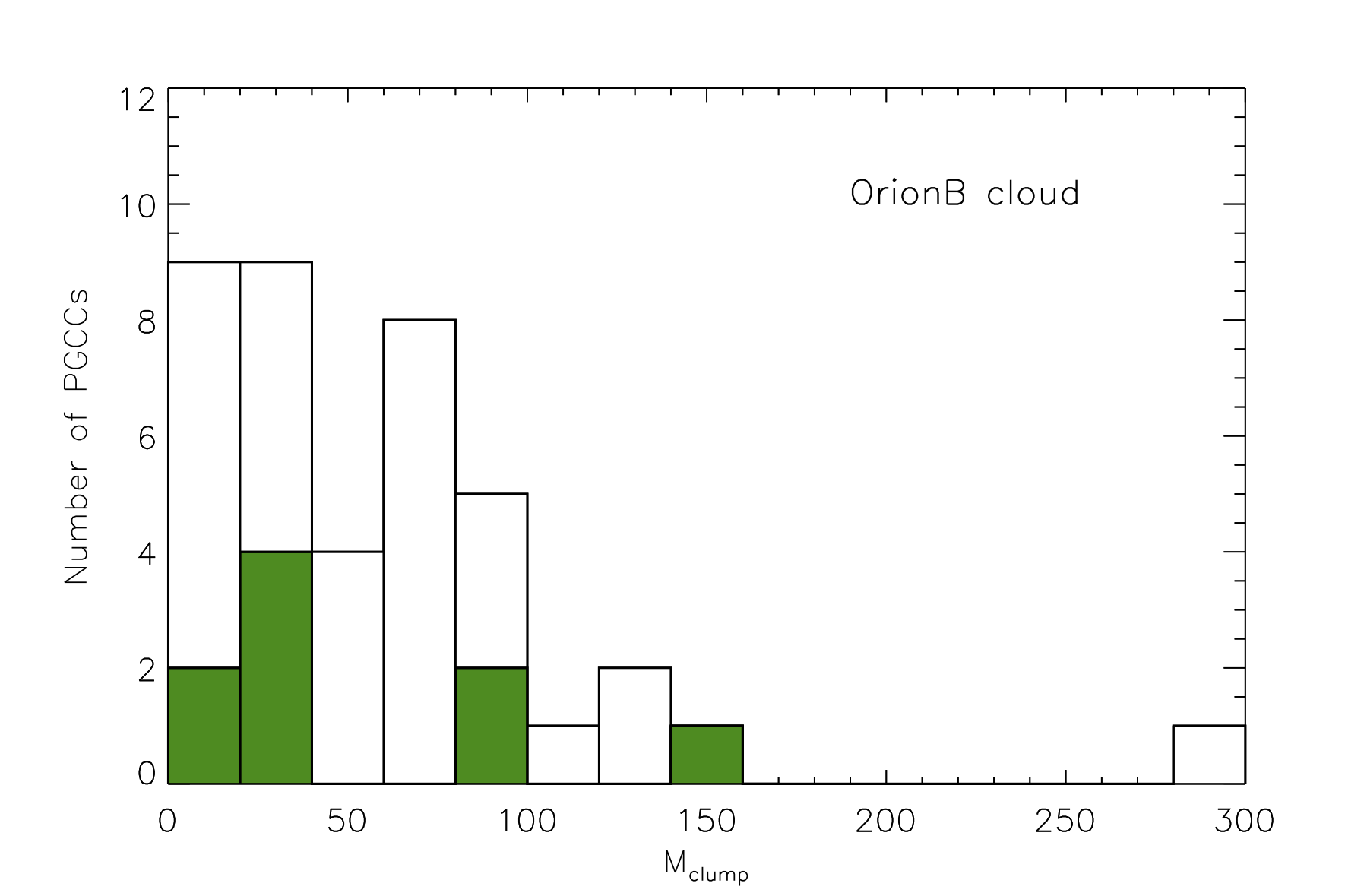} \\
\end{tabbing}
 \caption{\label{fig:histogram_PGCC} Histograms of the multiplicity (a),  column density (b), and mass (c) for the 40 PGCCs detected at SCUBA-2 850 $\micron$ dust continuum in this work. Different colors denote different regions as used above figures and open histograms indicate the total number of PGCCs in each bin.  }
\end{figure}

\clearpage
\startlongtable
\begin{deluxetable}{c|cccccc}
\tablecaption{96 PGCCs observed with SCUBA-2 in the Orion complex  \label{tab:list_PGCCs}}
\tablehead{
\colhead{Cloud} & \colhead{PGCC} & \colhead{R.A.(J2000)} & \colhead{Dec.(J2000)} &\colhead{Observation} &\colhead{Detection} &\colhead{rms} \\
\colhead{} & \colhead{} & \colhead{($^h$ $^m$ $^s$)} &\colhead{($^\circ$ $\arcmin$ $\arcsec$)} &\colhead{} &\colhead{} &\colhead{(mJy/beam)} 
}
\startdata
$\lambda$ Orionis&G188.25-12.97 &  05:17:25.20   &  +15:04:19.60  &  SCOPE & No &13.8 \\
 	{ }	&G188.85-13.42 &  05:17:08.63   &  +14:19:54.30  &  SCOPE& No & 7.8 \\ 
 	{ }	&G189.46-10.34 &  05:29:13.66   &  +15:29:35.60  &  SCOPE & No & 19.0 \\
{ }	&G189.92-14.58 &  05:15:28.66   &  +12:49:58.90  &  SCOPE & No & 14.1\\
{ }	&G190.0-13.5A1 &  05:19:28.09   &  +13:15:45.30  &  SCOPE & No & 12.0\\
{ }	&G190.12-14.47 &  05:16:17.78   &  +12:44:2.60  & SCOPE & No & 14.7\\
{ }	&G190.15-13.75 &  05:18:49.36   &  +13:05:11.80  &  SCOPE& Yes & 8.0 \\
{ }	&G190.1-13.7A1 &  05:18:54.66   &  +13:02:23.40  &  SCOPE& No & 11.4\\
{ }	&G190.1-14.3A1 &  05:16:47.16   &  +12:45:42.50  &  SCOPE&No & 12.8\\
 	{ }	&G190.36-13.38 &  05:20:33.62   &  +13:06:48.90  &  SCOPE&No &7.8\\
 	{ }	&G191.03-16.74 &  05:10:18.90   &  +10:47:24.00  &  SCOPE&No &13.1\\
 	{ }	&G191.04-16.44 &  05:11:26.74   &  +10:54:54.60  &  SCOPE&No &8.5 \\
{ }	&G191.18-16.70 &  05:10:50.72   &  +10:39:06.80  &  SCOPE&No &8.5\\
{ }	&G191.69-11.62 &  05:29:32.36   &  +12:57:15.40  &  SCOPE&No &12.9\\
{ }	&G191.90-11.21 &  05:31:24.92   &  +12:59:27.00  & SCOPE& Yes & 12.5\\
{ }	&G192.1-10.9A1 &  05:32:57.69   &  +12:57:25.10  & SCOPE& Yes & 12.3\\
{ }	&G192.12-11.10 &  05:32:15.03   &  +12:52:00.20  &  SCOPE&Yes & 13.2\\
{ }	&G192.2-11.3A1 &  05:31:45.05   &  +12:34:59.00  &  SCOPE&No &12.9\\
{ }	&G192.32-11.88 &  05:29:55.84   &  +12:17:34.00  &  SCOPE&Yes & 17.0\\
{ }	&G192.36-11.56 &  05:31:08.65   &  +12:15:46.20  &  SCOPE&No & 12.8\\
{ }	&G192.45-12.23 &  05:29:0.18   &  +12:00:27.60  &  SCOPE&No & 12.2\\
 	{ }	&G192.5-11.5A1 &  05:31:30.97   &  +12:12:12.00  &  SCOPE& No&13.5\\
 	{ }	&G192.5-11.5A2 &  05:31:36.03   &  +12:18:09.00  & SCOPE& No& 13.0\\
 	{ }	&G194.34-16.38 &  05:18:29.40   &  +08:15:44.00  & SCOPE& No & 11.6\\
 	{ }	&G194.69-16.84 &  05:17:36.40   &  +07:44:14.00  &  SCOPE& No & 10.1\\
 	{ }	&G194.81-15.54 &  05:22:21.55   &  +08:19:25.70  & SCOPE& No & 12,4\\
 	{ }	&G194.94-16.74 &  05:18:26.67   &  +07:34:12.00  &  SCOPE& No & 11.5\\
 	{ }	&G194.98-16.96 &  05:17:47.84   &  +07:26:22.60  & SCOPE& No & 8.6\\
 	{ }	&G195.00-16.95 &  05:17:47.46   &  +07:22:08.00  &  SCOPE& No & 9.2\\
 	{ }	&G195.0-16.4A1 &  05:20:1.58   &  +07:39:58.90  & SCOPE& No & 12.8\\
 	{ }	&G195.09-16.41 &  05:20:0.67   &  +07:39:10.00  &  SCOPE& No & 12.2\\
 	{ }	&G195.29-15.88 &  05:22:7.73   &  +07:44:44.30  &  SCOPE& No & 11.2\\
 	{ }	&G195.32-16.34 &  05:20:35.30   &  +07:28:48.70  &  SCOPE& No & 8.5\\
 	{ }	&G196.21-15.50 &  05:25:11.06   &  +07:10:26.00  &  SCOPE& No & 9.14\\
 	{ }	&G196.43-16.01 &  05:23:57.10   &  +06:43:50.30  &  SCOPE& No & 8.92\\
 	{ }	&G196.89-16.11 &  05:24:31.87   &  +06:18:33.60  &  SCOPE& No & 9.11\\
 	{ }	&G196.92-10.37 &  05:44:37.59   &  +09:10:55.70  &  SCOPE& Yes & 13.8\\
 	{ }	&G197.12-15.96 &  05:25:30.06   &  +06:11:48.90  & SCOPE& No & 6.72\\
 	{ }	&G197.13-10.17 &  05:45:44.66   &  +09:6:22.70  & SCOPE& No & 12.4\\
 	{ }	&G197.18-12.98 &  05:36:0.38   &  +07:39:37.70  & SCOPE& No & 9.28\\
 	{ }	&G197.46-15.39 &  05:28:9.30   &  +06:12:23.20  &  SCOPE& No & 6.71\\
 	{ }	&G198.03-15.24A &  05:29:33.60   &  +05:52:22.00  &SCOPE& No & 14.5\\  
 	{ }	&G198.03-15.24B &  05:29:49.61   &  +05:45:50.00  &  SCOPE& No & 11.8\\
{ }	&G198.36-8.88 &  05:52:41.39   &  +08:41:16.90  & SCOPE& No & 12.0\\
{ }	&G198.69-09.12 &  05:52:29.61   &  +08:16:45.00  &  SCOPE& Yes & 10.8 \\
{ }	&G199.11-10.32 &  05:49:6.28   &  +07:20:26.00  & SCOPE&  No & 12.1\\
 	{ }	&G199.15-10.46 &  05:48:42.18   &  +07:14:3.90  &  SCOPE& No & 10.4\\
 	{ }	&G200.34-10.97 &  05:49:9.61   &  +05:55:50.00  & SCOPE& Yes & 10.9\\
 	{ }	&G200.47-10.36 &  05:51:35.27   &  +06:08:56.70  &  SCOPE& No & 8.4\\
 	{ }	&G200.48-10.59 &  05:50:46.63   &  +06:01:18.20  &  SCOPE& No & 6.6\\
Orion B	&  G201.52-11.08 & 05:51:04.70  &  +04:53:23.10 & SCOPE & Yes & 9.4\\
{ }	&  G201.72-11.22 &  05:50:56.64 &  +04:39:32.20 & SCOPE & Yes & 8.6\\
{ }	&  G203.21-11.20 &  05:53:46.07 &  +03:23:41.40 & Archive & Yes & 15.0\\
{ }	&  G204.4-11.3A2 &  05:55:37.50 &  +02:11:15.30 & SCOPE & Yes & 13.0\\
{ }	&  G204.61-13.58 &  05:48:01.64 &  +01:02:31.20 & Archive  & No  & 13.8\\
{ }	&  G204.65-14.16 &  05:46:02.38 &  +00:43:50.80 & Archive & No  & 13.8\\
{ }	&  G204.80-14.35 &  05:45:39.28 &  +00:30:34.70 & Archive & No  & 13.8\\
{ }	&  G204.8-13.8A1&  05:47:26.03 &  +00:42:47.20 & SCOPE  & No  & 11.7\\
{ }	&  G204.85-14.07 &  05:46:43.02 &  +00:36:13.60 & Archive & No  & 13.8\\
{ }	&  G204.97-14.23 &  05:46:24.24 &  +00:25:27.40 & Archive & No  & 11.7\\ 
{ }	&  G205.46-14.56 &  05:46:08.42 &  -00:11:55.42 & Archive  & Yes & 44.5\\
{ }	&  G205.88-16.18 &  05:41:08.70 &  -01:16:06.70 & Archive  & No  & 13.6 \\
{ }	&  G206.12-15.76 &  05:42:45.63 &  -01:16:14.50 & Archive  & Yes & 13.6\\
{ }	&  G206.21-16.17 &  05:41:40.64 &  -01:34:42.97 & Archive  & Yes & 25.3\\
{ }	&  G206.69-16.60 &  05:40:58.88 &  -02:08:36.30 & Archive  & Yes & 18.7\\
{ }	&  G206.93-16.61 &  05:41:33.44 &  -02:18:04.30 & Archive  & Yes & 24.7\\
Orion A	&  G207.36-19.82 &  05:30:47.07 &  -04:09:45.60 & SCOPE  & Yes & 10.2\\
{ }	&  G207.3-19.8A2 &  05:31:04.07 &  -04:15:48.00 & Archive  & Yes & 9.8\\
{ }	&  G208.68-19.20 &  05:35:21.95 &  -05:01:59.60 & Archive  & Yes & 64.2\\
{ }	&  G208.89-20.04 &  05:32:36.76 &  -05:35:01.46 & Archive  & Yes & 34.1\\
{ }	&  G209.05-19.73 &  05:34:06.48 &  -05:34:30.40 & Archive  & Yes & 34.1\\
{ }	&  G209.29-19.65 &  05:34:55.50 &  -05:43:17.70 & Archive  & Yes & 31.3\\
{ }	&  G209.55-19.68 &  05:35:08.71 &  -05:57:19.80 & Archive  & Yes & 27.6\\
{ }	&  G209.77-19.40 &  05:36:29.17 &  -06:02:24.75 & Archive  & Yes & 13.9\\
{ }	&  G209.77-19.61 &  05:35:45.13 &  -06:09:48.77 & Archive  & Yes & 13.9\\
{ }	&  G209.79-19.80 &  05:35:14.59 &  -06:13:54.77 & Archive  & Yes & 13.9\\
{ }	&  G209.94-19.52 &  05:36:20.04 &  -06:12:34.86 & Archive  & Yes & 13.9\\
{ }	&  G210.20-20.05 &  05:34:55.10 &  -06:41:06.50 & Archive  & No & 20.0\\
{ }	&  G210.37-19.53 &  05:37:00.20 &  -06:36:01.22 & Archive  & Yes & 14.9\\
{ }	&  G210.49-19.79 &  05:36:22.02 &  -06:45:31.22 & Archive  & Yes & 22.0\\
{ }	&  G210.78-19.85 &  05:36:36.57 &  -07:05:30.20 & Archive & No  & 16.6 \\
{ }	&  G210.82-19.47 &  05:38:00.64 &  -06:57:17.83 & Archive  & Yes & 15.6\\
{ }	&  G210.82-19.70 &  05:37:13.72 &  -07:03:12.60 & Archive  & No  & 15.6\\
{ }	&  G210.97-19.33 &  05:38:45.67 &  -06:59:56.82 & Archive  & Yes & 14.3\\
{ }	&  G211.01-19.54 &  05:38:03.19 &  -07:07:51.18 & Archive  & Yes & 14.9\\
{ }	&  G211.16-19.33 &  05:38:58.55 &  -07:11:23.12 & Archive  & Yes & 10.3\\
{ }	&  G211.47-19.27 &  05:39:56.46 &  -07:30:24.31 & Archive  & Yes & 26.1\\
{ }	&  G211.72-19.25 &  05:40:17.99 &  -07:34:23.00 & Archive  & Yes & 11.0\\
{ }	&  G212.10-19.15 &  05:41:21.44 &  -07:53:54.70 & Archive  & Yes & 16.4\\
{ }	&  G212.84-19.45 &  05:41:29.52 &  -08:41:28.00 & Archive  & Yes & 20.0\\
{ }	&  G212.90-19.64 &  05:40:54.33 &  -08:47:18.90 & Archive  & No  & 20.0\\
{ }	&  G213.09-19.23 &  05:42:42.66 &  -08:46:08.60 & Archive  &  No  & 20.0\\
{ }	&  G213.21-18.31 &  05:46:16.68 &  -08:28:26.00 & SCOPE  &  No  & 15.4\\
{ }	&  G213.21-19.84 &  05:40:41.87 &  -09:08:32.50 & Archive  &  No  & 21.1\\
{ }	&  G215.44-16.38 &  05:57:02.78 &  -09:32:32.80 & SCOPE & Yes  & 9.9\\
{ }	&  G215.87-17.62 &  05:53:14.85 &  -10:26:33.10  &SCOPE & Yes & 9.0\\
\enddata
\end{deluxetable}

\clearpage

\begin{longrotatetable}
\begin{deluxetable}{llcccccccccccc}
\tablecaption{Properties of cores detected at 850 $\micron$ and associated YSOs in the $\lambda$ Orionis cloud \label{tab:list_cores_Lambda}}
\tablewidth{900pt}
\tabletypesize{\scriptsize}
\tablehead{
\colhead{PGCC} & \colhead{ID} & \colhead{Classification} &
\colhead{R.A.(J2000)} & \colhead{Dec.(J2000)} & \colhead{size} & 
\colhead{$ N_{\rm H_{2}}$} & \colhead{$ n_{\rm H_{2}}$} & 
\colhead{$M$} & \colhead{$M_{\rm J}$} &\multicolumn{4}{c}{associated YSOs}\\ 
\cline{2-14}\\
\colhead{} & \colhead{} & \colhead{} &  
\colhead{($^h$ $^m$ $^s$)} & \colhead{($^\circ$ $\arcmin$ $\arcsec$)} & \colhead{(pc)} &
\colhead{ (10$^{22}$ cm$^{-2}$)} & \colhead{(10$^{5}$ cm$^{-3}$)} & \colhead{(M$_{\sun}$)} & \colhead{(M$_{\sun}$)} &
\colhead{$L_{\rm bol}$ ($L_{\sun}$)} & \colhead{$T_{\rm bol}$ (K)} & \colhead{$\alpha_{3.6-22}$} & \colhead{Class}
} 
\startdata
G190.15-13.75	& North	&starless 		&05:19:01.11 & 13:08:05.78 	& 0.13	&10.6$\pm$1.8		& 3.3$\pm$0.6		&1.20$\pm$0.37	& 0.19$\pm$0.03	 &\nodata&\nodata&\nodata&\nodata \\  
{ }			& South	&starless 		&05:18:40.55 & 13:01:41.64	& 0.06	&10.2$\pm$1.3		& 3.1$\pm$0.5		&0.25$\pm$0.03	& 0.21$\pm$0.05	 &\nodata&\nodata&\nodata&\nodata\\
G191.90-11.21	&North 	&starless 		&05:31:28.99 & 12:58:55.00	&0.10	&2.5$\pm$0.3		& 0.8$\pm$0.2    	&0.34$\pm$0.07	&2.66$\pm$0.36	&\nodata&\nodata&\nodata&\nodata \\ 
{ }			&South	&starless 		&05:31:31.73 & 12:56:14.99	&0.10	&7.2$\pm$0.5		&2.2$\pm$0.5		&0.89$\pm$0.09 	&0.55$\pm$0.14 	&\nodata&\nodata&\nodata&\nodata \\ 
G192.12-10.90	&North	&starless 		&05:33:02.64 & 12:57:53.59 	&0.18	&6.1$\pm$0.2		&1.9$\pm$0.3		&0.97$\pm$0.30	&0.59$\pm$0.16	 &\nodata&\nodata&\nodata&\nodata\\
{ }			&South	&protostellar 	&05:32:52.52 & 12:55:08.60	&0.04	&5.6$\pm$0.3		&1.7$\pm$0.7		&0.07$\pm$0.01	&\nodata			 & 2.47$\pm$ 0.17 & 220.6 $\pm$7.5 & -0.28 & flat\\
G192.12-11.10	&		&starless 		&05:32:19.54 & 12:49:40.19	&0.12	&17.5$\pm$1.6		&5.4$\pm$0.2		&2.21$\pm$0.33	&0.13$\pm$0.05 	&\nodata&\nodata&\nodata&\nodata \\ 
G192.32-11.88	&North	&starless 		&05:29:54.47 & 12:16:56.00	&0.06	&11.0$\pm$1.9		&3.4$\pm$0.4		&0.49$\pm$0.05	&0.37$\pm$0.09	 &\nodata&\nodata&\nodata&\nodata\\
{ }			&South	&starless 		&05:29:54.74 & 12:16:32.00	&0.05	&9.5$\pm$1.2		&2.9$\pm$0.2		&0.22$\pm$0.02	&0.46$\pm$0.13	&\nodata&\nodata&\nodata&\nodata \\ 
G196.92-10.37	&		&protostellar 	&05:44:29.56 & 09:08:50.20	&0.22	&18.2$\pm$1.3		&5.6$\pm$0.1		&5.41$\pm$0.32	&\nodata 			& 14.87 $\pm$ 2.53 & 67.9$\pm$ 5.0 & 1.48 & 0	\\
G198.69-09.12	&North1	&starless 		&05:52:29.61 & 08:15:37.04	&0.09	&6.4$\pm$0.4		&2.0$\pm$0.4		&0.43$\pm$0.07	&0.46$\pm$0.13 	 &\nodata&\nodata&\nodata&\nodata \\ 
{ }			&North2	&starless 		&05:52:25.30 & 08:15:08.78	&0.06	&6.7$\pm$0.1		&2.1$\pm$0.4		&0.21$\pm$0.02	&0.43$\pm$0.13	&\nodata&\nodata&\nodata&\nodata \\ 
{ }			&South	&protostellar 	&05:52:23.66 & 08:13:37.18	&0.12	&15.5$\pm$1.2		&4.8$\pm$0.9		&1.78$\pm$0.25	&\nodata 			& 18.44 $\pm$ 0.39	& 101.7 $\pm$ 2.4& 0.82  & I\\
G200.34-10.97	&North	&protostellar 	&05:49:03.71 & 05:57:55.74	&0.09	&8.2$\pm$1.2		&2.5$\pm$0.6		&0.77$\pm$0.06	&\nodata 			& 1.15 $\pm$ 0.14 & 64.8 $\pm$ 5.2 & 1.41  & 0\\
{ }			&South	&protostellar 	&05:49:07.74 & 05:55:36.22	&0.11	&6.9$\pm$0.3		&2.1$\pm$0.5		&0.84$\pm$0.11	&\nodata 			& 0.13 $\pm$ 0.04 & 161.3 $\pm$ 11.3 & 0.18  & flat\\
\enddata
\end{deluxetable}
\end{longrotatetable}

\clearpage

\begin{longrotatetable}
\begin{deluxetable}{llcccccccccccc}
\tablecaption{Properties of cores detected at 850 $\micron$ and associated YSOs in the Orion A cloud \label{tab:list_cores_OrionA}}
\tablewidth{900pt}
\tabletypesize{\scriptsize}
\tablehead{
\colhead{PGCC} & \colhead{ID} & \colhead{Classification} &
\colhead{R.A.(J2000)} & \colhead{Dec.(J2000)} & \colhead{size} & 
\colhead{$ N_{\rm H_{2}}$} & \colhead{$ n_{\rm H_{2}}$} & 
\colhead{$M$} & \colhead{$M_{\rm J}$} &\multicolumn{4}{c}{associated YSOs}\\ 
\cline{2-14}\\
\colhead{} & \colhead{} & \colhead{} &  
\colhead{($^h$ $^m$ $^s$)} & \colhead{($^\circ$ $\arcmin$ $\arcsec$)} & \colhead{(pc)} &
\colhead{ (10$^{22}$ cm$^{-2}$)} & \colhead{(10$^{5}$ cm$^{-3}$)} & \colhead{(M$_{\sun}$)} & \colhead{(M$_{\sun}$)} &
\colhead{$L_{\rm bol}$ ($L_{\sun}$)} & \colhead{$T_{\rm bol}$ (K)} & \colhead{$\alpha_{3.6-22}$} & \colhead{Class}
} 
\startdata
G207.36-19.82		&North1	&starless		&05:30:50.94 & -04:10:35.60 	& 0.06	&35.3$\pm$3.5		& 5.7$\pm$1.1		&1.13$\pm$0.46	& 0.08$\pm$0.03		&\nodata&\nodata&\nodata&\nodata \\ 
{ }				&North2	&starless		&05:30:50.67 & -04:10:15.60 	& 0.04	&30.0$\pm$3.1		& 4.9$\pm$0.9		&0.44$\pm$0.14	& 0.11$\pm$0.02		&\nodata&\nodata&\nodata&\nodata\\
{ }				&North3	&starless 		&05:30:46.40 & -04:10:27.60 	& 0.04	&25.0$\pm$1.8		& 4.1$\pm$0.8		&0.47$\pm$0.15 	& 0.14$\pm$0.08		&\nodata&\nodata&\nodata&\nodata\\ 
{ }				&North4	&starless   	&05:30:44.81 & -04:10:27.62 	& 0.03	&25.8$\pm$1.9		& 4.2$\pm$0.8		&0.15$\pm$0.05 	& 0.13$\pm$0.08		&\nodata&\nodata&\nodata&\nodata\\
{ }				&South	&starless		&05:30:46.81 & -04:12:29.39	& 0.19	&6.5$\pm$0.4		& 1.1$\pm$0.2		&2.01 $\pm$0.91	& 1.05$\pm$0.13 		& \nodata&\nodata&\nodata&\nodata \\ 
G207.3-19.8A2		&North1 	&starless		&05:31:03.40 & -04:15:46.00	&0.06	&7.0$\pm$0.5		& 1.1$\pm$0.2		&0.16 $\pm$0.05	&0.93$\pm$0.40		&\nodata&\nodata&\nodata&\nodata\\
{ }				&North2 	&starless		&05:31:02.06 & -04:14:57.00	&0.08	&7.3$\pm$0.5		& 1.2$\pm$0.2		&0.36 $\pm$0.18	&0.87$\pm$0.27		&\nodata&\nodata&\nodata&\nodata\\
{ }				&North3 	&starless		&05:30:59.99 & -04:15:39.00	&0.04	&6.4$\pm$0.4		&1.0$\pm$0.2		&0.10 $\pm$0.03	&1.07$\pm$0.53		&\nodata&\nodata&\nodata&\nodata\\
{ }				&South	&starless		&05:31:03.27 & -04:17:00.00	&0.16	&3.3$\pm$0.2		&0.5$\pm$0.1		&0.73 $\pm$0.08	&2.89$\pm$1.12		&\nodata&\nodata&\nodata&\nodata\\
G208.68-19.20		&North1	&starless		&05:35:23.37 & -05:01:28.70 	&0.09	&108.7$\pm$10.8	&17.7$\pm$3.4		&7.60 $\pm$2.53	&0.03$\pm$0.02		& \nodata&\nodata&\nodata&\nodata \\ 
{ }				&North2	&starless   	&05:35:20.45 & -05:00:52.95 	&0.05	&116.3$\pm$19.8	&18.9$\pm$3.6		&2.22 $\pm$1.15	&0.03$\pm$0.01		&\nodata&\nodata&\nodata&\nodata	\\
{ }				&North3	&starless 		&05:35:18.03 & -05:00:20.64 	&0.05	&116.6$\pm$10.2	&18.9$\pm$3.7		&2.59 $\pm$1.50	&0.03$\pm$0.01		&\nodata&\nodata&\nodata&\nodata \\ 
{ }				&South	&starless 		&05:35:26.33 & -05:03:56.70	&0.09	&46.5$\pm$4.4		&7.6$\pm$1.4		&3.47 $\pm$1.03	&0.12$\pm$0.03		& \nodata&\nodata&\nodata&\nodata \\ 
G208.89-20.04		&East	&protostellar 	&05:32:48.40 & -05:34:47.14	&0.13	&26.3$\pm$2.4		&4.3$\pm$0.3		&3.86 $\pm$0.29	&\nodata				& 0.60$\pm$ 0.02 & 308.9 $\pm$ 2.6& 1.31 & I\\
{ }				&West	&starless	  	&05:32:42.22 & -05:35:58.95	&0.18	&27.5$\pm$2.1		&4.5$\pm$0.8		&8.22 $\pm$0.15	&0.13$\pm$0.04		&\nodata&\nodata&\nodata&\nodata\\				
G209.05-19.73		&North	&starless 		&05:34:03.96 & -05:32:42.48	&0.23	&5.7$\pm$0.8		&0.9$\pm$0.6		&2.83$\pm$0.15 	&1.90$\pm$1.52		&\nodata&\nodata&\nodata&\nodata	\\
{ }				&South	&starless		&05:34:03.12 & -05:34:10.98	&0.14	&9.6$\pm$0.9		&1.6$\pm$0.4		&1.65$\pm$0.29 	& 0.88$\pm$0.07		&\nodata&\nodata&\nodata&\nodata	\\
G209.29-19.65		&North1	&starless		&05:35:00.25 & -05:40:02.40	&0.07	&14.7$\pm$1.3		&2.4$\pm$0.8		&0.70$\pm$0.10 	& 0.54$\pm$0.42		&\nodata&\nodata&\nodata&\nodata	\\
{ }				&North2	&starless		&05:34:57.30 & -05:41:44.40	&0.16	&12.6$\pm$1.5		&2.1$\pm$0.3		&3.05 $\pm$0.11	& 0.68$\pm$0.23		&\nodata&\nodata&\nodata&\nodata	\\
{ }				&North3	&starless		&05:34:54.75 & -05:43:34.40	&0.19	&12.7$\pm$1.2		&2.1$\pm$0.6		&4.55 $\pm$0.57	& 0.67$\pm$0.44		&\nodata&\nodata&\nodata&\nodata	\\
{ }				&South1	&starless		&05:34:55.99 & -05:46:03.20	&0.07	&31.3$\pm$2.3		&5.1$\pm$0.4		&1.49$\pm$0.26 	& 0.17$\pm$0.02		&\nodata&\nodata&\nodata&\nodata	\\
{ }				&South2	&starless		&05:34:53.81 & -05:46:12.80	&0.08	&38.3$\pm$6.3		&6.2$\pm$0.5		&2.31$\pm$0.97 	& 0.13$\pm$0.06		&\nodata&\nodata&\nodata&\nodata	\\
{ }				&South3	&starless		&05:34:49.87 & -05:46:11.60	&0.06	&16.4$\pm$2.5		&2.7$\pm$0.2		&0.50$\pm$0.05 	& 0.46$\pm$0.40		&\nodata&\nodata&\nodata&\nodata	\\
G209.55-19.68		&North1	&starless		&05:35:08.76 & -05:55:50.39	&0.16	&4.1$\pm$0.2		&0.7$\pm$0.1		&0.74$\pm$0.04 	&2.67$\pm$ 1.38		&\nodata&\nodata&\nodata&\nodata\\
{ }				&North2	&starless		&05:35:07.01 & -05:56:38.39	&0.05	&19.0$\pm$1.2		&3.1$\pm$0.6		&0.28 $\pm$0.07	&0.27$\pm$ 0.07		&\nodata&\nodata&\nodata&\nodata\\
{ }				&North3	&protostellar	&05:35:01.45 & -05:55:27.39	&0.12	&23.3$\pm$5.4		&3.8$\pm$0.5		&2.20$\pm$0.38 	&\nodata				&0.60$\pm$ 0.01& 337.6 $\pm$ 3.7& 0.42 & I\\
{ }				&South1	&protostellar	&05:35:13.25 & -05:57:54.38	&0.23	&7.6$\pm$0.3		&1.2$\pm$0.2		&2.67 $\pm$0.16	&\nodata		 		&1.26$\pm$ 0.06 &175.3 $\pm$ 6.0& 2.67 & I\\
{ }				&South2	&starless		&05:35:11.73 & -05:57:02.65	&0.12	&30.1$\pm$13.2	&4.9$\pm$0.9		&1.71 $\pm$0.16	&0.14$\pm$0.07		&\nodata&\nodata&\nodata&\nodata\\
{ }				&South3	&starless	 	&05:35:08.96 & -05:58:26.38	&0.16	&10.0$\pm$1.5		&1.6$\pm$0.5		&1.73$\pm$0.21 	&0.70 $\pm$0.01		&\nodata&\nodata&\nodata&\nodata\\				
G209.77-19.40		&East1	&protostellar	&05:36:32.45 & -06:01:16.68	&0.05	&27.8$\pm$3.7		&4.5$\pm$0.6		&0.35$\pm$0.09 	& \nodata				&0.65$\pm$ 0.02& 338.3 $\pm$ 4.0& 0.56 & I\\
{ }				&East2	&starless 	 	&05:36:32.19 & -06:02:04.68	&0.07	&39.4$\pm$4.8		&6.4$\pm$0.9		&1.20$\pm$0.56 	& 0.09$\pm$0.02		& \nodata&\nodata&\nodata&\nodata \\
{ }				&East3	&starless 		&05:36:35.94 & -06:02:44.66	&0.04	&33.3$\pm$1.8		&5.4$\pm$0.7		&0.26$\pm$0.12 	& 0.11$\pm$0.06		&\nodata&\nodata&\nodata&\nodata \\
{ }				&West	&starless		&05:36:21.19 & -06:01:32.73	&0.09	&10.6$\pm$2.9		&1.7$\pm$0.2		&0.50$\pm$0.07 	&0.64$\pm$0.36 		&\nodata&\nodata&\nodata&\nodata\\	
G209.77-19.61		&East	&protostellar 	&05:35:52.23 & -06:10:00.80	&0.11	&14.8$\pm$3.6		&2.4$\pm$0.2		&1.09$\pm$0.13 	&\nodata 				&17.18$\pm$ 0.26 & 350.6 $\pm$ 8.3 & 0.22 & flat\\			
{ }				&West	&starless		&05:35:37.21 & -06:09:44.80	&0.06	&14.7$\pm$1.1		&2.4$\pm$0.5		&0.31$\pm$0.17 	&0.36 $\pm$0.11		&\nodata&\nodata&\nodata&\nodata\\
G209.79-19.80		&East	&protostellar	&05:35:21.92 & -06:13:08.77	&0.10	&9.8$\pm$2.1		&1.6$\pm$0.2		&0.77$\pm$0.39	&\nodata				& 0.24$\pm$ 0.05 & 107.7$\pm$ 9.5 & 1.01 & I\\			
{ }				&West	&starless		&05:35:11.19 & -06:14:00.73	&0.27	&26.9$\pm$2.9		&4.4$\pm$0.4		&7.06$\pm$2.56 	&0.15 $\pm$0.11		&\nodata&\nodata&\nodata&\nodata \\	
G209.94-19.52		&North	&starless		&05:36:11.55 & -06:10:44.76	&0.13	&19.1$\pm$4.5		&3.1$\pm$0.5		&2.81$\pm$0.28	&0.33 $\pm$0.07		&\nodata&\nodata&\nodata&\nodata \\				
{ }				&South1	&starless		&05:36:24.96 & -06:14:04.71	&0.16	&15.1$\pm$2.9		&2.5$\pm$0.2		&3.52$\pm$0.09	&0.47 $\pm$0.08		&\nodata&\nodata&\nodata&\nodata \\	
{ }				&South2	&protostellar	&05:36:37.03 & -06:15:00.64	&0.07	&12.0$\pm$1.7		&1.9$\pm$0.8		&0.53$\pm$0.08 	&\nodata 				&0.17$\pm$ 0.02 & 137.1 $\pm$ 16.4 & 0.05 & flat\\
G210.37-19.53		&North	&starless		&05:36:55.03 & -06:34:33.19	&0.06	&6.7$\pm$0.3		&1.1$\pm$0.2		&0.28$\pm$0.02 	&1.26$\pm$0.17		&\nodata&\nodata&\nodata&\nodata \\
{ }				&South	&starless	 	&05:37:00.55 & -06:37:10.16	&0.15	&11.1$\pm$0.9		&1.8$\pm$0.7		&2.45 $\pm$0.23	&0.60$\pm$0.09 		&\nodata&\nodata&\nodata&\nodata \\
G210.49-19.79		&East1	&protostellar	&05:36:25.28 & -06:44:42.79	&0.10	&56.6$\pm$10.4	&9.2$\pm$0.1		&2.62$\pm$0.13 	&\nodata				&9.35 $\pm$ 0.27 & 699.6 $\pm$ 6.8 & 0.00 & flat \\			
{ }				&East2	&protostellar 	&05:36:23.13 & -06:46:10.79	&0.16	&83.9$\pm$10.1	&13.6$\pm$0.6		&10.36$\pm$0.43	&\nodata 				& 1.32 $\pm$ 0.5  & 168.1 $\pm$ 6.1 & 1.27 & I \\
{ }				&West	&protostellar 	&05:36:18.40 & -06:45:26.79	&0.11	&113.4$\pm$13.7	&18.4$\pm$0.2		&7.12$\pm$0.25 	&\nodata 				& 41.62 $\pm$ 7.63 & 51.0 $\pm$ 8.8 & 1.29 & 0\\
G210.82-19.47		&North1	&protostellar	&05:37:56.56 & -06:56:35.15	&0.14	&3.8$\pm$0.6		&0.6$\pm$0.6		&0.72$\pm$0.08 	&\nodata 				& 1.10 $\pm$ 0.06 & 213.1 $\pm$ 8.4 & 1.22 & I \\			
{ }				&North2	&starless	 	&05:37:59.84 & -06:57:09.86	&0.06	&4.5$\pm$0.8		&0.7$\pm$0.1		&0.15$\pm$0.05 	&2.92$\pm$0.23 		&\nodata&\nodata&\nodata&\nodata \\
{ }				&South	&starless	 	&05:38:04.09 & -06:58:17.70	&0.08	&4.1$\pm$0.8		&0.7$\pm$0.8		&0.22$\pm$0.03 	&3.29$\pm$0.47 		&\nodata&\nodata&\nodata&\nodata \\					
G210.97-19.33		&North	&protostellar	&05:38:40.36 & -06:58:21.90	&0.12	&8.2$\pm$0.5		&1.3$\pm$0.2		&0.88 $\pm$0.22	&\nodata 				& 22.31 $\pm$ 0.32 & 283.7 $\pm$ 6.3& -0.28 & flat\\		
{ }				&South1	&protostellar 	&05:38:49.46 & -07:01:17.90	&0.17	&21.4$\pm$3.4		&3.5$\pm$0.2		&4.63 $\pm$1.01	&\nodata 				& 0.73 $\pm$ 0.03 & 208.3 $\pm$ 5.5 & 2.02 & I\\
{ }				&South2	&starless	 	&05:38:45.30 & -07:01:04.41	&0.07	&12.5$\pm$2.8		&2.0$\pm$0.1		&0.40$\pm$0.09 	&0.44$\pm$0.07		&\nodata&\nodata&\nodata&\nodata \\
G211.01-19.54		&North	&starless		&05:37:57.23 & -07:06:56.72	&0.10	&25.3$\pm$1.2		&4.1$\pm$0.2		&2.33$\pm$0.36 	&0.19$\pm$0.03 		&\nodata&\nodata&\nodata&\nodata \\					
{ }				&South	&protostellar 	&05:37:59.04 & -07:07:32.73	&0.07	&24.5$\pm$3.3		&4.0$\pm$0.1		&1.13$\pm$0.32 	&\nodata		 		& 0.43 $\pm$ 0.02  & 48.3 $\pm$ 2.8 & 1.17 & 0 \\
G211.16-19.33		&North1	&protostellar 	&05:39:11.80 & -07:10:29.87	&0.11	&9.3$\pm$0.7		&1.5$\pm$0.2		&0.79 $\pm$0.05	&\nodata 				& 0.71 $\pm$ 0.05& 363.3 $\pm$ 21.7 & 0.54 & I\\
{ }				&North2	&protostellar 	&05:39:05.89 & -07:10:37.88	&0.10	&7.5$\pm$0.6		&1.2$\pm$0.2		&0.50$\pm$0.20 	&\nodata				& 3.93$\pm$ 0.03 & 69.9 $\pm$ 0.5 & 2.41 & 0\\
{ }				&North3	&starless		&05:39:02.26 & -07:11:07.89	&0.10	&6.2$\pm$0.5		&1.0$\pm$0.1		&0.41$\pm$0.12 	&1.20$\pm$0.17		&\nodata&\nodata&\nodata&\nodata \\
{ }				&North4	&starless	 	&05:38:55.67 & -07:11:25.93	&0.13	&4.3$\pm$0.3		&0.7$\pm$0.1		&0.51$\pm$0.04 	&2.12$\pm$0.98 		&\nodata&\nodata&\nodata&\nodata \\
{ }				&North5	&starless		&05:38:46.00 & -07:10:41.90	&0.12	&7.1$\pm$0.5		&1.2$\pm$0.2		&0.64$\pm$0.18 	&0.98$\pm$0.16		&\nodata&\nodata&\nodata&\nodata \\
{ }				&South	&protostellar 	&05:39:02.94 & -07:12:49.89	&0.06	&10.3$\pm$1.8		&1.7$\pm$0.3		&0.22$\pm$0.10 	&\nodata 				& 4.27 $\pm$ 0.14 & 83.3 $\pm$ 2.3 & 1.14 & I\\			G211.47-19.27		&North	&starless		&05:39:57.27 & -07:29:38.28	&0.07	&40.9$\pm$2.6		&6.6$\pm$0.3		&1.66$\pm$0.43 	&0.07$\pm$0.01 		&\nodata&\nodata&\nodata&\nodata \\
{ }				&South	&starless	 	&05:39:55.92 & -07:30:28.28	&0.01	&71.8$\pm$4.7		&11.7$\pm$0.3		&12.25$\pm$3.18 	&0.03$\pm$0.01		&\nodata&\nodata&\nodata&\nodata \\
G211.72-19.25		&North	&protostellar	&05:40:13.72 & -07:32:16.80	&0.03	&9.1$\pm$1.3		&1.5$\pm$0.5		&0.07$\pm$0.06 	& \nodata 				& 35.94 $\pm$ 0.01 & 56.4 $\pm$ 0.1 & 0.17 & flat \\
{ }				&South1	&starless		&05:40:19.04 & -07:34:28.79	&0.12	&8.3$\pm$1.5		&1.3$\pm$0.5		&1.15$\pm$0.82 	&0.79$\pm$0.09		&\nodata&\nodata&\nodata&\nodata \\
{ }				&South2	&starless		&05:40:21.18 & -07:34:04.00	&0.10	&6.8$\pm$0.5		&1.1$\pm$0.4		&0.62 $\pm$0.21 	&1.07$\pm$ 0.11		&\nodata&\nodata&\nodata&\nodata\\
G212.10-19.15		&North1	&starless		&05:41:21.56 & -07:52:27.66	&0.18	&19.0$\pm$1.4		&3.1$\pm$0.4		&3.52$\pm$1.68 	&0.18$\pm$0.03		&\nodata&\nodata&\nodata&\nodata \\
{ }				&North2	&starless 		&05:41:23.98 & -07:53:48.49	&0.12	&18.1$\pm$1.4		&2.9$\pm$0.4		&1.65 $\pm$0.72	&0.19$\pm$0.04		&\nodata&\nodata&\nodata&\nodata \\
{ }				&North3	&protostellar	&05:41:24.82 & -07:55:08.48	&0.11	&20.2$\pm$1.5		&3.3$\pm$0.4		&1.36$\pm$0.78 	&\nodata 				&9.06 $\pm$ 0. 05 & 85.7 $\pm$ 0.9 & 1.87 & I\\	
{ }				&South	&protostellar	&05:41:26.39 & -07:56:51.81	&0.15	&18.6$\pm$1.3		&3.0$\pm$0.4		&2.41$\pm$1.00 	&\nodata 				&1.66 $\pm$ 0.16 & 50.6 $\pm$ 1.3 & 0.87 & 0\\
G212.84-19.45		&North	&protostellar	&05:41:32.14 & -08:40:10.94	&0.10	&19.7$\pm$1.4		&3.2$\pm$0.5		&1.36 $\pm$0.50	&\nodata				& 3.63 $\pm$ 0.02 & 48.2 $\pm$ 0.3 & 0.64 & 0\\
{ }				&South	&protostellar 	&05:41:29.70 & -08:43:00.22	&0.09	&6.8$\pm$0.4		&1.1$\pm$0.2		&0.32$\pm$0.11 	&\nodata				& 7.09 $\pm$ 1.30 & 56.9 $\pm$ 4.1 & 1.60 & 0\\
G215.44-16.38		&		&starless		&05:56:58.45 & -09:32:42.30	&0.06	&6.0$\pm$0.4		&1.0$\pm$0.2		&0.19$\pm$0.04 	&1.21 $\pm$0.16		&\nodata&\nodata&\nodata&\nodata\\	
G215.87-17.62		&North	&starless		&05:53:41.91 & -10:24:02.00	&0.26	&6.1$\pm$0.4		&1.0$\pm$0.2		&4.05 $\pm$0.84	&1.18 $\pm$0.16 		&\nodata&\nodata&\nodata&\nodata \\				
{ }				&Middle	&starless		 &05:53:32.41& -10:25:06.07	&0.24	&4.9$\pm$0.3		&0.8$\pm$0.1		&2.69$\pm$0.64 	&1.68$\pm$0.18		&\nodata&\nodata&\nodata&\nodata \\
{ }				&South	&starless		 &05:53:26.43& -10:27:26.02	&0.20	&5.8$\pm$0.4		&0.9$\pm$0.1		&2.23$\pm$0.69 	&1.28 $\pm$0.16		&\nodata&\nodata&\nodata&\nodata \\\enddata
\end{deluxetable}
\end{longrotatetable}

\clearpage

\begin{longrotatetable}
\begin{deluxetable}{llcccccccccccc}
\tablecaption{Properties of cores detected at 850 $\micron$ and associated YSOs in the Orion B cloud  \label{tab:list_cores_OrionB}}
\tablewidth{900pt}
\tabletypesize{\scriptsize}
\tablehead{
\colhead{PGCC} & \colhead{ID} & \colhead{Classification} &
\colhead{R.A.(J2000)} & \colhead{Dec.(J2000)} & \colhead{size} & 
\colhead{$ N_{\rm H_{2}}$} & \colhead{$ n_{\rm H_{2}}$} & 
\colhead{$M$} & \colhead{$M_{\rm J}$} &\multicolumn{4}{c}{associated YSOs}\\ 
\cline{2-14}\\
\colhead{} & \colhead{} & \colhead{} &  
\colhead{($^h$ $^m$ $^s$)} & \colhead{($^\circ$ $\arcmin$ $\arcsec$)} & \colhead{(pc)} &
\colhead{ (10$^{22}$ cm$^{-2}$)} & \colhead{(10$^{5}$ cm$^{-3}$)} & \colhead{(M$_{\sun}$)} & \colhead{(M$_{\sun}$)} &
\colhead{$L_{\rm bol}$ ($L_{\sun}$)} & \colhead{$T_{\rm bol}$ (K)} & \colhead{$\alpha_{3.6-22}$} & \colhead{Class}
} 
\startdata
G201.52-11.08		& 		&protostellar 	&05:50:59.01 & 04:53:53.10 	& 0.05	&7.9$\pm$0.5		& 3.3$\pm$0.8		&0.14$\pm$0.04&	 \nodata 			& 3.03 $\pm$ 0.37 & 66.4 $\pm$ 3.9 & 1.00 & 0\\ 
G201.72-11.22		& 		&starless		&05:50:54.53 & 04:37:42.60 	& 2.25	&3.2$\pm$0.2		& 1.3$\pm$0.4		&1.83$\pm$0.21& 	0.87$\pm$0.37 		&\nodata&\nodata&\nodata&\nodata\\
G203.21-11.20		& East1	&starless 		&05:53:51.11 & 03:23:04.90 	& 0.12	&23.9$\pm$3.5		& 9.8$\pm$0.6		&2.50$\pm$0.43 &	 0.03$\pm$0.01		&\nodata&\nodata&\nodata&\nodata\\ 
				& East2	&starless   	&05:53:47.90 & 03:23:08.90 	& 0.12	&25.0$\pm$3.7		& 10.3$\pm$0.7	&2.65$\pm$1.73 &	 0.03$\pm$0.01		&\nodata&\nodata&\nodata&\nodata\\
				& West1	&starless		&05:53:42.83 & 03:22:32.90	& 0.12	&26.8$\pm$2.7		& 11.1$\pm$1.9		&2.81$\pm$0.13 &	 0.03$\pm$0.01		&\nodata&\nodata&\nodata&\nodata\\
				& West2	&starless		&05:53:39.62 & 03:22:24.90	& 0.10	&32.5$\pm$2.8		& 13.4$\pm$1.5	&2.51$\pm$0.19 &	 0.02$\pm$0.01		&\nodata&\nodata&\nodata&\nodata\\
G204.4-11.3A2		& East	&starless 		&05:55:38.43 & 02:11:33.30 	& 0.08	&47.4$\pm$3.4		& 19.6$\pm$1.6	&2.97$\pm$0.23 &	 0.01$\pm$0.01		&\nodata&\nodata&\nodata&\nodata\\
				& West	&starless		 &05:55:35.49 & 02:11:01.30 	& 0.05	&35.5$\pm$1.5		& 14.7$\pm$1.5	&0.99$\pm$0.24 &	 0.02$\pm$0.01		&\nodata&\nodata&\nodata&\nodata\\
G205.46-14.56		& North1	&starless		&05:46:05.49 &-00:09:32.35	& 0.03	&41.4$\pm$2.8		& 17.0$\pm$1.9	&0.39$\pm$0.06 &	 0.02$\pm$0.01		&\nodata&\nodata&\nodata&\nodata\\
				& North2	&protostellar	&05:46:07.89 &-00:10:01.97	& 0.06	&54.9$\pm$3.3		& 22.6$\pm$1.5	&1.79 $\pm$0.38&	 \nodata			& 15.74 $\pm$ 0.13 & 117.4 $\pm$ 1.6 & 1.08 & I\\				
				& North3	&protostellar	&05:46:08.06 &-00:10:43.57	& 0.09	&87.0$\pm$5.5		& 35.9$\pm$1.7	&6.21 $\pm$1.33& 	\nodata			& 0.50 $\pm$ 0.04 & 182.8  $\pm$ 16.6 & 1.39 & I\\
				& Middle1	&starless		&05:46:09.65 &-00:12:12.92	& 0.05	&54.5$\pm$3.3		& 22.5$\pm$1.1	&1.04 $\pm$0.16& 	0.01$\pm$0.01 		&\nodata&\nodata&\nodata&\nodata\\
				& Middle2	&starless		&05:46:07.49 &-00:12:22.42	& 0.03	&46.6$\pm$2.9		& 19.2$\pm$1.9	&0.48$\pm$0.09 & 	0.01$\pm$0.01 		&\nodata&\nodata&\nodata&\nodata\\
				& Middle3	&starless		&05:46:07.37 &-00:11:53.35	& 0.06	&47.6$\pm$3.1		& 19.6$\pm$1.9	&1.36 $\pm$0.25& 	0.01$\pm$0.01 		&\nodata&\nodata&\nodata&\nodata\\
				& South1	&protostellar	&05:46:07.11 &-00:13:34.57	& 0.13	&66.5$\pm$4.3		& 27.4$\pm$1.7	&11.35$\pm$2.18& 	\nodata			&9.66 $\pm$ 0.36&19.7 $\pm$ 0.7& 1.18 & 0\\
				& South2	&protostellar	&05:46:04.49 &-00:14:18.87	& 0.03	&48.0$\pm$3.0		& 19.8$\pm$2.9	&0.46$\pm$0.10 & 	\nodata 			& 23.31 $\pm$ 4.19 & 273.2 $\pm$ 5.2 & -0.11 & flat\\
				& South3	&protostellar	&05:46:03.54 &-00:14:49.34	& 0.04	&49.5$\pm$3.1		& 20.4$\pm$1.9	&0.86 $\pm$0.16& 	\nodata 			& 2.95 $\pm$ 0.05 & 346.6  $\pm$ 3.1 &  0.85 & I\\
G206.12-15.76		&		&starless		&05:42:45.27 &-01:16:11.37	&0.14	&26.3$\pm$2.1		&10.8$\pm$1.3		&3.86$\pm$1.73 &	0.03$\pm$0.01		 &\nodata&\nodata&\nodata&\nodata\\				
G206.21-16.17		&North	&starless		&05:41:39.28 &-01:35:52.86	&0.21	&11.5$\pm$2.8		&4.7$\pm$0.8		&4.97$\pm$0.05 &	0.17$\pm$0.09 		&\nodata&\nodata&\nodata&\nodata\\
				&South	&starless		&05:41:34.23 &-01:37:28.76	&0.11	&8.7$\pm$0.2		&3.6$\pm$0.8		&1.05$\pm$0.46 &	0.25$\pm$0.09		&\nodata&\nodata&\nodata&\nodata\\
G206.69-16.60		&North	&starless		&05:40:58.08 &-02:07:28.29	&0.10	&8.0$\pm$0.6		&3.3$\pm$0.8		&0.75$\pm$0.03 &	0.32$\pm$0.08		&\nodata&\nodata&\nodata&\nodata\\
				&South	&starless		&05:40:58.62 &-02:08:40.72	&0.17	&12.4$\pm$1.5		&5.1$\pm$0.5		&3.48 $\pm$0.11&	0.17$\pm$0.10		&\nodata&\nodata&\nodata&\nodata\\
G206.93-16.61		&East1	&starless		&05:41:40.54 &-02:17:04.29	&0.12	&41.3$\pm$6.6		&17.1$\pm$2.9		&5.37$\pm$0.54 &	0.03$\pm$0.01		&\nodata&\nodata&\nodata&\nodata\\
				&East2	&protostellar	&05:41:37.32 &-02:17:16.29	&0.10	&45.5$\pm$3.4		&18.8$\pm$2.4		&4.29$\pm$0.16 &	\nodata		&38.23 $\pm$ 3.79 & 250.9$\pm$3.9 &1.48& I\\
				&West1	&starless		&05:41:25.57 &-02:16:04.30	&0.08	&24.4$\pm$3.8		&10.1$\pm$2.3		&1.55$\pm$0.23 &	0.06$\pm$0.02		&\nodata&\nodata&\nodata&\nodata\\
				&West2	&starless		&05:41:26.39 &-02:18:16.40	&0.04	&64.2$\pm$9.3		&26.5$\pm$2.3		&1.37$\pm$0.23&	0.01$\pm$0.01		&\nodata&\nodata&\nodata&\nodata\\
				&West3	&starless		&05:41:25.04 &-02:18:08.11	&0.06	&99.1$\pm$1.9		&40.8$\pm$4.7		&3.38 $\pm$0.15&	0.01$\pm$0.01		&\nodata&\nodata&\nodata&\nodata\\
				&West4	&starless		&05:41:25.84 &-02:19:28.42	&0.08	&21.9$\pm$1.5		&9.0$\pm$0.2		&1.25 $\pm$0.67&	0.07$\pm$0.02		&\nodata&\nodata&\nodata&\nodata\\
				&West5	&starless		&05:41:28.77 &-02:20:04.30	&0.11	&57.0$\pm$3.6		&23.5$\pm$1.6		&7.10$\pm$0.72 &	0.02$\pm$0.01		 &\nodata&\nodata&\nodata&\nodata\\
				&West6	&starless		&05:41:29.57 &-02:21:16.06	&0.08	&20.1$\pm$1.5		&8.3$\pm$0.9		&1.15$\pm$0.15 &	0.08$\pm$0.02		&\nodata&\nodata&\nodata&\nodata\\
\enddata
\end{deluxetable}
\end{longrotatetable}

\clearpage

\begin{deluxetable}{cc cccc c cccc c cccc c cccc } 
\tablecaption{Statistics of core properties \label{tab:statistics_cores}} 
\tabletypesize{\scriptsize}\tablecolumns{21}
\tablewidth{0pc} \setlength{\tabcolsep}{0.05in}
 \tablehead{
\colhead{ Cloud} &  \colhead{} &\multicolumn{4}{c}{$N_{\rm H_{2}}$    (10$^{22}$ cm$^{-2})$} && \multicolumn{4}{c}{$n_{\rm H_{2}}$  (10$^{5}$ cm$^{-3})$} &&
\multicolumn{4}{c}{core mass ($M_{\sun}$)} && \multicolumn{4}{c}{core size (pc)}   \\
 \cline{3-6} \cline{8-11} \cline{13-16} \cline{18-21} 
 \colhead{}          &  \colhead{}  & \colhead{min} & \colhead{max} & \colhead{mean} & \colhead{median} & \colhead{} & \colhead{min} & \colhead{max} & \colhead{mean} & \colhead{median} &\colhead{}  & \colhead{min} & \colhead{max} & \colhead{mean} & \colhead{median}  & \colhead{} & \colhead{min} & \colhead{max} & \colhead{mean} & \colhead{median} 
}\startdata
$\lambda$ Orionis & & 2.5  & 18.2  & 9.5 &  8.2 && 0.7 & 5.6 &  2.9 & 2.5 & & 0.06 & 5.41 &  1.07 & 0.77 & & 0.03& 0.19 &  0.08 & 0.09 \\ 
Orion A      	& & 3.3  & 116.6  & 23.4 & 14.7  && 0.5 & 18.9 & 3.8 & 3.4 & & 0.07 & 12.25  & 2.39 & 1.18 & &  0.02 & 0.26 &  0.11& 0.11 \\
Orion B		& & 3.2  & 99.1  & 38.4 & 38.4    && 1.3 & 40.8 & 15.6 & 15.8 & & 0.14  & 11.36  & 2.66 & 1.81	& & 0.03 & 2.25 & 0.16 & 0.10 \\
\enddata
\end{deluxetable}

\begin{deluxetable}{cc ccccc c ccccc} 
\tablecaption{Physical properties of PGCCs from the PGCC catalog \label{tab:PGCCs}} 
\tabletypesize{\scriptsize}\tablecolumns{13}
\tablewidth{0pc} \setlength{\tabcolsep}{0.05in}
 \tablehead{
 \colhead{Cloud} &   \colhead{number of PGCCs} &\multicolumn{5}{c}{median} & \colhead{} & \multicolumn{5}{c}{mean}\\
\cline{3-7} \cline{9-13}
 \colhead{}&   \colhead{}  &  
 \colhead{$N_{\rm H_{2}}$ } &  \colhead{ $T_{\rm d}$ } &  \colhead{$\beta$ } &  \colhead{$n_{\rm H_{2}}$ } &  \colhead{$M_{clump}$ } &  
 \colhead{} & \colhead{ $N_{\rm H_{2}}$} &  \colhead{$T_{\rm d}$} &  \colhead{$\beta$} &  \colhead{ $n_{\rm H_{2}}$} &  \colhead{$M_{clump}$} \\
 \colhead{} &  \colhead{} &   
 \colhead{(10$^{20}$ cm$^{-2}$) }  &  \colhead{(K)} &  \colhead{} &  \colhead{(10$^{2}$ cm$^{-3}$)} &  \colhead{(M$_{\sun}$)} & \colhead{}&
  \colhead{ (10$^{20}$ cm$^{-2}$) }  &  \colhead{(K)} & \colhead{} &  \colhead{(10$^{2}$ cm$^{-3}$)} & \colhead{ (M$_{\sun}$) }
}
\startdata
$\lambda$ Orionis &  177  & 3.2   & 16.1  & 1.7  & 2.2  & 4.9 & & 6.4 & 16.0 & 1.6 & 4.7 & 8.5  \\
Orion A &        135  & 10.9  & 13.4  & 2.1  & 6.6  & 13.8 & & 28.1 & 13.8 & 2.0 & 18.3 & 30.8  \\
Orion B &       154  & 6.4   & 13.9  & 2.0  & 4.1  & 7.7 &  & 13.4 & 14.0 & 1.9 & 9.1 & 16.8  \\
\enddata
\end{deluxetable}

\begin{deluxetable}{lcccccccc}
\tabletypesize{\scriptsize}\tablecolumns{9}
\tablewidth{0pc} \setlength{\tabcolsep}{0.05in}
\tablecaption{ Statistics of YSOs   }
 \tablehead{
\colhead{Cloud} &  \multicolumn{4}{c}{number of YSOs}& \colhead{} & \colhead{$L_{\rm bol}$  \tablenotemark{a}} & \colhead{$T_{\rm bol}$  \tablenotemark{a}} & \colhead{$\alpha$\tablenotemark{a}} \\
\cline{2-5}  
\colhead{} & \colhead{total}  & \colhead{Class 0} &  \colhead{Class I} & \colhead{Flat} & \colhead{ } & \colhead{  ($L_{\sun}$)} & \colhead{(K) } & \colhead{}
}\startdata
$\lambda$ Orionis     & 5       & 2  & 1 & 2 && 2.47 & 101.7 & 0.82\\
Orion A  			& 22	    & 6 & 11 & 5 && 1.49 & 152.6 & 1.14 \\
Orion B  			& 7       & 2 & 4 & 1 && 9.66 & 182.8  & 1.08\\
\enddata
\tablecomments{}
\tablenotetext{a}{The median value of all YSOs in a given cloud. }
\label{tab:YSOs}
\end{deluxetable}

\begin{deluxetable}{llcccc}
\tabletypesize{\scriptsize}\tablecolumns{6}
\tablewidth{0pc} \setlength{\tabcolsep}{0.05in}
\tablecaption{  Two types of sub-clouds in the $\lambda$ Orionis cloud  }
 \tablehead{
\colhead{sub-cloud} & \colhead{Region} & \colhead{Detection rate of PGCCs} & \colhead{Dense gas fraction}  
&\colhead{ core mass  \tablenotemark{a}} & \colhead{$N_{\rm H_{2}}$ \tablenotemark{a} }   \\
\colhead{} & \colhead{} & \colhead{($\%$)} & \colhead{}  &  \colhead{(M$_{\sun}$)} & 
\colhead{ (10$^{22}$ cm$^{-2}$) }
}\startdata
Dense and massive    & B30, B35       & 38  & 0.19 & 1.68 & 11.3  \\
Diffuse and low-mass & rest of the cloud & 8   & 0.05  & 0.96 & 5.7  \\
\enddata
\tablecomments{}
\tablenotetext{a}{ The median value of cores in each sub-cloud.}
\label{tab:Lambda_clouds}
\end{deluxetable}



\clearpage



\end{document}